\documentclass[epjST]{svjour}
\usepackage{graphicx}
\usepackage{wrapfig}
\begin{document} 
\title{Spiral wave chimeras for coupled oscillators with inertia }

\author{ Volodymyr Maistrenko\inst{1} \thanks{\email{maistren@nas.gov.ua}} 
\and  Oleksandr Sudakov\inst{1,2} 
  \and Yuri Maistrenko\inst{1,3}}
\institute{Scientific Center for Medical and Biotechnical Research, NAS of Ukraine, \\   Volodymyrska Str. 54, 01030 Kyiv, Ukraine \and  Taras Shevchenko National University of Kyiv,  Volodymyrska Str. 60, 01030 Kyiv , Ukraine \and Institute of Mathematics, NAS of Ukraine, Tereshchenkivska Str. 3, 01024 Kyiv, Ukraine} 

\date{\today}
 
\abstract
{ We report the appearance and the metamorphoses of spiral wave chimera states in coupled phase oscillators with inertia. First, when the coupling strength is small enough, the system behavior resembles classical two-dimensional (2D) Kuramoto-Shima spiral chimeras with bell-shape frequency characteristic of the incoherent cores [1,2]. As the coupling increases, the cores acquire concentric regions of constant time-averaged frequencies, the chimera becomes quasiperiodic. Eventually, with a subsequent increase in the coupling strength, only one such region is left, i.e., the whole core becomes frequency-coherent. An essential modification of the system behavior occurs, when the parameter point enters the so-called {\it solitary} region. Then, isolated oscillators are normally present on the spiral core background of the chimera states. These solitary oscillators do not participate in the common spiraling around the cores; instead, they start to oscillate with different time-averaged frequencies (Poincar\'e winding numbers). The number and the disposition of solitary oscillators can be any, given by the initial conditions. At a further increase in the coupling, the spiraling disappears, and the system behavior passes to a sort of spatiotemporal chaos.}

\maketitle
\section {Introduction}
Spiral wave chimeras are fascinating two-dimensional (2D) patterns first reported in 2003 by Kuramoto \& Shima~\cite{ks2003,sk2004}.  Manifesting the regular 2D spiraling, this kind of patterns possesses,  nevertheless, finite-sized incoherent cores with a bell-shaped frequency distribution of individual oscillators. Since that, spiral wave chimeras have been intensively studied both numerically and analytically
\cite{kkjm2004,mls2010,OWYYS2012,Nkomo2013,pa2013,XKK2015,L2017,OWK2018,TRT2018,OK2019,T2019}  and got  recently an experimental confirmation in~\cite{TRT2018}. 

In this paper, with the use of a detailed numerical study, we demonstrate the appearance of spiral wave chimeras for  Kuramoto model with inertia,  and we follow their transformations with increasing the coupling strength $\mu$. We begin with a standard chimeric pattern including a number of incoherent bell-shaped cores and then go to the situation where the cores become coherent. The latter means that all in-core oscillators start to rotate with the averaged frequency (Poincar\'e winding number) different from the frequency of the spiral rotation around the cores. The transition includes also intermediate patterns characterized by partially coherent cores (so-called quasiperiodic spiral chimeras, as in \cite{OWK2018}, and it ends eventually in a sort of spatiotemporal chaos.  The scenario is described in details for 4-core chimeras in Ch.2.  Examples of the spiral wave chimeras with a larger number of cores and more involved frequency characteristics are presented in Ch.3.
We find that, depending on the parameters,  spiral cores of the chimera states can be surrounded by the so-called  {\it solitary cloud} which is a collection of randomly distributed isolated oscillators characterized by different averaged frequencies.  Spiral chimeras including  a solitary cloud typically arise in the solitary parameter region \cite{JMK2015,JBLDKM2018}. This is a new type of the behavior  corresponding to weak chimera states \cite{ab2015}. Its peculiarity is that the number and the distribution of solitary oscillators in the cloud can be any,  which is controlled by the initial conditions.

Our model is a two-dimensional array of $N{\times}N$ identical phase oscillators of the Kuramoto--Sakaguchi type model with inertia, where phases $\varphi_{i,j}$ evolve according to the equation

\vspace*{-0.4cm}
\begin{equation}
m\ddot{\varphi}_{ij} + \epsilon \dot{\varphi}_{ij} = \frac{\mu}{|B_{P}(i,j)|} \sum\limits_{(i^{\prime},j^{\prime})\in B_{P}(i,j) }\sin(\varphi_{i^{\prime}j^{\prime}} - \varphi_{ij}- \alpha),
\end{equation}
and where indices $i,j=1, ... , N$ are periodic modulo $N$, $m$ is the mass, $\epsilon$ is the damping coefficient, and $\alpha$ is the phase lag. The network coupling is assumed to be non-local and isotropic:  each oscillator $\varphi_{ij}$ is coupled with equal strength $\mu>0$ to all its nearest neighbors $\varphi_{i^{\prime}j^{\prime}}$  within a range $P$, i.e., to those oscillators falling in the circular neighborhood 

\vspace*{-0.3cm}
$$  B_{P}(i,j):=\{ (i^{\prime},j^{\prime}){:} (i^{\prime}-i)^{2}+(j^{\prime}-j)^{2}\le P^{2}\},$$
where distances $|i^{\prime}-i|$ and  $|j^{\prime}-j|$ are calculated regarding the periodic boundary conditions of the network, and $|B_{P}(i,j)|$ denotes the cardinality of $B_{P}(i,j)$.
Without loss of generality, we put $m =1$ and $\epsilon=0.1$ and will explore the network dynamics varying the phase lag $\alpha$ and coupling strength $\mu$. 

Numerical simulations were performed on the basis of the Runge--Kutta solver DOPRI5 at the computer cluster CHIMERA, http://nll.biomed.kiev.ua/cluster. The Ukrainian Grid Infrastructure provided the distributed cluster resources and the parallel software with graphics processing units~\cite{sls2011,scm2017}.
At the beginning stage of the simulations, random initial conditions were chosen from  the phase range $[0,2\pi]$ and the frequency range $[-\mu/\epsilon, \mu/\epsilon]$, independently for each oscillator. After identifying a pattern, the parameter region for its existence was built by the standard continuation method with a parameter step of about 10\% of the presumed region size; the step was reduced up to 1\% near the boundary. The network size was mostly $N{\times}N=100{\times}100$ and was increased to $N{\times}N=800{\times}800$ in the cases of interest. The simulation time was between $10^3 $ and $10^5$, depending on the task. The averaging frequency of individual oscillators was calculated at the last $10^3$ time units of the simulation. In total, about 27000 network trajectories were computed and analyzed.

\vspace*{-0.4cm}
\section{Spiral wave chimeras with 2 and 4 cores}

Figure 1 illustrates two examples of spiral wave chimera states  with 2  and 4 incoherent cores, respectively.  We find that the states of this kind typically arise in model (1) at small enough values of the coupling parameter $\mu$ and at intermediate values of $\alpha$.  They resemble classical spiral wave chimeras first reported in the pioneering works \cite{ks2003,sk2004}, in particular, for the standard Kuramoto model without inertia (see  \cite{OWYYS2012,pa2013,XKK2015,OWK2018}).  A distinguishing property of the chimeras is the bell-shaped frequency profile of the incoherent cores.  We follow the transformations of a 4-core chimera state with an increase in the coupling strength $\mu$ at fixed $\alpha=0.42$, along the dashed vertical line depicted on the main bifurcation diagram in Fig.~2(a). 

\begin{figure}[ht! ]
\vspace*{-0.4cm}
\begin{center}
 \includegraphics[width=0.43\linewidth]{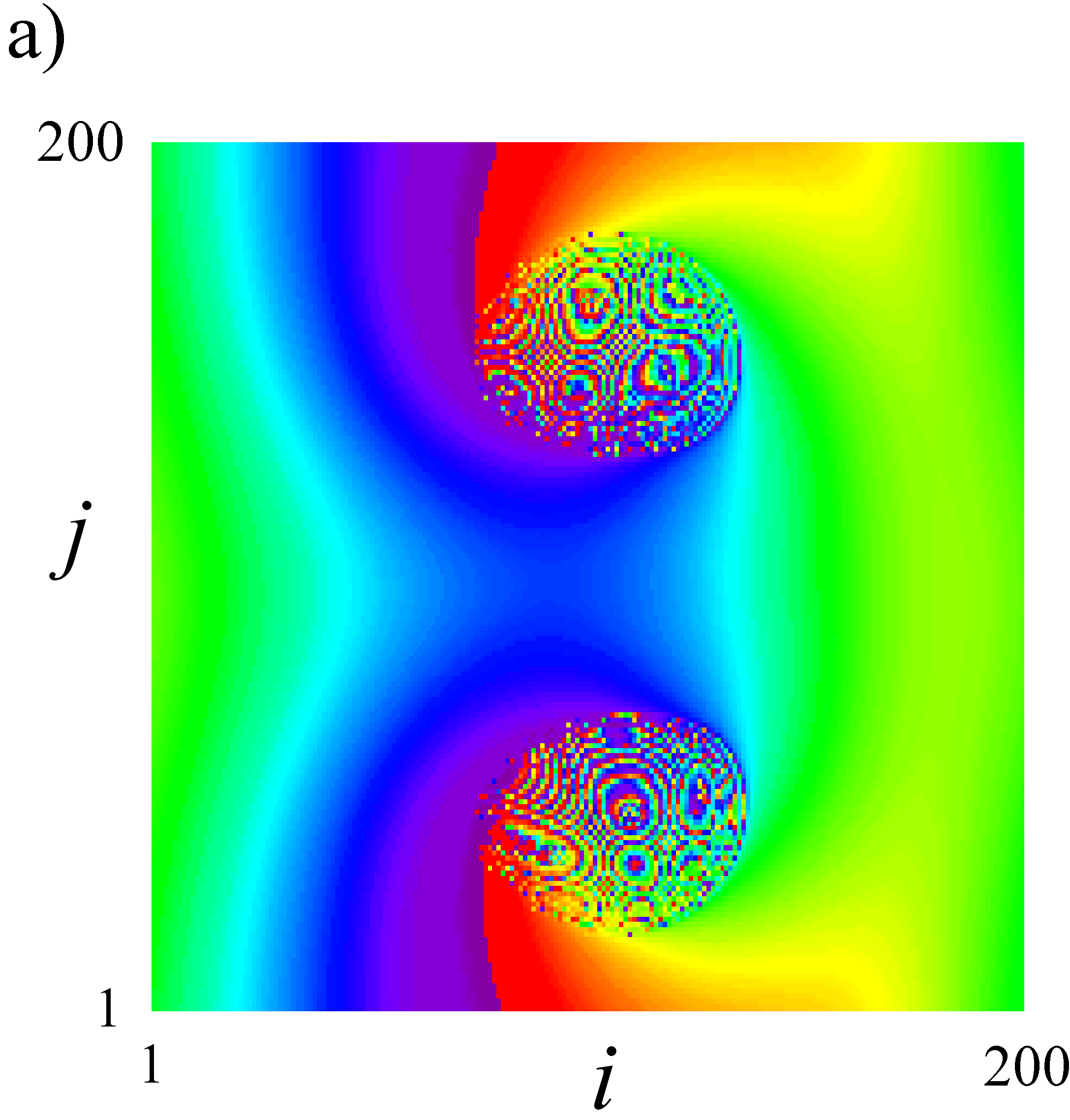}    \ \  \ 
\includegraphics[width=0.43\linewidth]{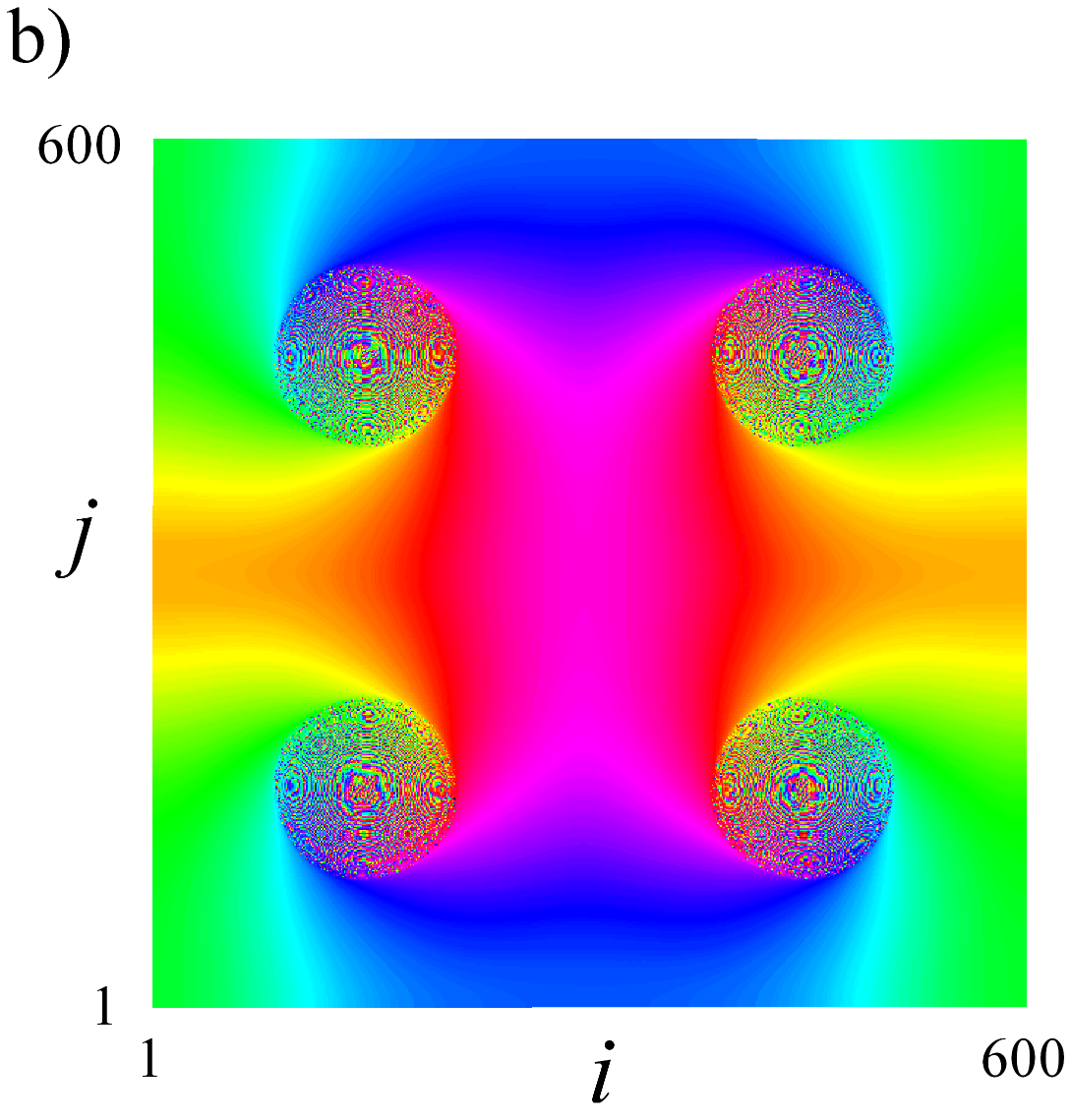} \ \ \  \
\includegraphics[width=0.06\linewidth]{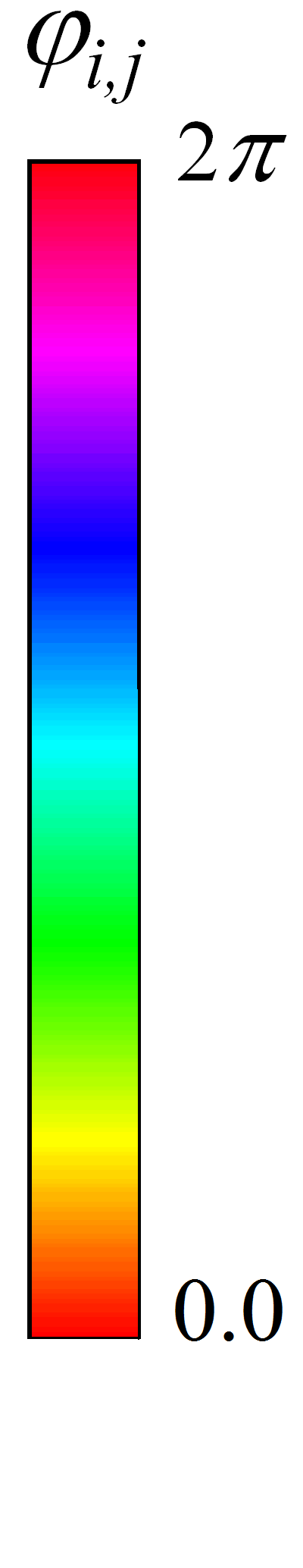}
\caption
{ Phase snapshots of spiral wave chimera states  with incoherent core in model (1).  \quad  (a) 2-core
 chimera at parameters  $\alpha=0.82, \mu=0.005, P=14, N=200$, (b) 4-core chimera 
at parameters $\alpha=0.77, \mu=0.012,  P = 42, N=600$.}
 \label{f1}
\end{center}
\end{figure}

Results of direct numerical simulations of model (1) in the two-parameter plane of the phase lag $\alpha$ and coupling strength $\mu$ are presented in Fig.2 for coupling radius $P = 7$ (a) and $P = 16$ (b). This figure reveals the appearance of regions of 2- and 4-core spiral wave chimera states (shown in blue and green, respectively) at intermediate values of phase lag. Typical shapes of the chimera states are shown in insets.  In the region shown in red color, chimera cores are frequency-coherent, i.e., all oscillators there rotates with the same average frequency. It has to be noted that only regions for 2- and 4-core chimeras are shown in Fig.2. Our simulations yield additional regions for chimeras with 6, 8, and more cores (not shown in Fig.2; illustrative examples can be found in the next chapter).

Alternatively, if the network interactions are not phase-lagged ($\alpha=0$) or the phase lag parameter $\alpha$ is small enough, the network displays full synchronization. The synchronous regime is Lyapunov-stable for all $|\alpha|<\pi/2$ and $\mu>0$.  It co-exists, however, with many other states (some of them are displayed in Fig.2). Then,  its basin of attraction, the so-called 'sync basin', shrinks promptly as $\alpha$ departs from $0$ and approaches $\pi/2$.   We have never obtained the  synchronous state in our simulations at intermediate and large values of  $\alpha$, when starting from random initial conditions; in contrary, numerous other states have been manifested. The 'sync basin' appears to be so tiny that it needs probably a special procedure to allocate it. An importance of this issue for multistable systems of different nature was first pointed out in ~\cite{wsg2006}, and our model (1) represents  a genuine example of this kind. 

In the opposite situation, when the phase lag $\alpha$ is close to $\pi/2$, the so-called {\it solitary states} \cite{JMK2015,JBLDKM2018} arise and occupy the major part of the system parameter space. The solitary state behavior means that some number of the oscillators start to rotate with a different time-averaged frequency as compared to the spiraling background. Parameter region for  the solitary state existence and stability is shown in gray.  As one can observe, the region is separated from the uncoupled $\mu=0$ level, which implies that solitary behavior should be a result of the network iteration and not a consequence of the uncoupled multistability. The solitary region is bounded from the left-down by a homoclinic bifurcation curve denoted by $HB$, where the very first, single-solitary state (i.e. that with only one solitary oscillator) is born in a homoclinic bifurcation (see \cite{JBLDKM2018}  for details). With an increase of the parameters beyond the bifurcation curve, more and more oscillators split up from the synchronous cluster and run into the solitary rotation. Typical shapes of the solitary states are illustrated in snapshots in the right-upper corner of both Fig.2 (a) and (b). 

Of a special interest is the transition between chimera and solitary regions, where our simulations show a complicated behavior resembling the spiral chaos \cite{mbca1996,sk2006}.  It is observed in a thin layer between the two blank regions in the right-down part of Fig.2 (the layer left blank in both (a) and (b) figures). Typical snapshots are shown in insets, see also examples in the next Chapter.   A pink dotted line in (a) indicates the upper boundary, where the spiral chimeric behavior was detected in the simulations. 

It should be noted that though Fig. 2 is obtained for model (1) with rather small $N=100$,  our simulations approve that similar parameter regions exist for larger $N$ and the same relative coupling radius $r = P/N$.  In the simulations, we have verified and confirmed  this fact for $N=200, 400,$ and $800$,  and the respective $P$ such that  $r=0.07$ and $r=0.16$. We expect that this invariance in the system behavior is preserved in the limit $N\to\infty$.

\begin{figure}[ht! ]
\vspace*{-0.6cm}
\begin{center}
  \includegraphics[width=0.6\linewidth]{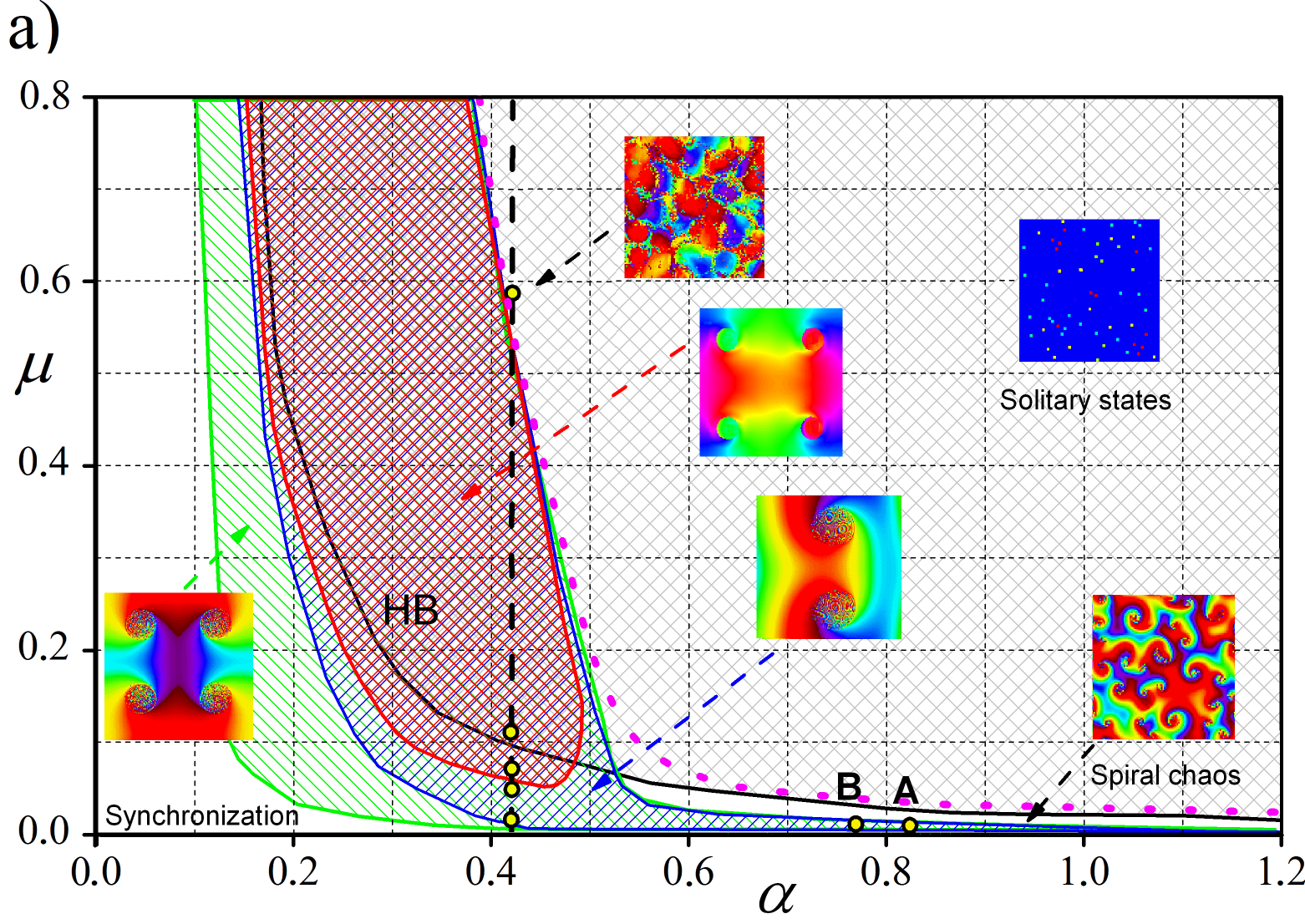} 
  \includegraphics[width=0.6\linewidth]{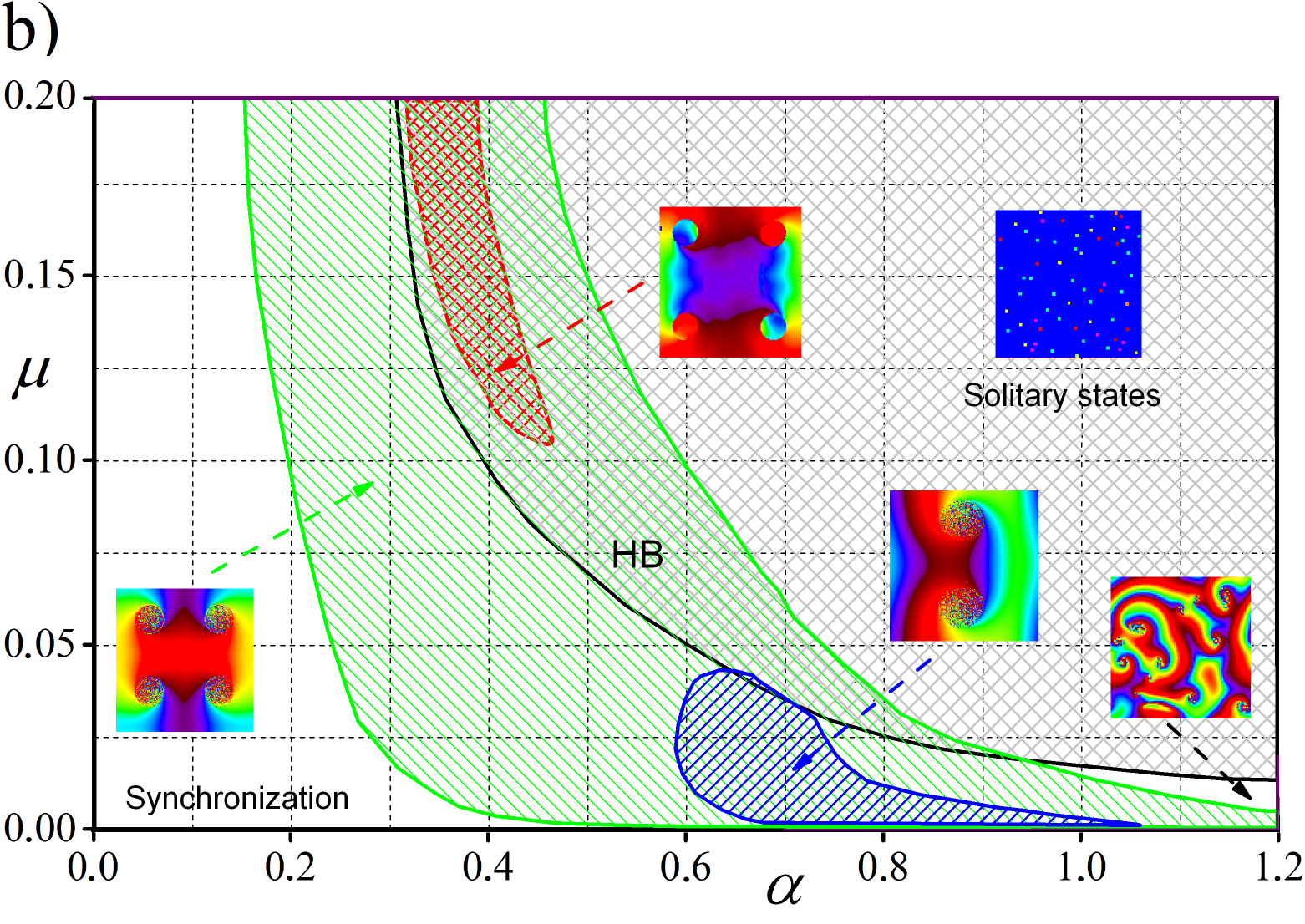} 
\caption{\label{fig:epsart} 
Phase diagram of model (1)  in  $(\alpha, \mu)$ parameter plane with $N=100$ and coupling radius $P=7$ (a) and  $P=17$ (b).  Regions of spiral wave chimeras with  2 and 4 cores are shown in blue and green, respectively. In the red region, the chimera cores are coherent.  Solitary states exist in the light gray region bounded by the bifurcation curve $HB$.  Parameter points $A$ and $B$ in (a) correspond to Fig. 1, left and right snapshots, respectively. Snapshots of typical states are shown in the insets. Note that vertical scale in (b) is four times larger than in (a). Other parameters $m=1$ and $\epsilon=0.1$.}
 \label{f2}
\end{center}
\end{figure}

\begin{wrapfigure}[17]{r}{0.47\linewidth} 
\vspace{-0.2cm}
\begin{center}
 \includegraphics[width=1.0\linewidth]{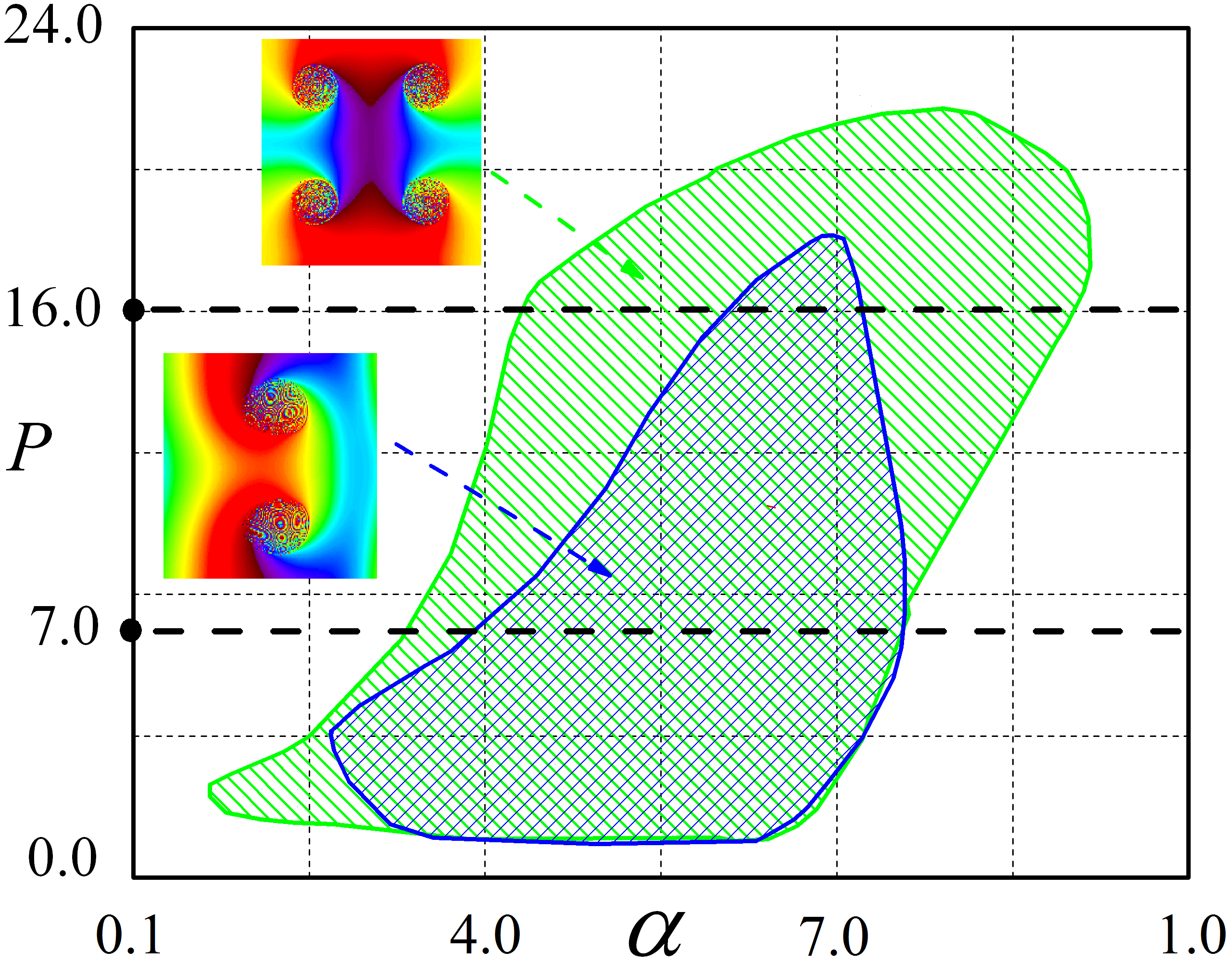}
\caption{ Stability regions of 2- (blue) and 4- (green) core chimera states in the $(\alpha, P)$ parameter plane. Dotted cross-sections correspond to Fig.~2 (a) and (b).  Parameters $ \mu=0.015$ and $N=100$. }
 \label{f3-reg}
\end{center}
\end{wrapfigure}

An important question is how the chimera regions depend on the coupling radius $P$.  Comparing Fig.2 (a) and (b), one can observe that blue region, i.e. that for the 2-core chimeras, shrinks essentially, as $P$ increases from 7 to 16, while  the shape and size of the green 4-core region remains approximately the same. To illustrate this issue  in more details, we present the slice of the chimera regions in the $(\alpha, P)$ parameter plane for the fixed $\mu=0.015$ in Fig.3.  
 As  can be seen, indeed, the 2-core region is rapidly  shrinking with an increase of $P$ and disappears at about $P=18$. The green 4-core region, in contrary, preserves approximately its width for the 
  $P$ interval from $1$  to $18$ and only then starts to rapidly shrink and disappears at about $P=22$.  

Figure 3 demonstrates, therefore, that spiral wave chimeras in model (1) arise at the local, nearest neighboring coupling and are preserved with an increase of the coupling radius $P$ up to some limiting value. This limiting $P$-value of the chimera existence depends on the other parameters and, as we observed in the simulations, never exceeds $25$ (at $N=100$). In other words, each network oscillator should interact with not more than half of its neighbors to get a chance for a spiral chimera to arise in the network (see \cite{OWYYS2012,msom2015}, where similar fact is reported for Kuramoto model without inertia, in both 2D and 3D cases). The other peculiarity
is given by the fact that, as can be seen from Figs. 2 and 3, spiral chimeras are
possible only at intermediate values of the phase-lag parameter $\alpha$. There are no such striking patterns, if $\alpha$ is close to $0$ or $\pi/2$.

\begin{figure}
\begin{center}
 \includegraphics[width=0.27\linewidth]{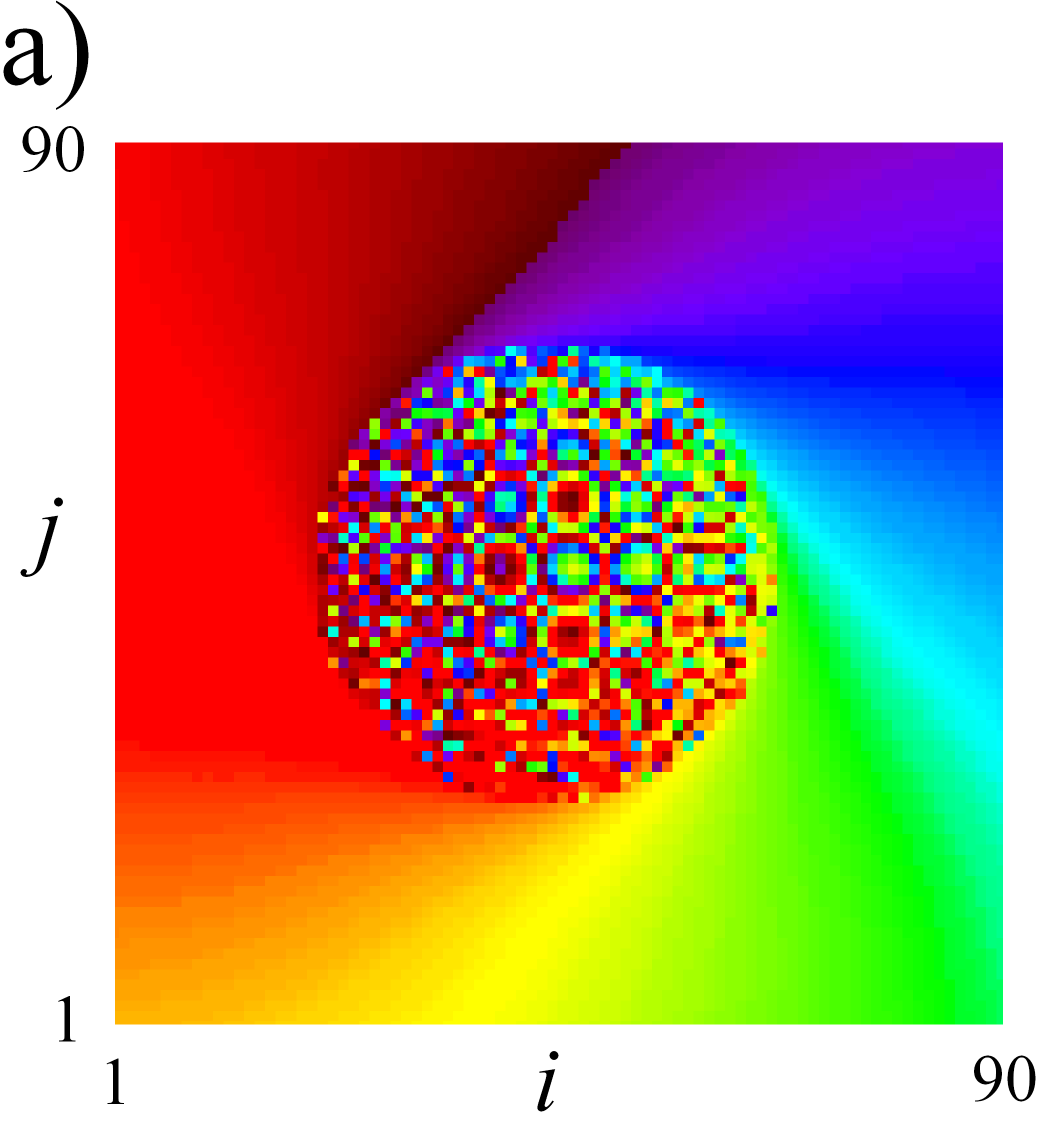} \hspace{0.1cm}
 \includegraphics[width=0.06\linewidth]{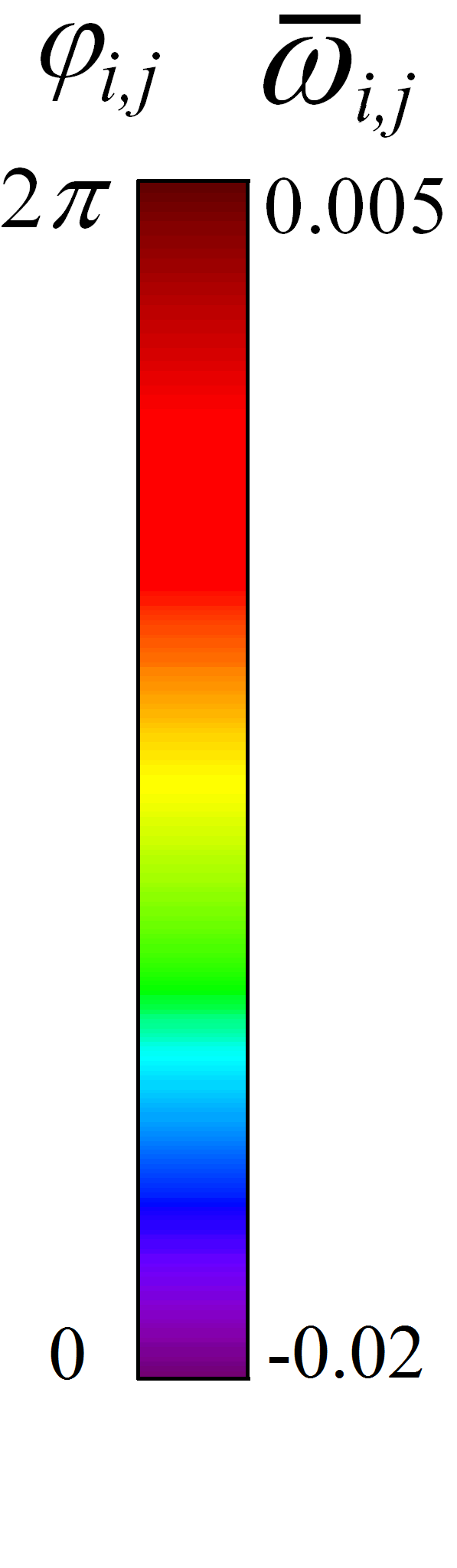}  
\includegraphics[width=0.27\linewidth]{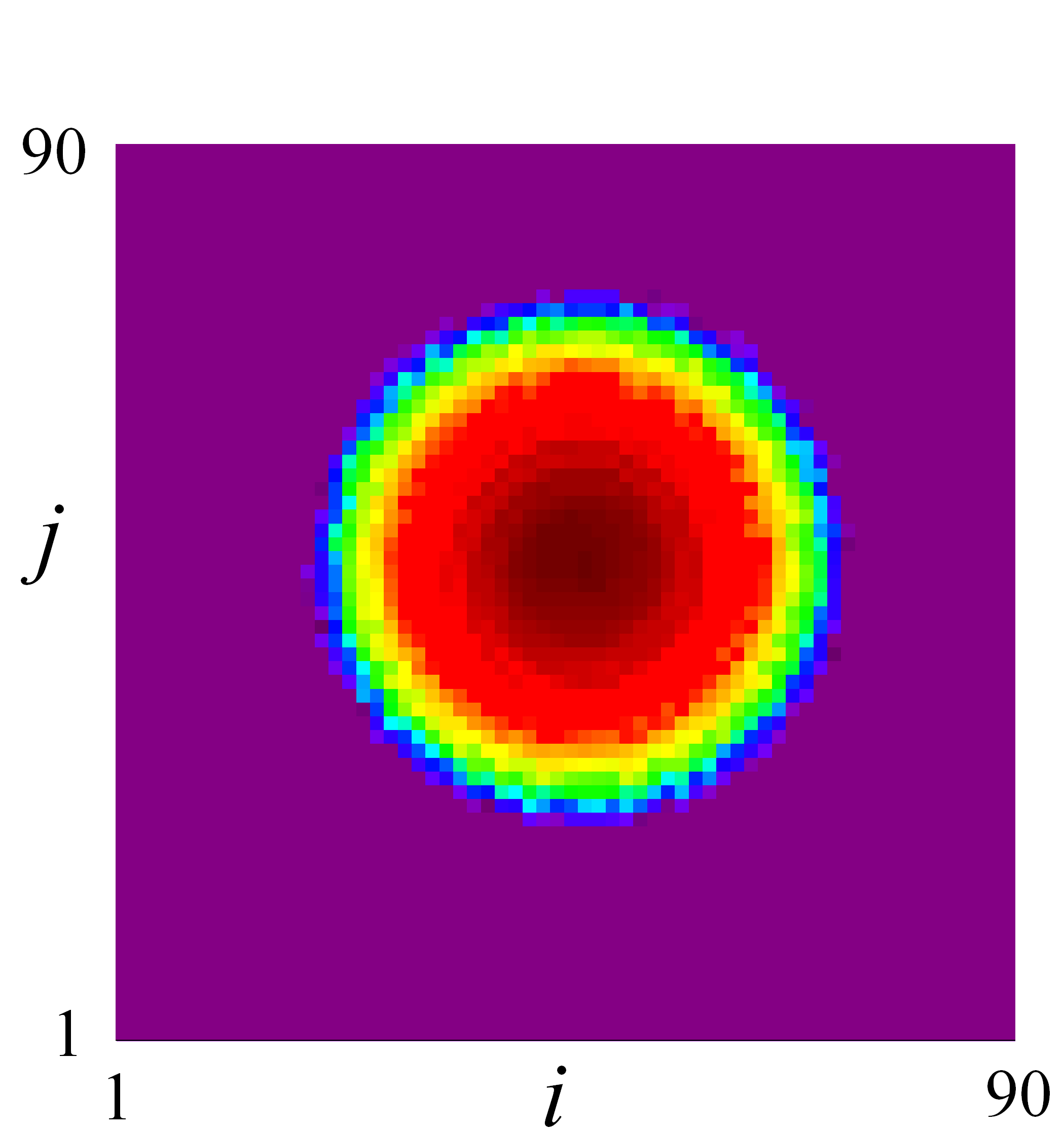} 
\includegraphics[width=0.28\linewidth]{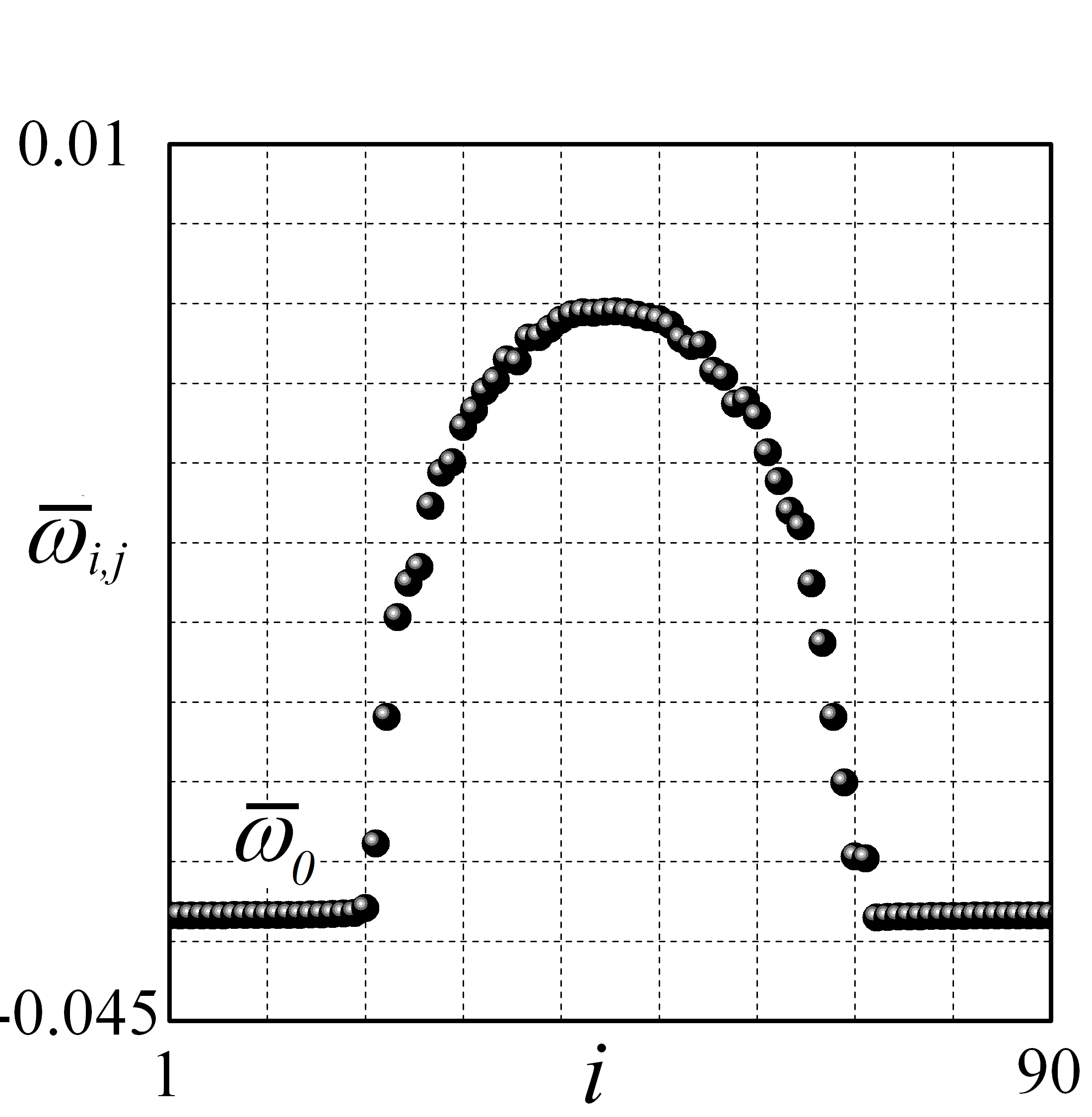} \\

 \includegraphics[width=0.27\linewidth]{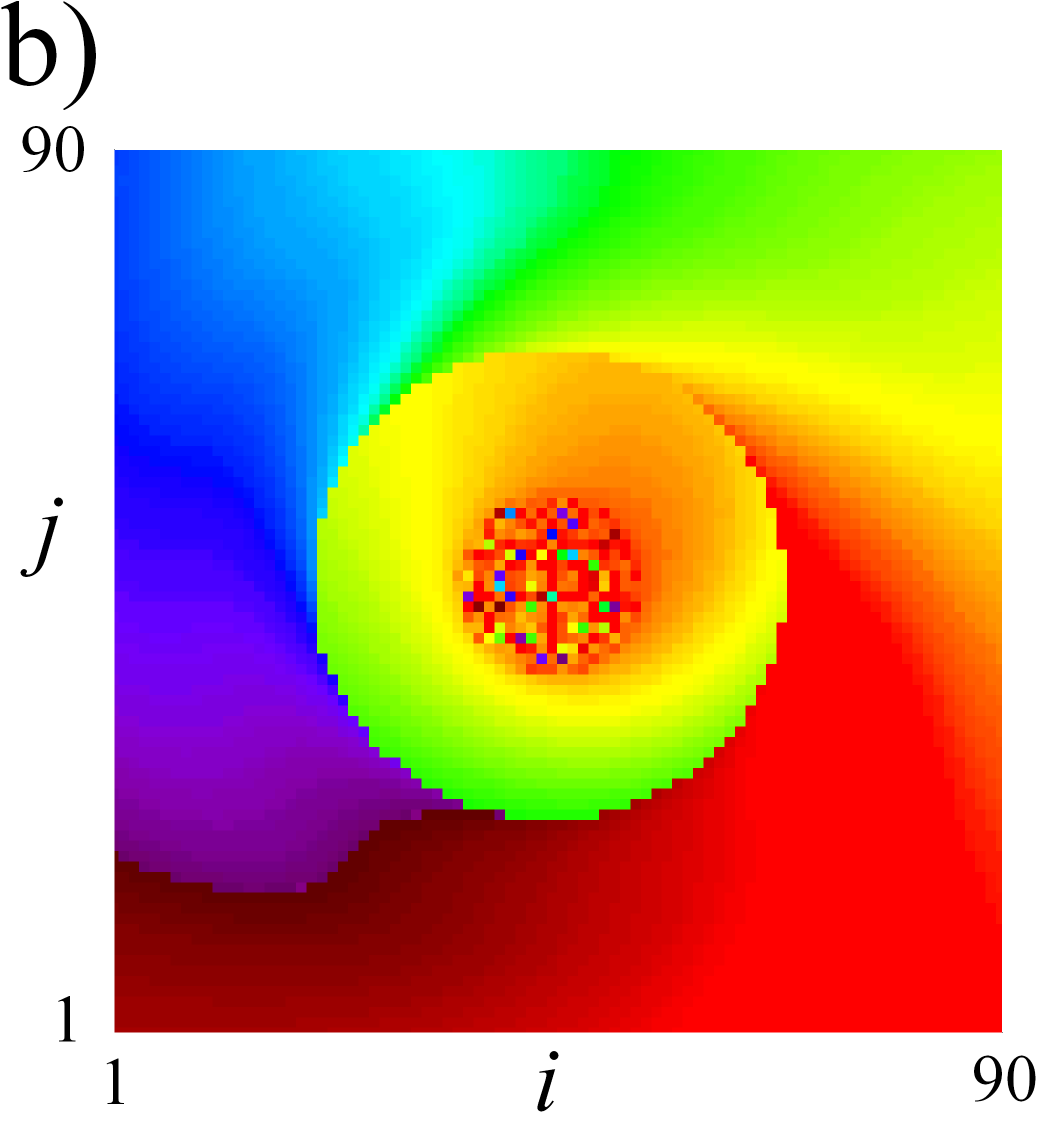}  \hspace{0.1cm}
 \includegraphics[width=0.06\linewidth]{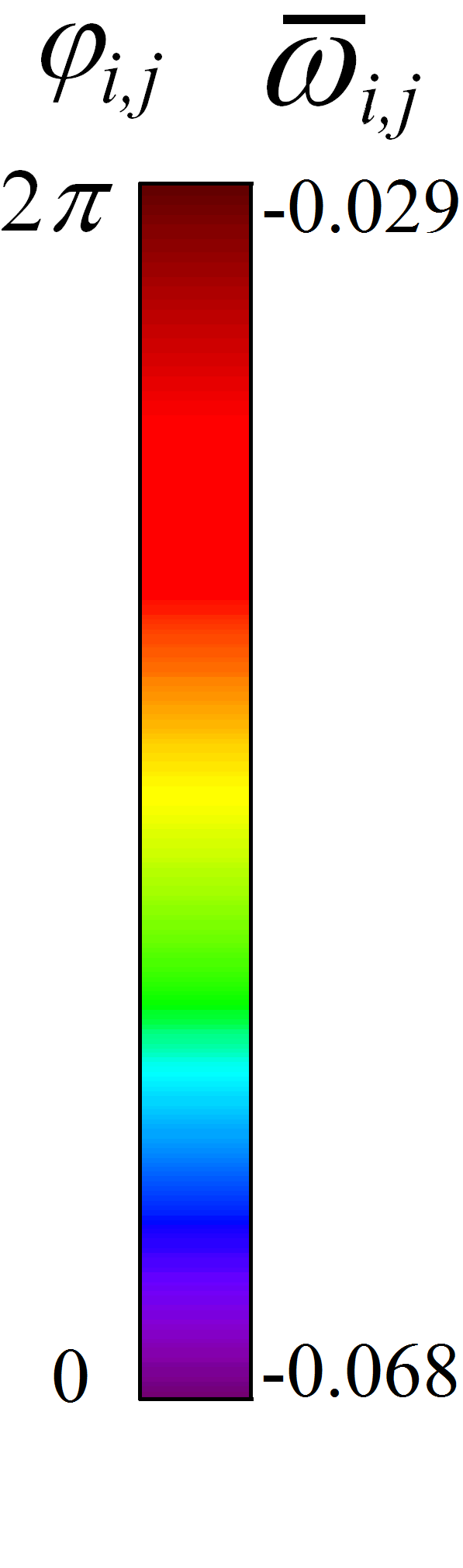}  
\includegraphics[width=0.27\linewidth]{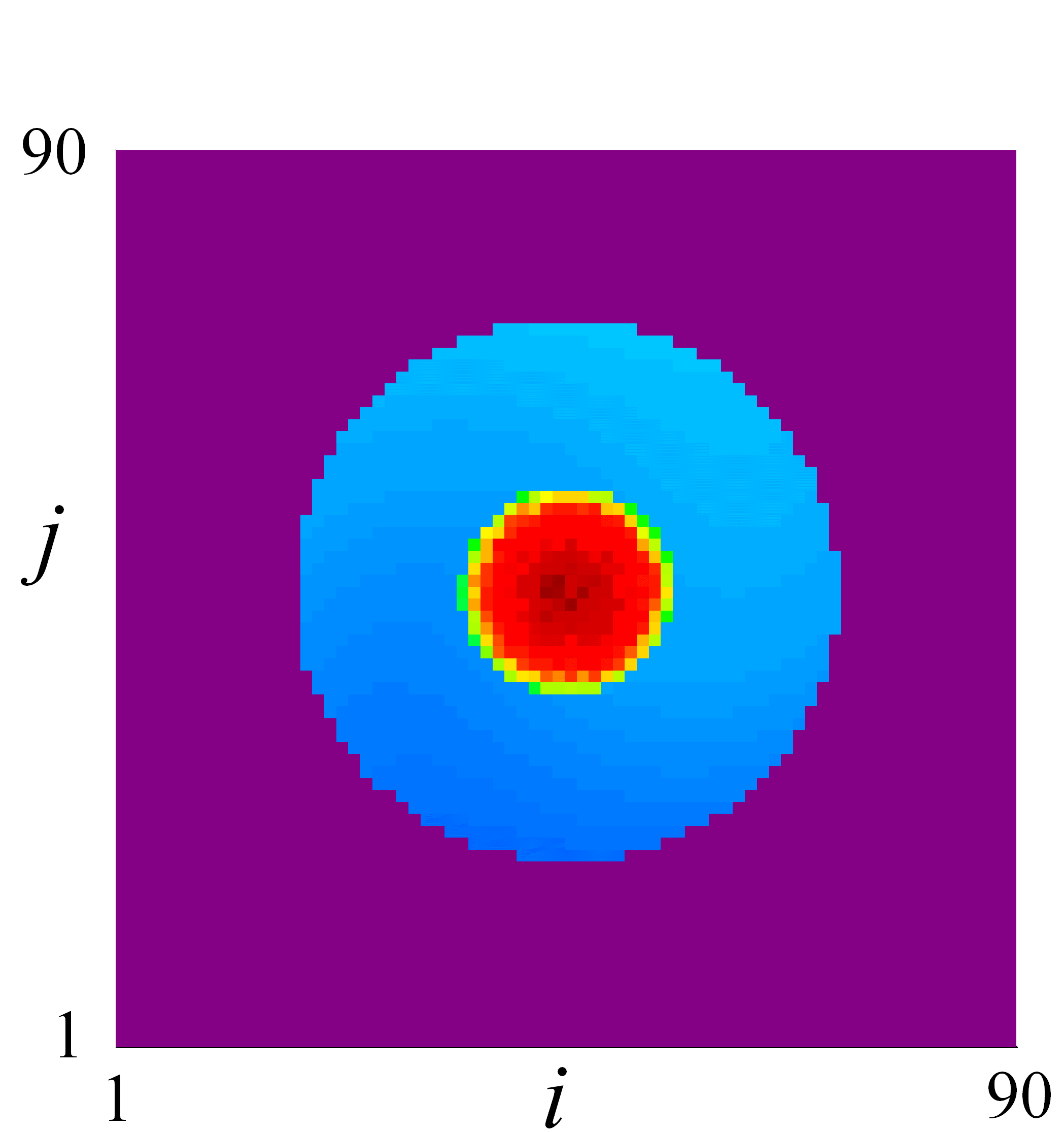}  
\includegraphics[width=0.27\linewidth]{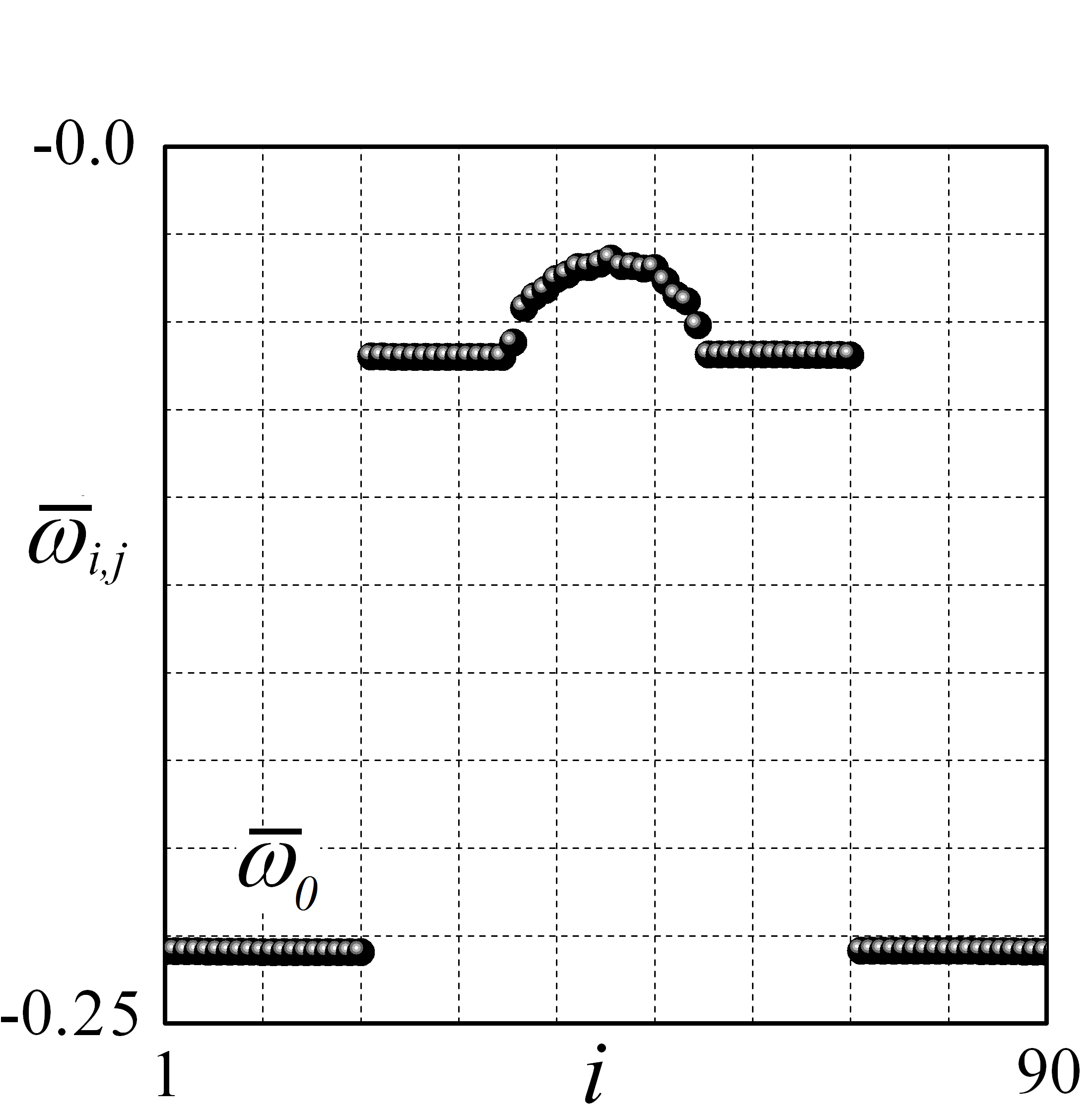} \\

 \includegraphics[width=0.27\linewidth]{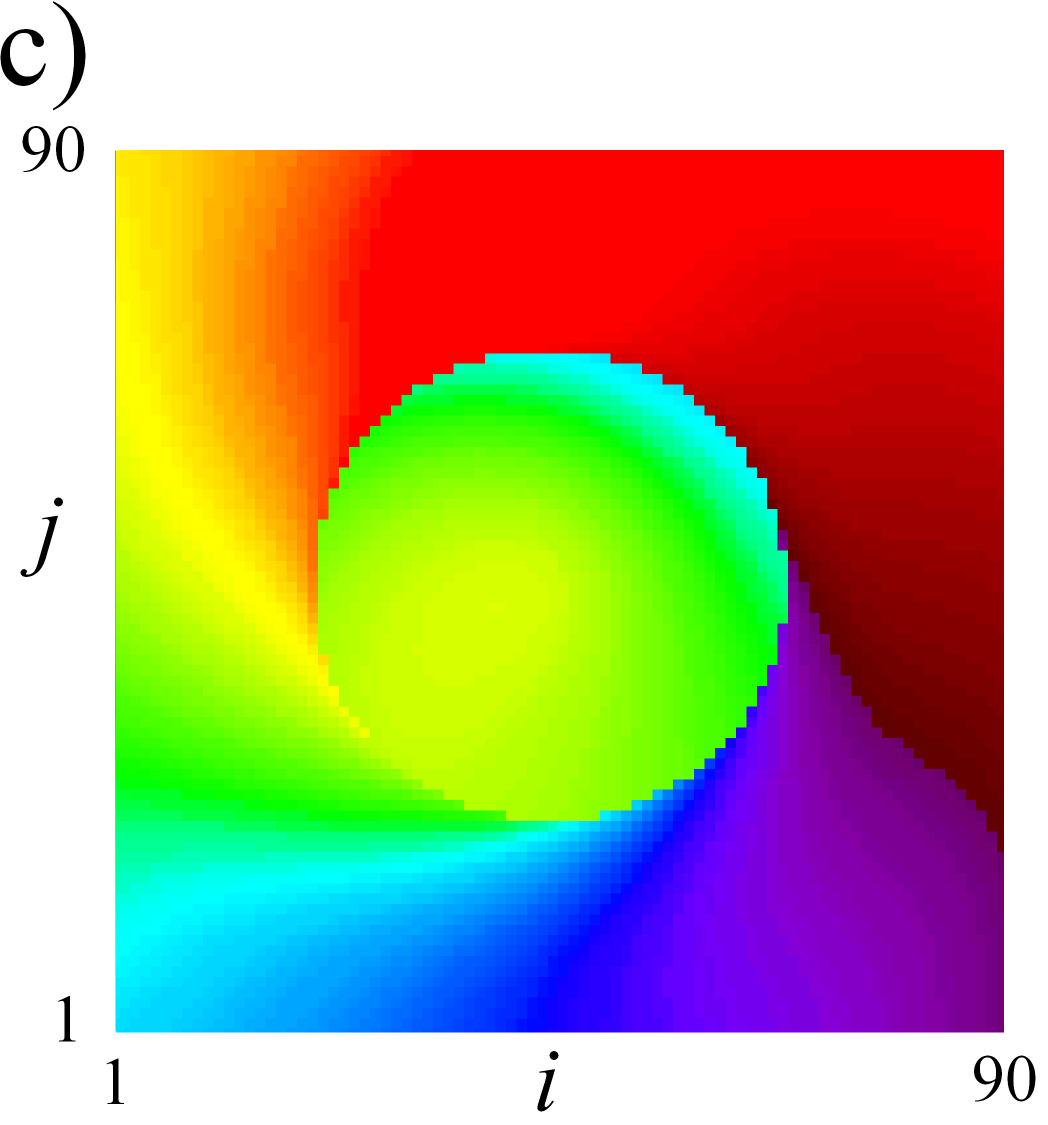}  \hspace{0.1cm}
 \includegraphics[width=0.06\linewidth]{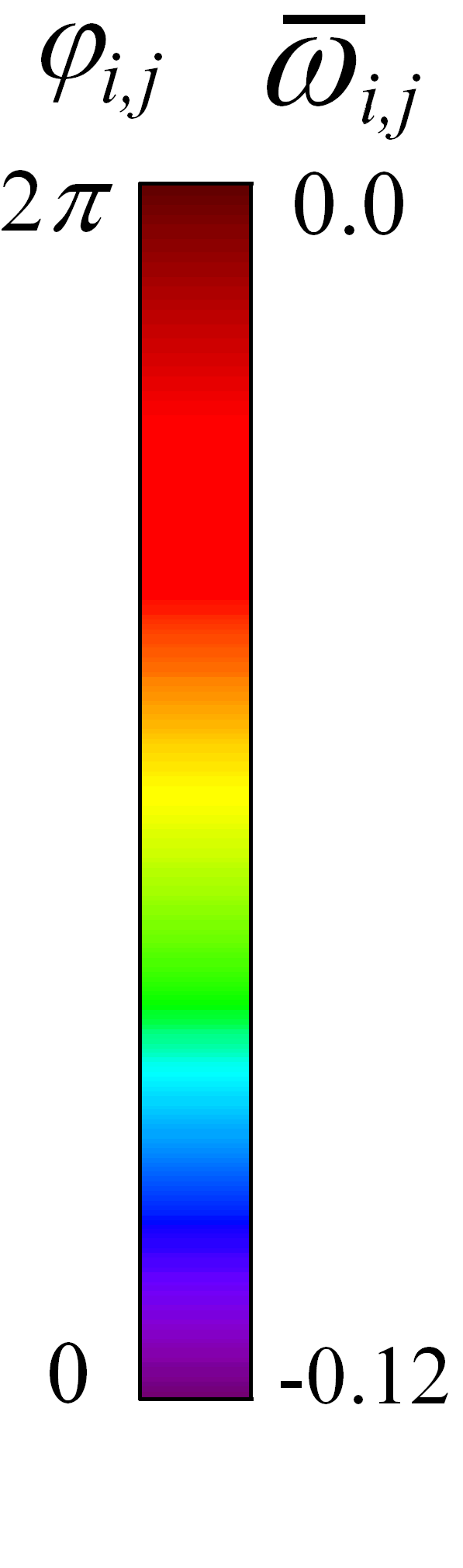}  
\includegraphics[width=0.27\linewidth]{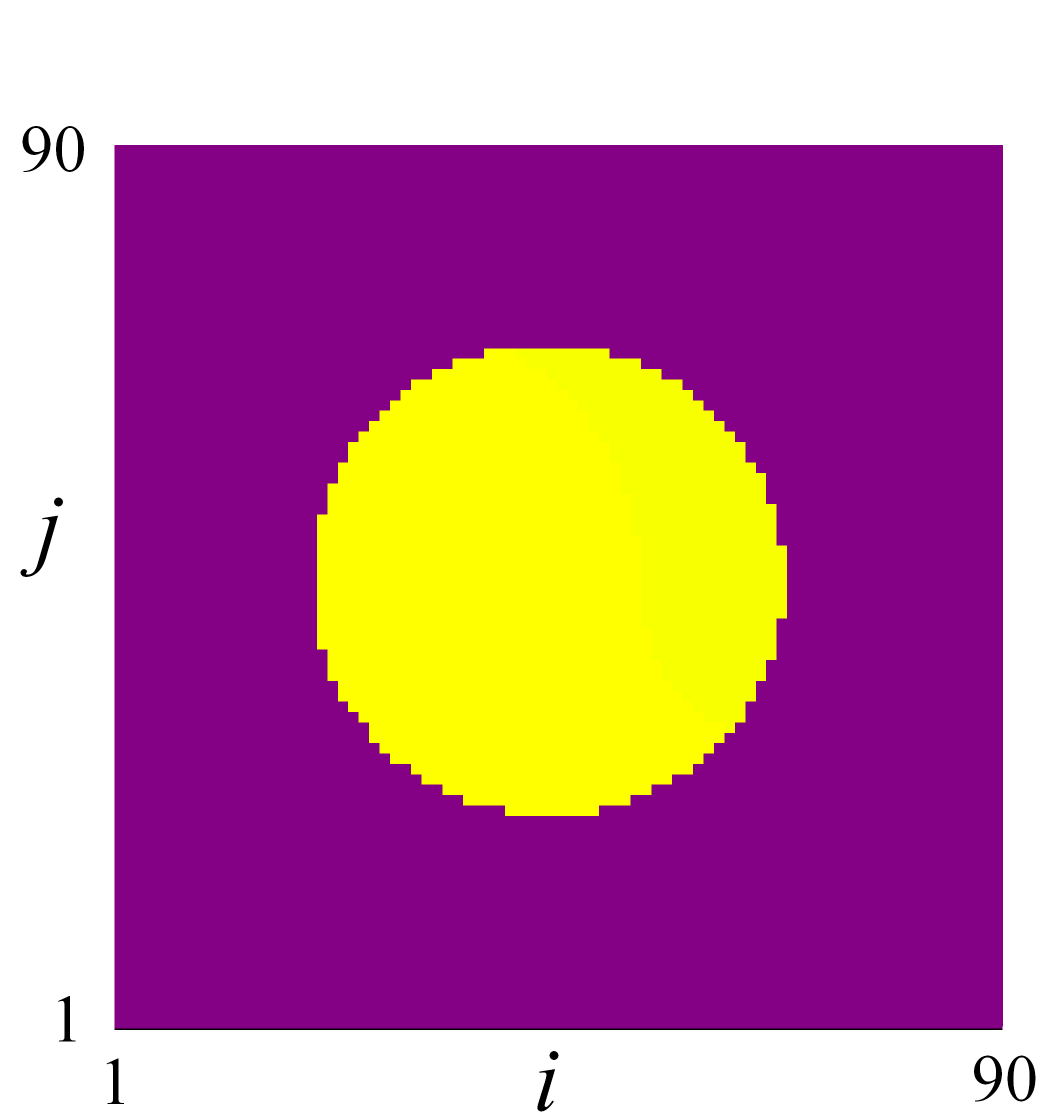}  
\includegraphics[width=0.28\linewidth]{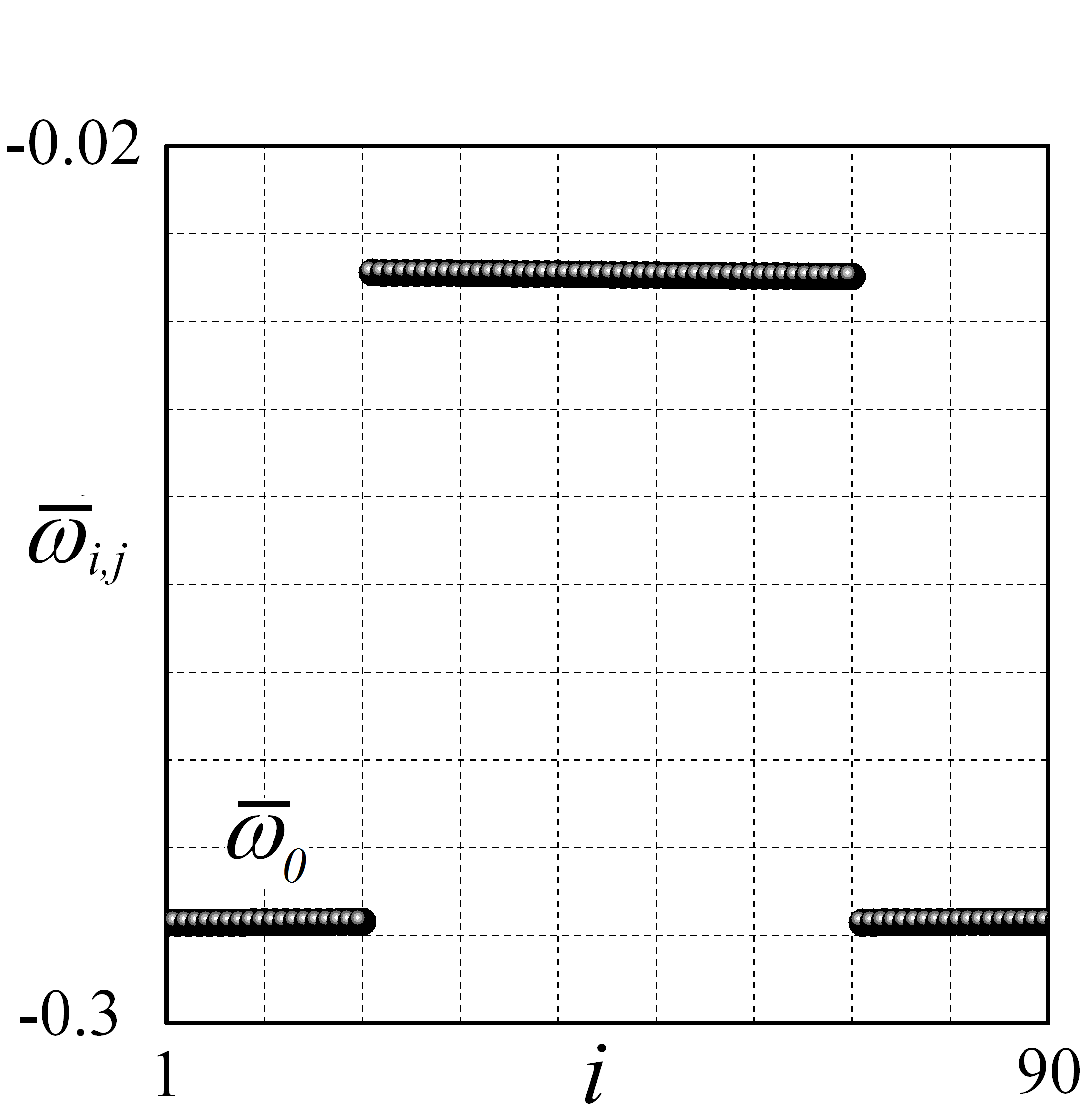} \\

 \includegraphics[width=0.27\linewidth]{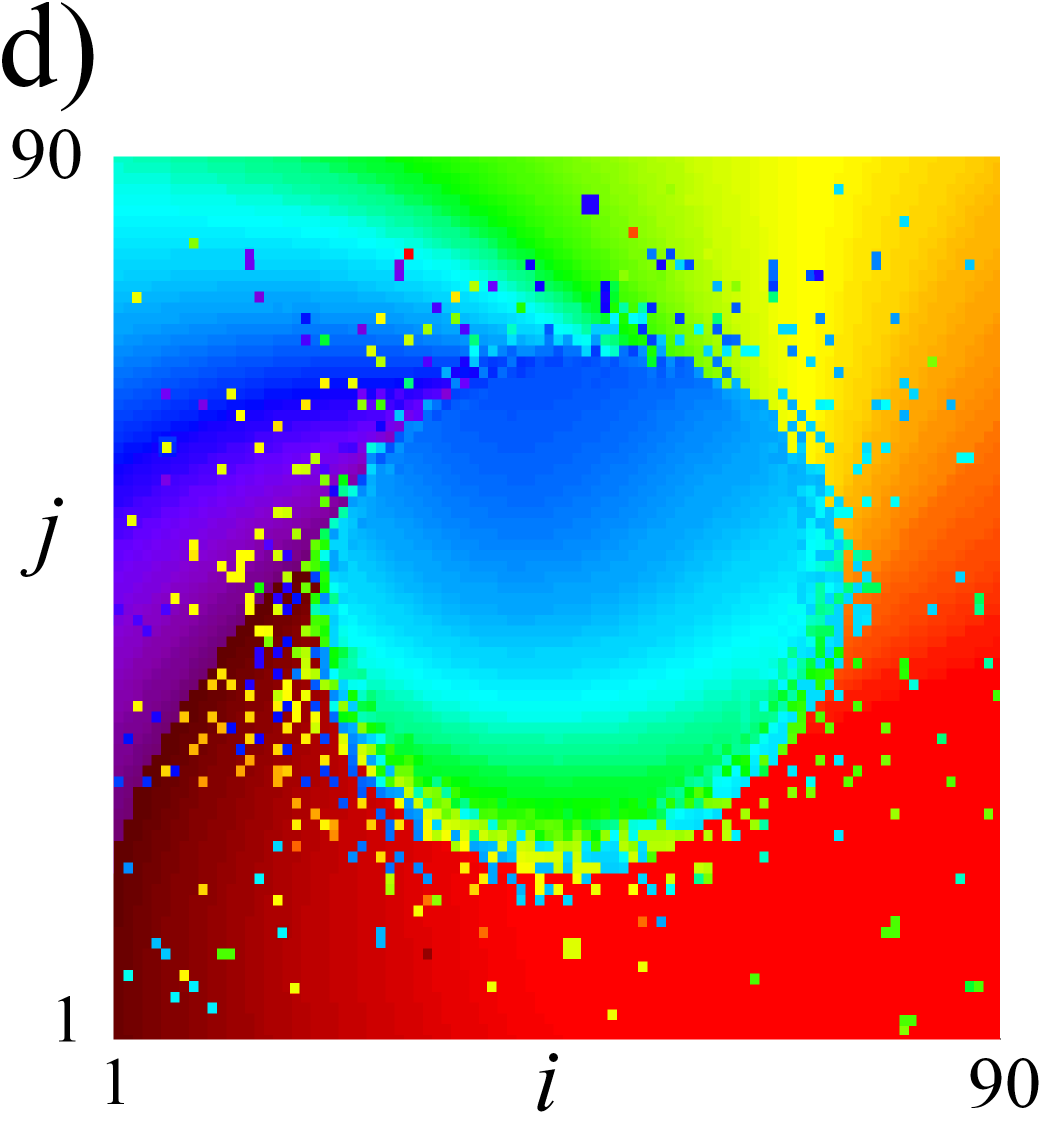}  \hspace{0.1cm}
 \includegraphics[width=0.06\linewidth]{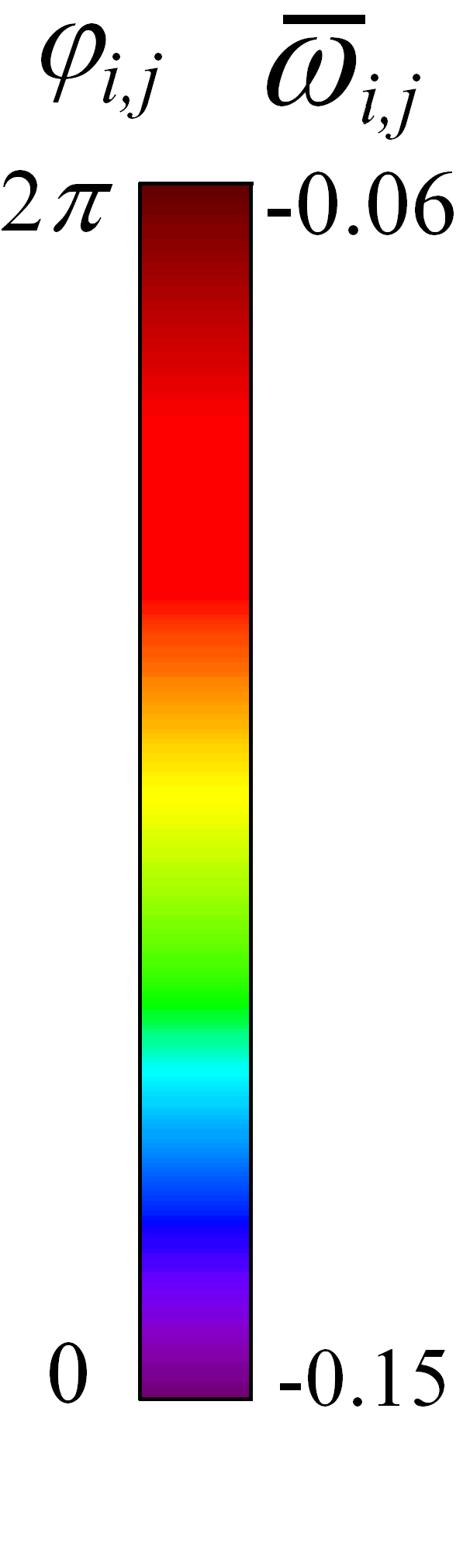}  
\includegraphics[width=0.27\linewidth]{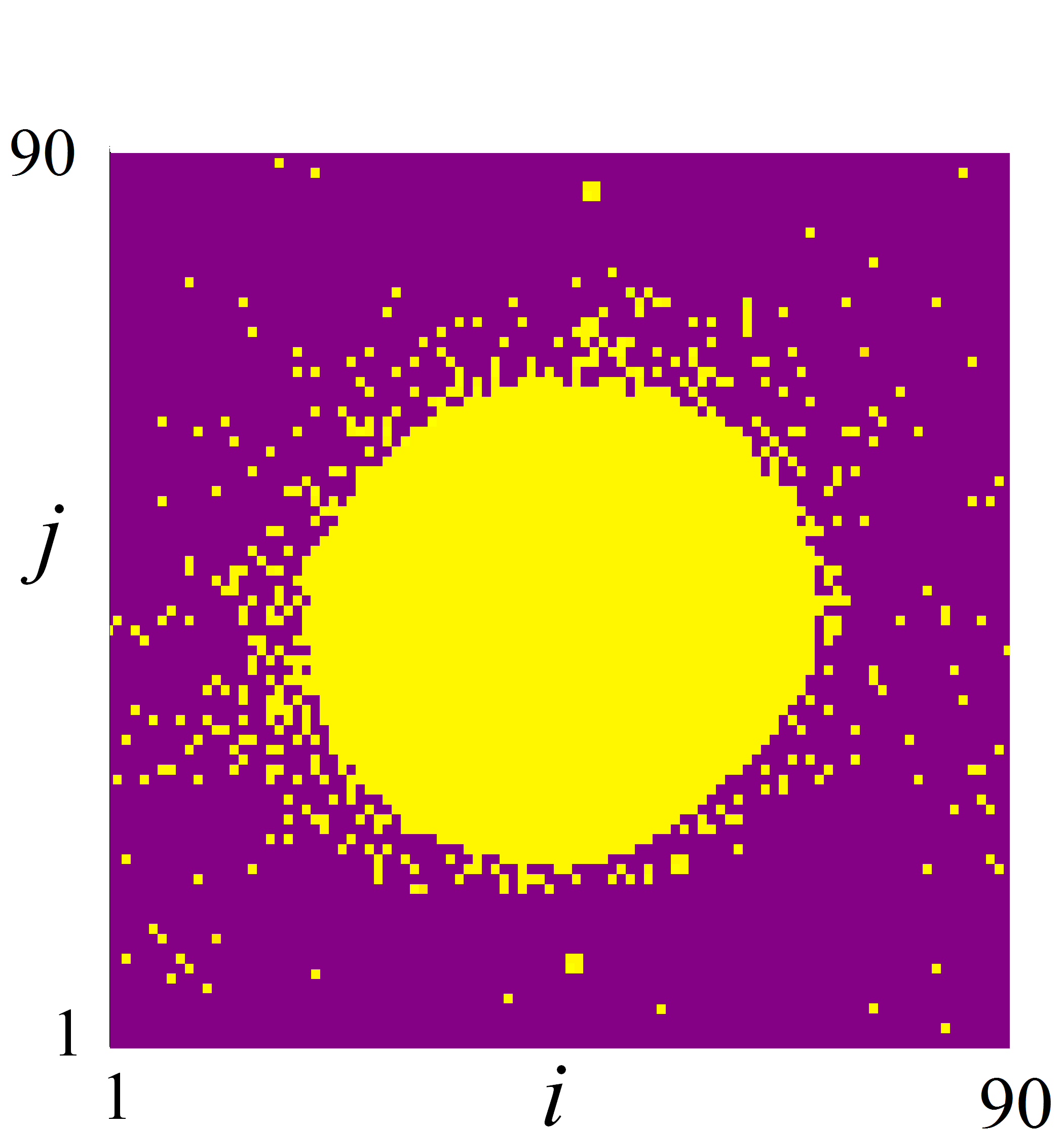}  
\includegraphics[width=0.28\linewidth]{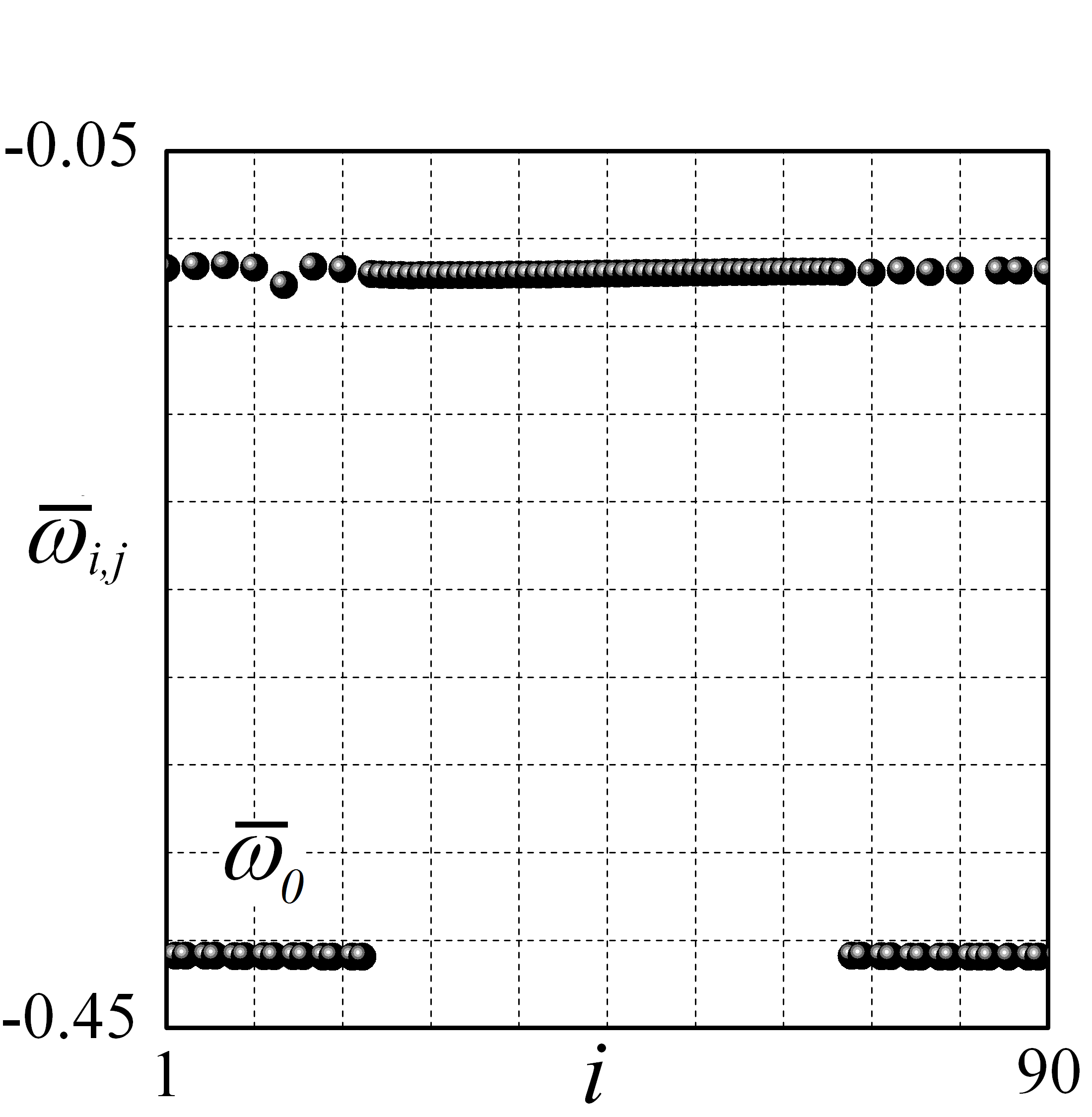} 

 \includegraphics[width=0.27\linewidth]{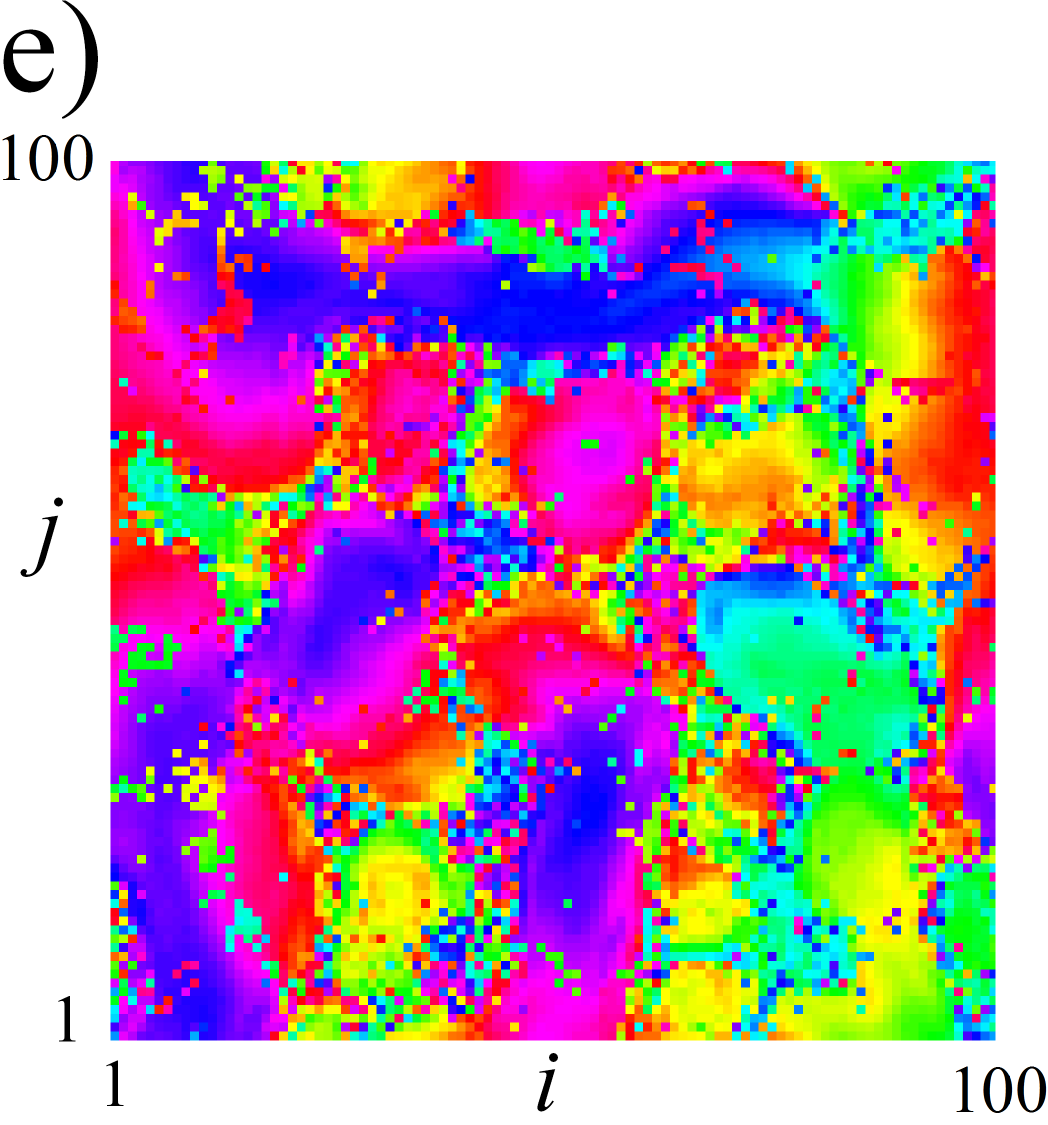}  \hspace{0.1cm}
 \includegraphics[width=0.06\linewidth]{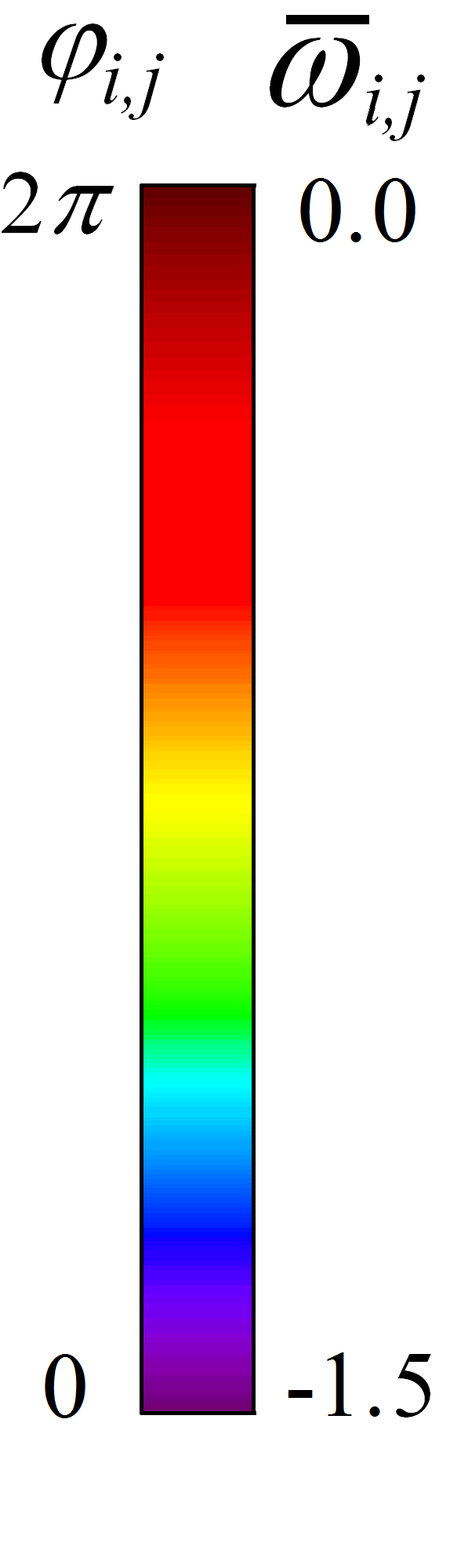}  
\includegraphics[width=0.27\linewidth]{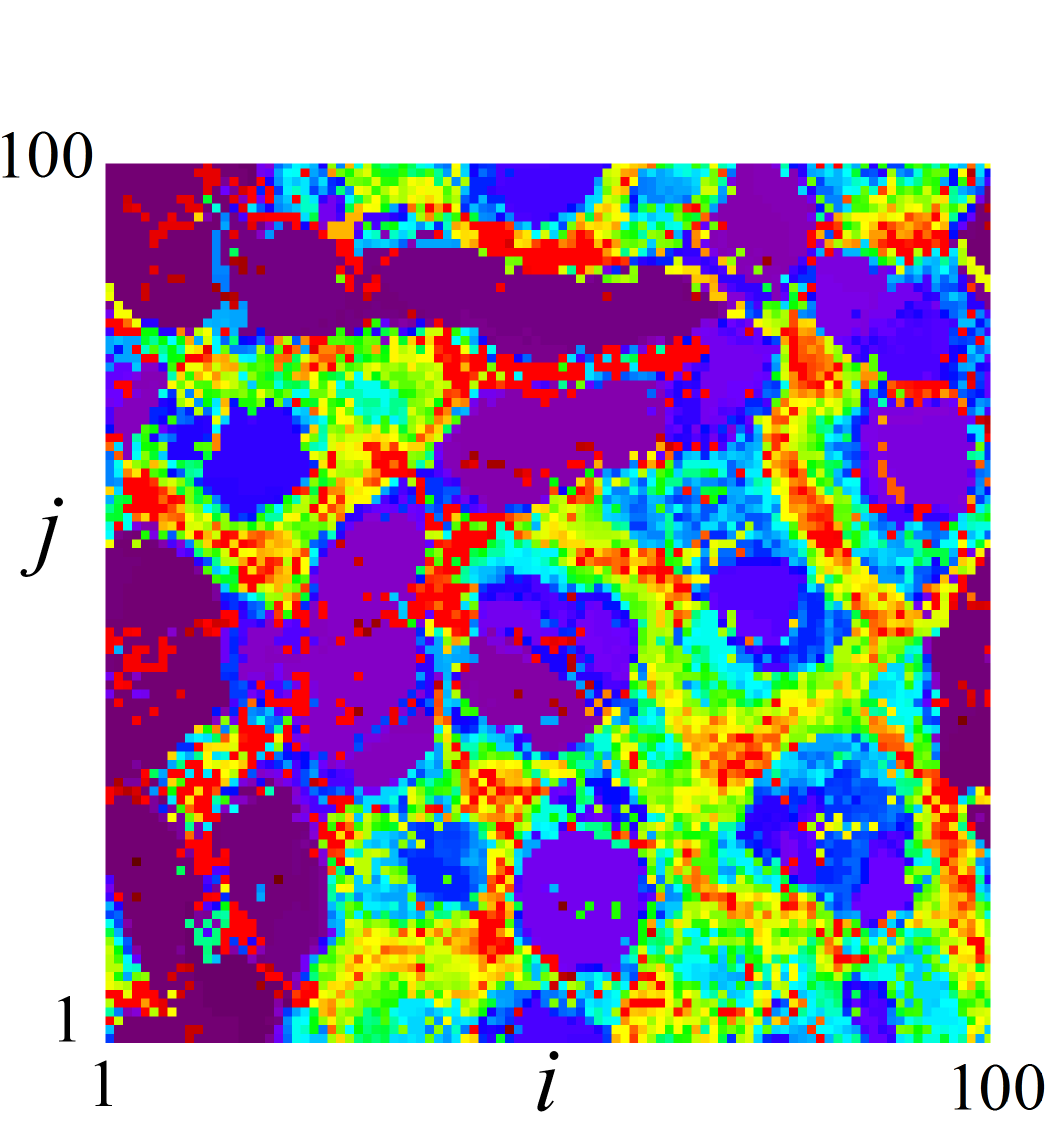}  
\includegraphics[width=0.28\linewidth]{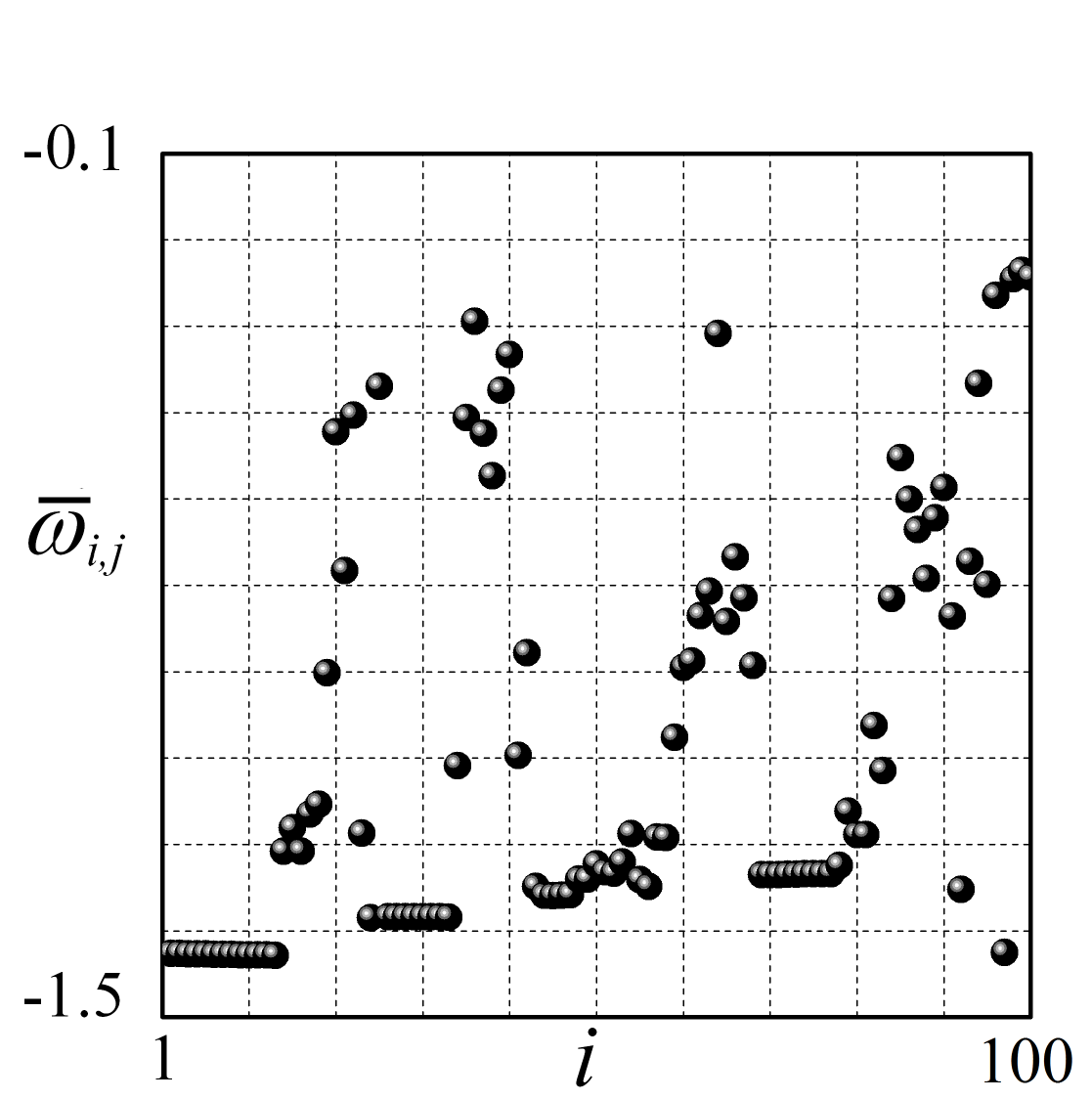} 

 \caption{Transitions of a 4-core spiral wave chimera state of model (1) for the fixed phase lag  $\alpha=0.42$  (small yellow circles along the dashed line in Fig.2a)). For each value of the coupling parameter $\mu$  (increasing from top to bottom, $\mu=0.01,  0.06, 0.07, 0.11, 0.58$, respectively), phase snapshots (left column) and time-averaged frequencies of individual oscillators (middle column) are shown.  The frequency cross-sections are illustrated in the right column. Other parameters:  (a-d) $N=800, P=56$, and only one core is shown in the grid 90x90; (e)  $N=100, P=7$. Simulation time $t=10^4$, time-averaged frequency interval $\Delta T = 1000$.}
\label{f4}
\end{center}
\end{figure}
 
A typical scenario for the transition from incoherent to coherent cores in the 4-core spiral wave chimera 
is illustrated in Fig. 4. We fix here $\alpha=0.42$ and increase the coupling strength $\mu$ along the vertical dashed line with small circles in Fig. 2(a). Only one of four chimera cores is shown, the others are characterized, as we observe in the simulations, by similar phase and frequency profiles. We will return to this question later in Fig. 5 after considering the transformation of just one chimera core in detail.

First, the solution in Fig. 4(a) represents a classical spiral wave chimera state for $\mu=0.01$. Thus, the frequency profile is smooth and bell-shaped (as it typically occurs for spiral chimeras in Kuramoto model without inertia). 
For larger $\mu,$ the frequency profile sharpens up and, at some value $\mu\approx0.05$, loses its smoothness. A concentric ring arises inside the core, where the oscillators are characterized by the same averaged frequency. A chimera core of this kind  is illustrated in Fig. 4(b), where one can also see that the bell-shape core profile is still preserved, but only close to the core center. Chimera patterns with a ring of a constant frequency may be referred, following \cite{pr2008,OWK2018,OK2019}, as {\it quasiperiodic} chimera states (assuming, clearly, that new appeared frequency is incommensurate with the basic spiraling around the core). A further increase in $\mu$ results in additional concentric rings of constant frequencies (not shown in Fig.4; examples can be found in the next chapter, as well as in ~\cite{XKK2015,OWK2018}).
Eventually, after a sequence of transformations, all oscillators inside the chimera core start to rotate with the same averaged frequency. The core becomes frequency-coherent, as illustrated in Fig. 4(c) for $\mu=0.07$. A chimera state with {\it coherent core} is born,  respective parameter region is shown in red in Fig. 2(a).

 As can be seen in Fig.2(a), the upper part of the chimera region belongs to the solitary region. In the solitary region, isolated oscillators can arise everywhere, depending on the initial conditions, and they create the so-called {\it solitary cloud} around the core, see Fig. 4(d) for $\mu=0.11$. Solitary oscillators in the cloud do not participate in the common spiraling around the core; instead, each of them oscillates with their own average frequency which coincides (more often) with the core frequency, as one can conclude from Fig.4(d). However, in more developed situations (i.e., when the parameter point enters dipper into the solitary region), solitary oscillators can obey different frequencies; see examples in the next chapter. Our simulations show that,  by a simple manipulation of the initial conditions,  the number and the disposition of solitary oscillators in the cloud can be obtained any  (but not more than half of the total number of oscillators in the network).

In particular, spiral chimeras with only one or a few oscillators in the cloud can be easily obtained, as well as those without solitary oscillators at all. See an example in Fig.4(c) with no solitary cloud, despite the parameters belong to the solitary region. However, such patterns with minimal or zero number of solitary oscillators are rather exceptional and obtained usually by parameter continuations from the outside of the solitary region. In most cases, solitary cloud appears to be quite extensive, as soon as simulations involve random initial conditions (like that shown in Fig. 4(d)). Eventually, with a further increase of $\mu$, the considered 4-core chimera ceases to exist, and the system dynamics drops, in a crisis bifurcation, to a sort of spatiotemporal chaos. Typical phase and frequency snapshots for the spatiotemporal behavior are illustrated in Fig. 4(e). A hyper-chaotic character of the state is confirmed by the huge number, more than one hundred, of positive Lyapunov exponents, where the maximal one equals $\approx0.156$.

With a more increase of $\mu$ along the dashed line in Fig. 2(a), only solitary states are left in the system dynamics. A typical snapshot can be seen in the upper-right corner of the phase diagram. 
Our simulations approve that combinatorially many solitary states exist and are stable inside the solitary region. Up to $N^2/2,$ network oscillators can be forced, by specially prepared initial conditions, to fall into the solitary rotation with a frequency different from those of the main synchronized cluster. 
Based on this, we expect that the number of stable solitary states can grow exponentially in the limit $N\to\infty$. 
If so, the network dynamics obeys a property of the {\it spatial chaos} (see~\cite{omps2011} and references there).

Transformations of only one spiral core  are shown in Fig. 4 for. Is the scenario the same for other cores? To shed light on the question we present, in Fig.5,  the graphs of time-averaged frequencies of all oscillators belonging to  four chimera cores, denoting them  $\bar{\omega}_{1}, \bar{\omega}_{2}, \bar{\omega}_{3}$, and $ \bar{\omega}_{4}$, respectively. 

\begin{figure*}[ht! ]
\vspace*{-0.6cm}
\begin{center}
   \includegraphics[width=0.43\linewidth]{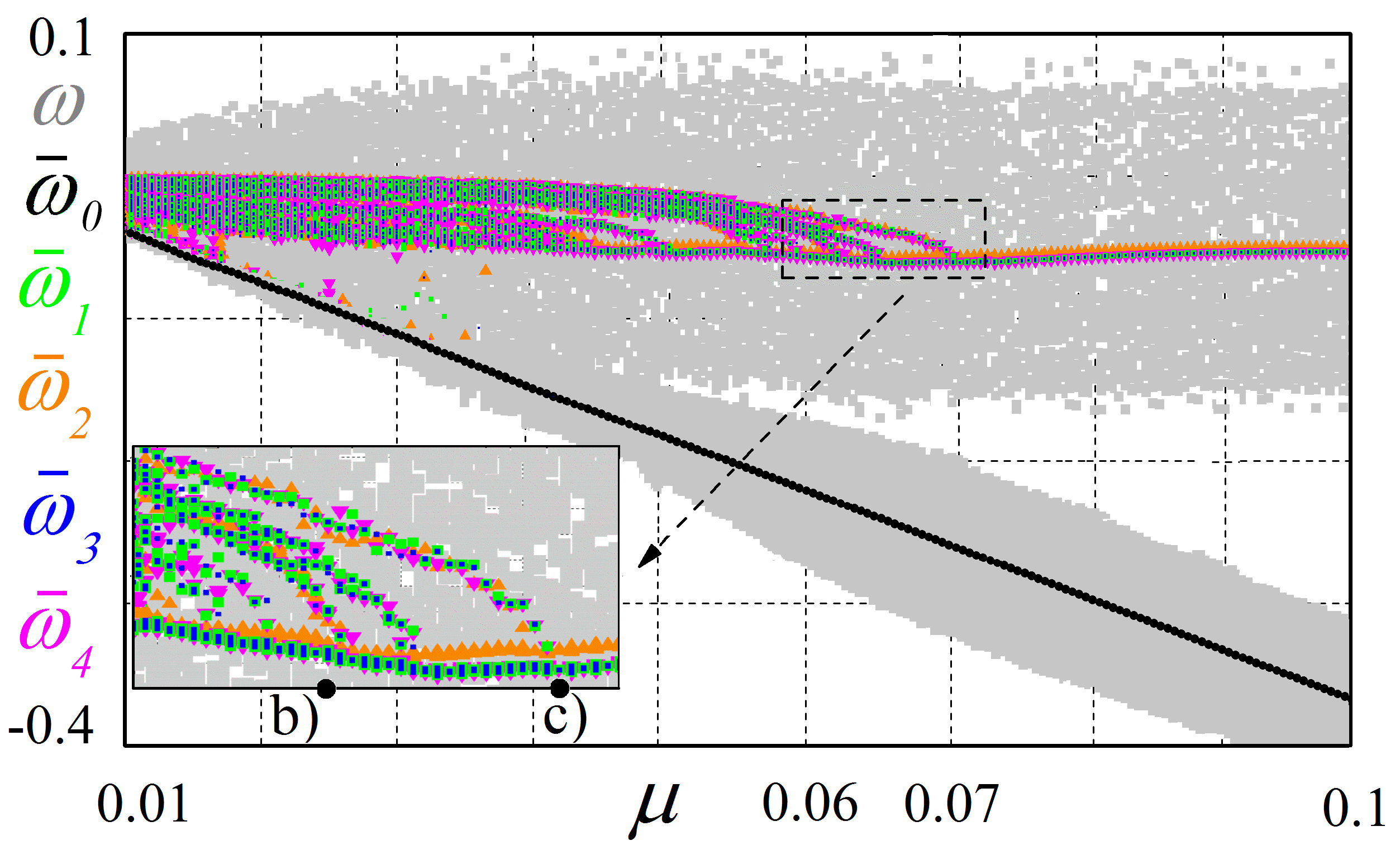} 
\vspace*{-0.2cm}
 \caption
{Frequency plot of a 4-core chimera state from Fig.4 as a function of the coupling parameter $\mu$. Averaged frequencies of all oscillators are shown,  different colors correspond to oscillators from different cores denoted by $\bar{\omega}_{1}, \bar{\omega}_{2}, \bar{\omega}_{3}$, and $ \bar{\omega}_{4}$, respectively. Basic spiral frequency $\bar{\omega}_{0}$ is plotted in black. Light grey background shows the spreading of the instant frequencies. Points b) and c) indicate respective parameter values from Fig. 4 b): $\mu=0.06$ and c): $\mu=0.07$.  Zoom of the key bifurcation transitions is shown in the inset,  see the description in the text.  Other parameters $\alpha=0.42$,  $P=7$,  $N=100$.  Simulations time $ t=2\times10^{4}$.  Averaging at the last 1000 time units of the simulation.}
\label{f5}
\end{center} 
\end{figure*}

In the figure, the averaged frequencies are mapped in different colors corresponding to different cores.  Basic spiraling frequency $\bar{\omega}_{0}$ is plotted in black, and the light grey background stands for spreading of the instant frequencies.  Parameters as in Fig. 4: $\alpha=0.42$, $P=7$, $N=100$, and we vary the coupling parameter $\mu$ between $0.01$ to $0.1$.
First, at smaller $\mu$, the oscillator frequencies all together fill some interval, reflecting the bell-shape profile in each core (as in Fig. 4(a)). With an increase of $\mu$, the frequencies get closer to one another, and they become practically identical beyond $\mu\approx0.07$. The more detailed inspection of the graphs shows, nevertheless, that only three frequencies coincide, while the 4th one slightly deviates (see the inset), and such picture is preserved in our simulations up to $\mu=0.1$. 

To finalize the chapter, we present four more examples of a 4-core
chimera state in Fig.6 to illustrate a variability of the system dynamics at different parameters. 
Among others, an important question concerns the number of different frequencies (plateaus) possible in the
chimera cores; we observe at most five such frequencies in Fig.6(d). 

\begin{figure*}[ht! ]
\begin{center}
\vspace*{-0.4cm}
   \includegraphics[width=0.23\linewidth]{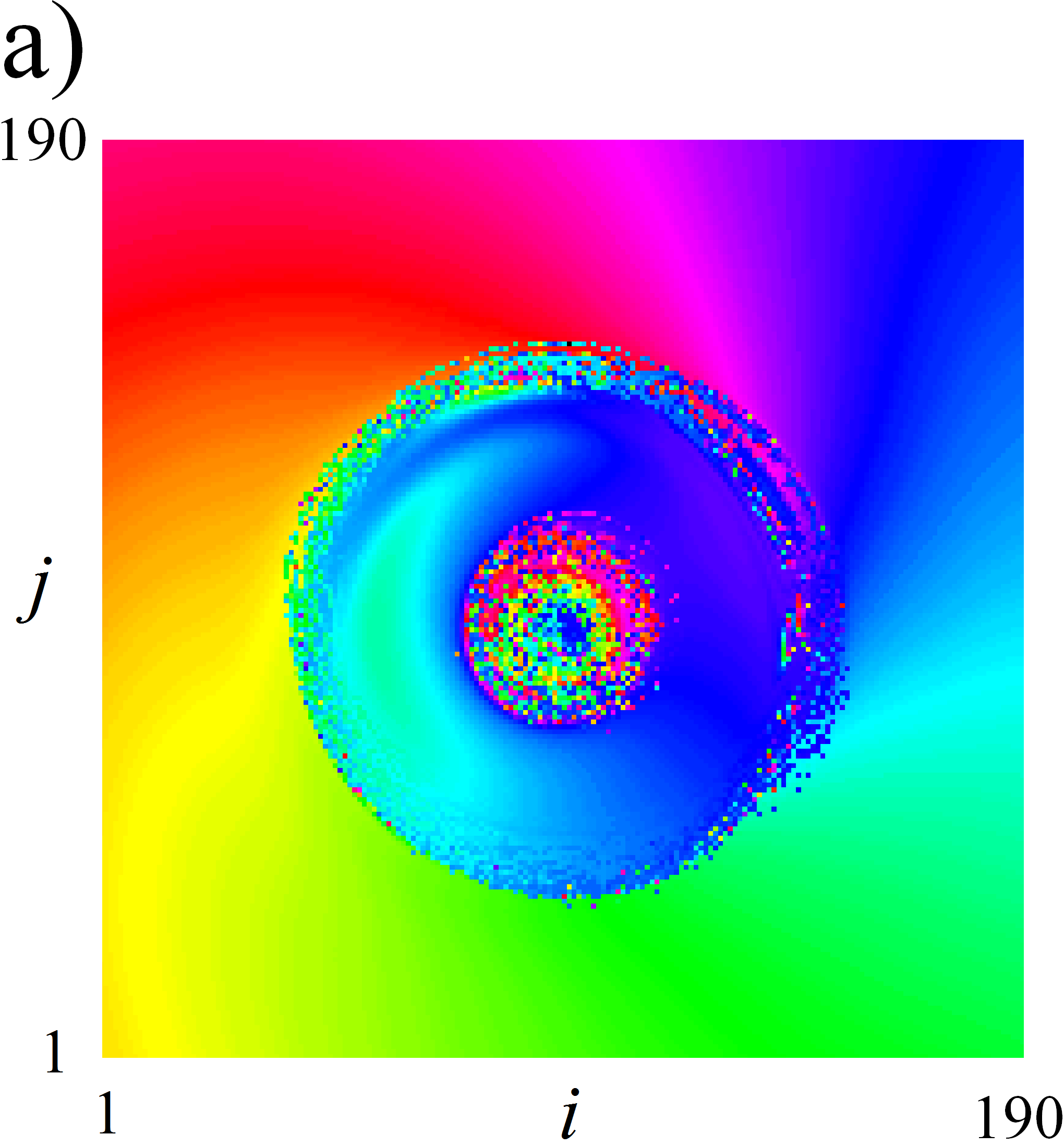} \includegraphics[width=0.23
\linewidth]{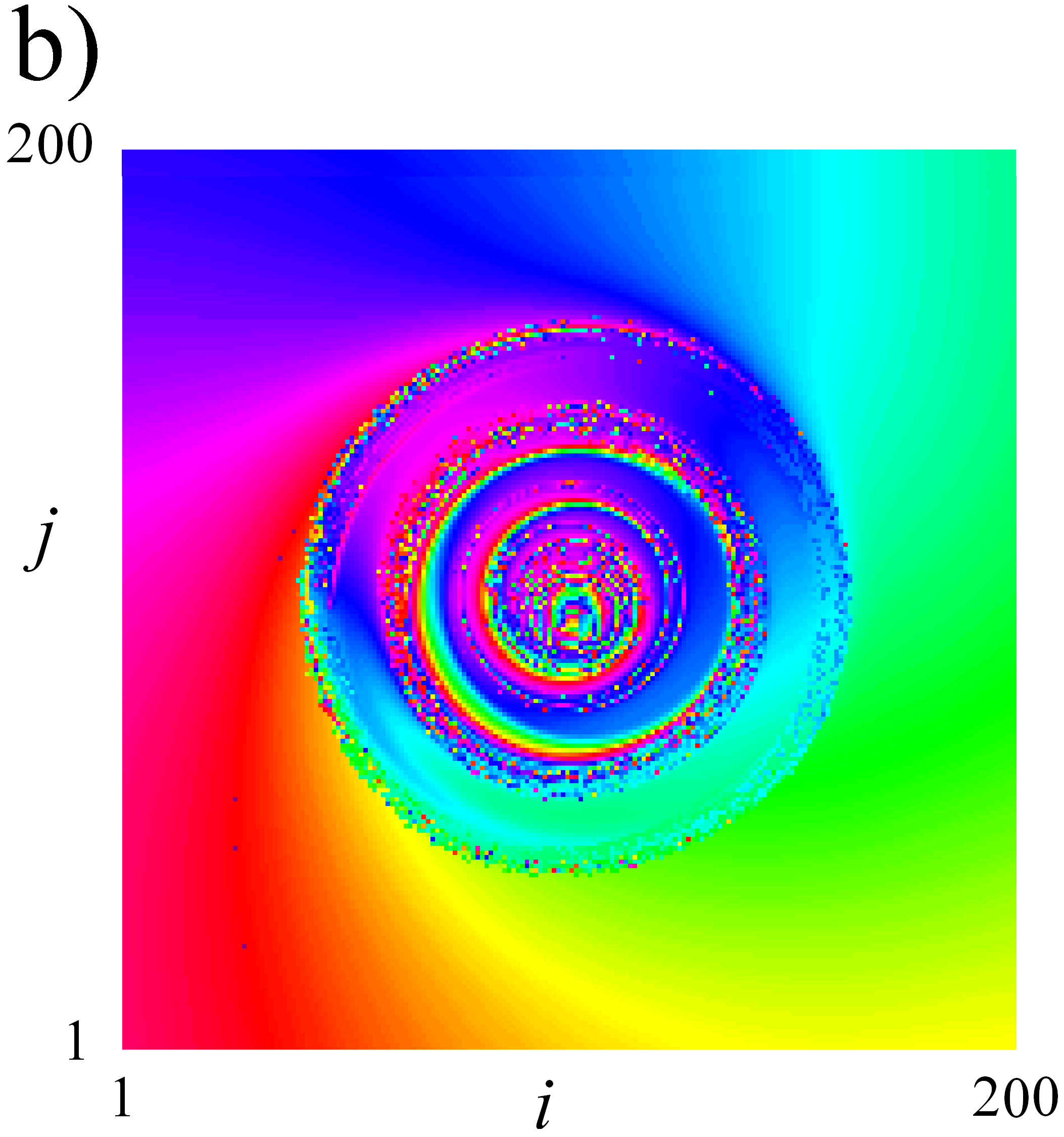} \includegraphics[width=0.23\linewidth]{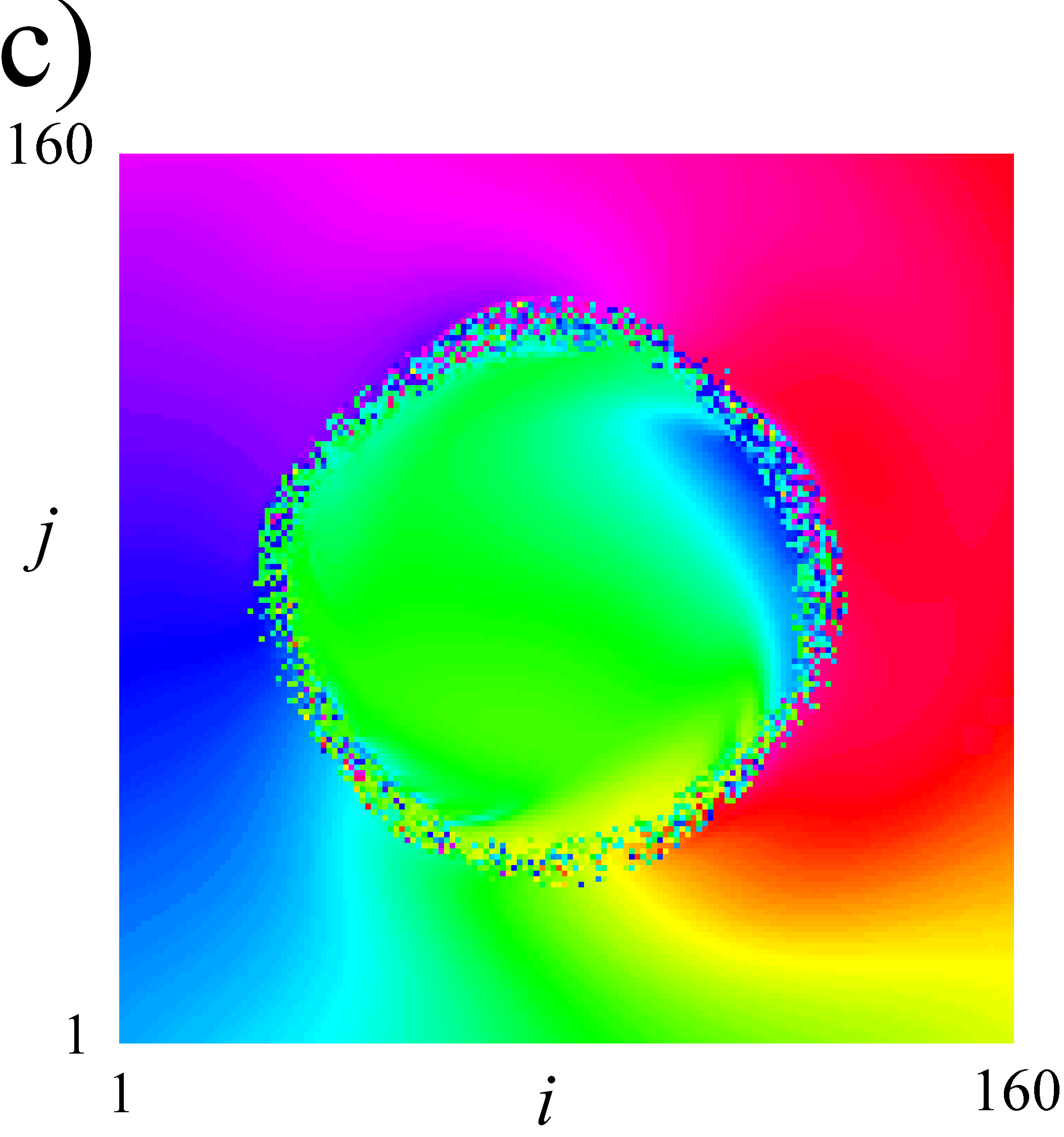} \includegraphics[width=0.23\linewidth]{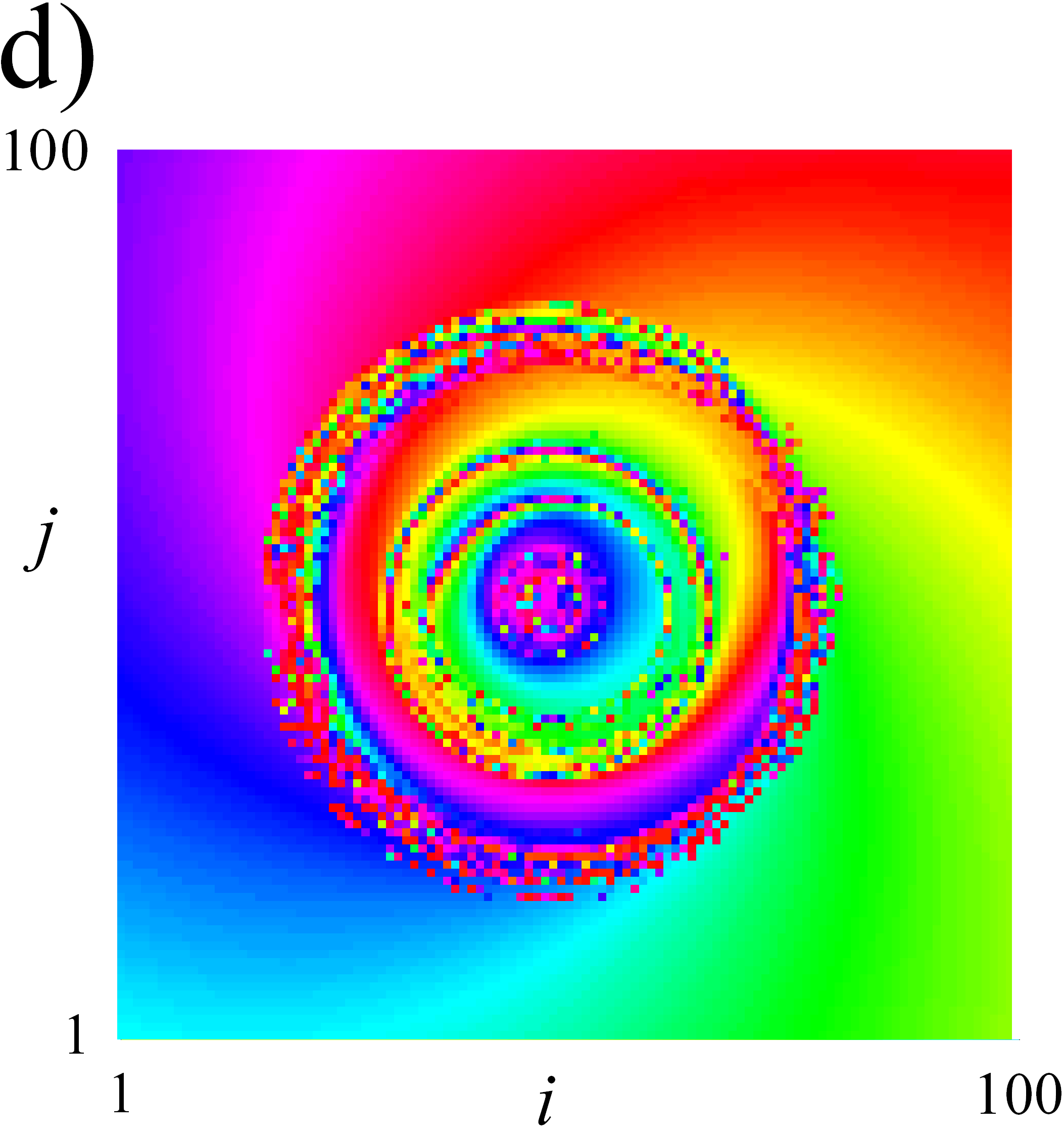} \includegraphics[width=0.035
\linewidth]{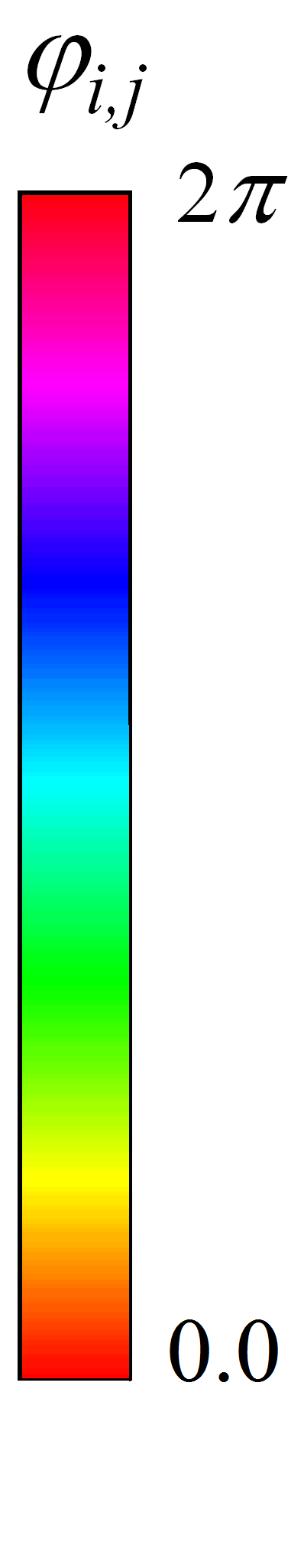} 

\vspace*{0.2cm}
 \includegraphics[width=0.23\linewidth]{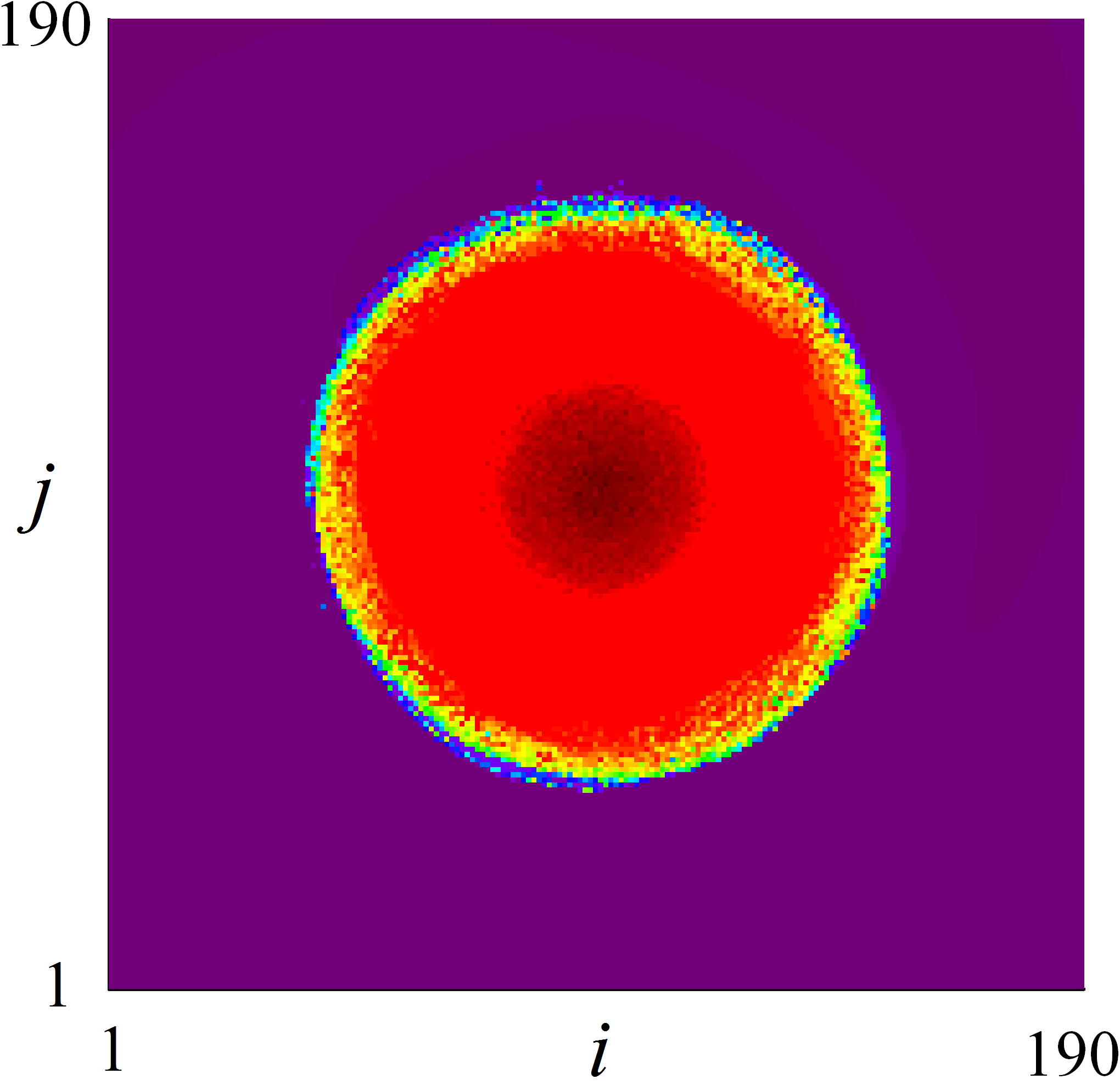} \includegraphics[width=0.23
\linewidth]{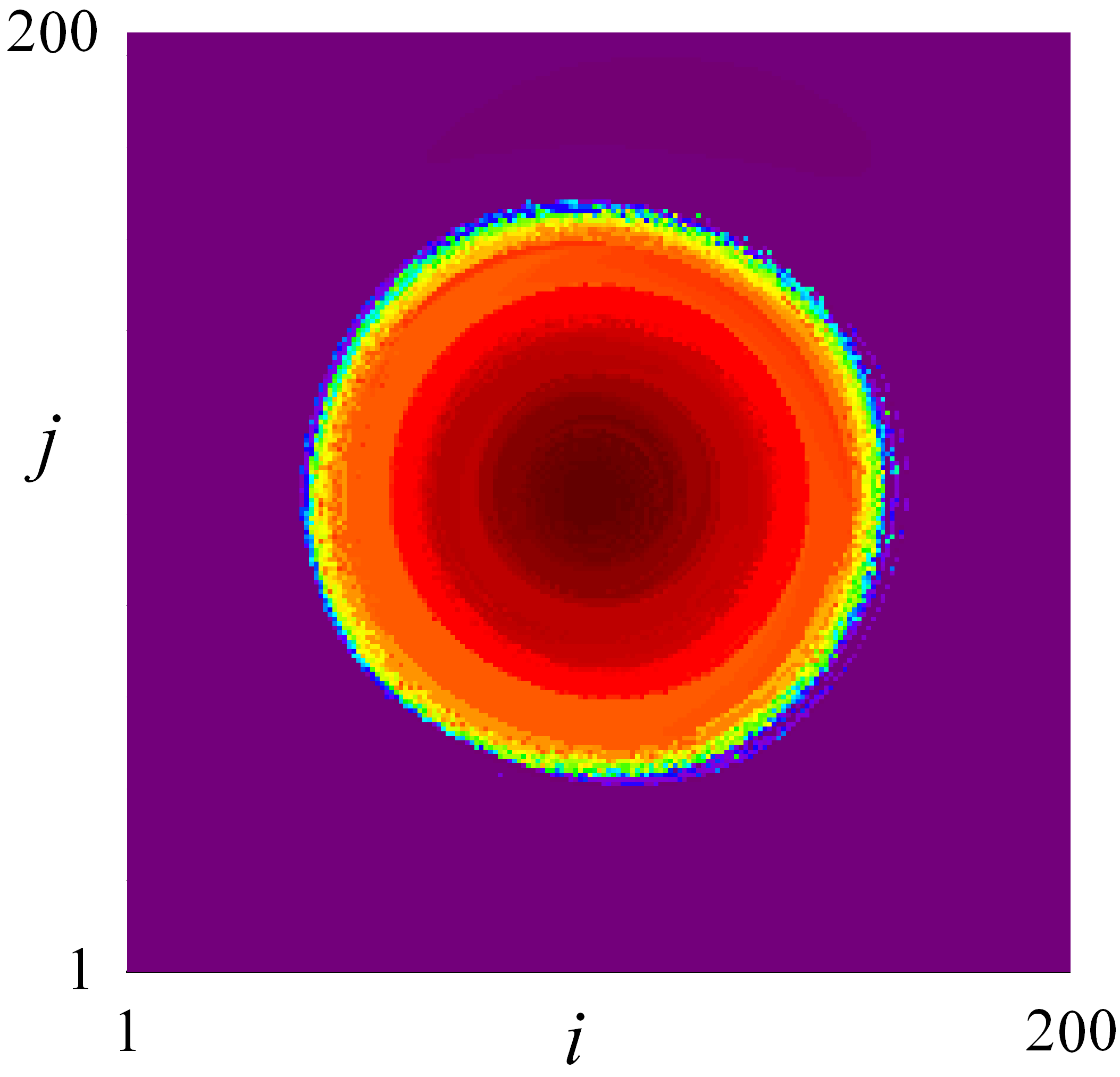} \includegraphics[width=0.23\linewidth]{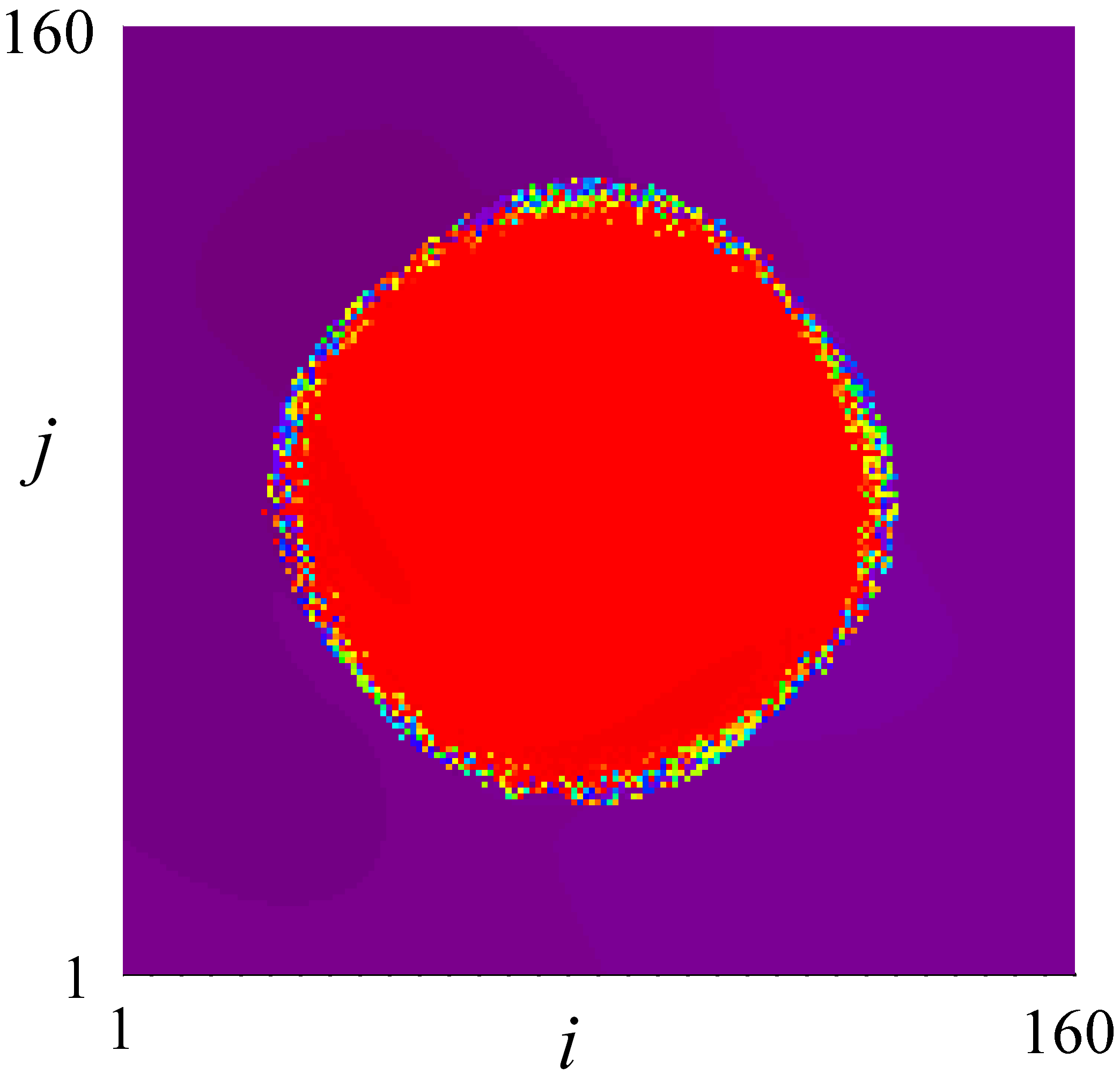}  \includegraphics[width=0.23\linewidth]{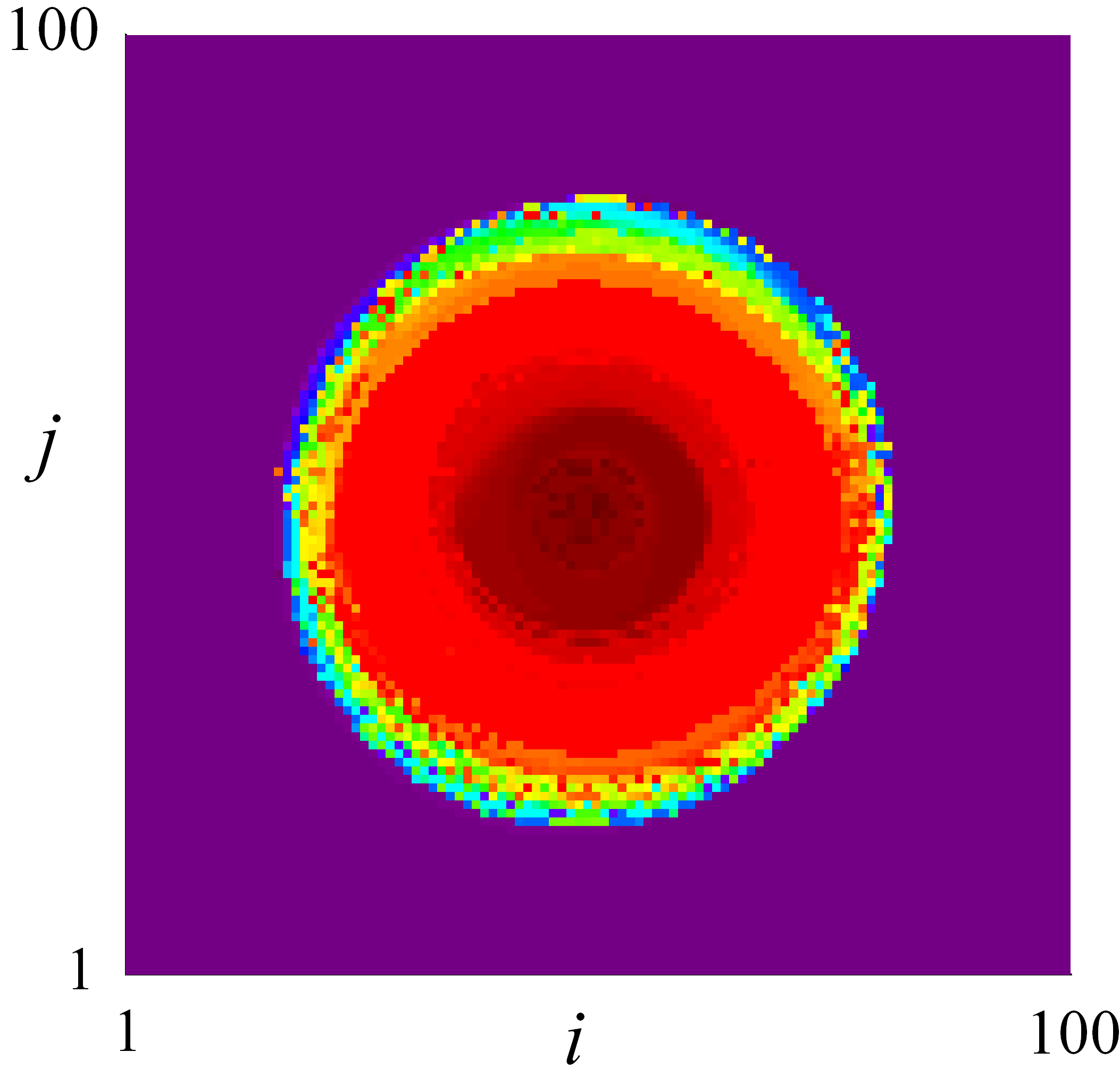} \includegraphics[width=0.035
\linewidth]{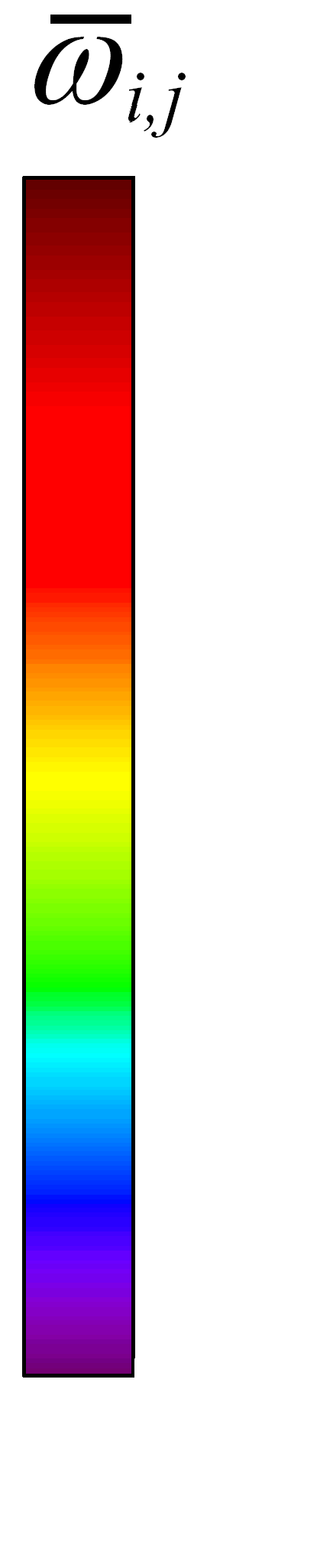} 

\vspace*{0.2cm}
 \includegraphics[width=0.23\linewidth]{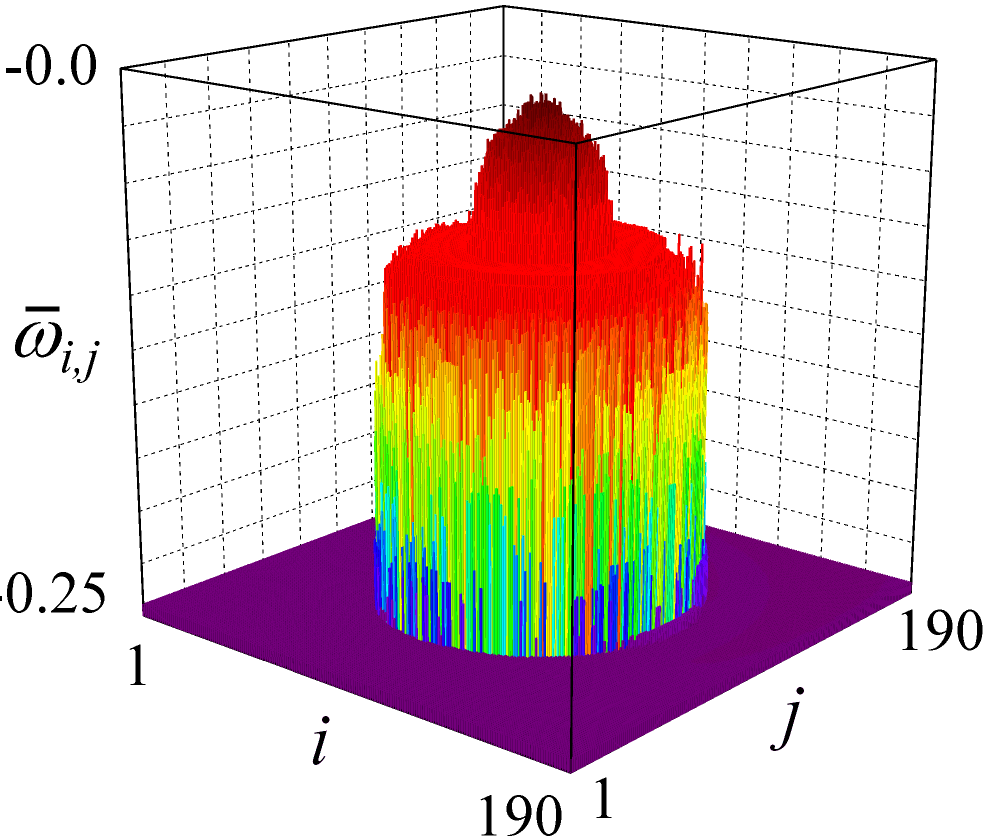}  \includegraphics[width=0.23
\linewidth]{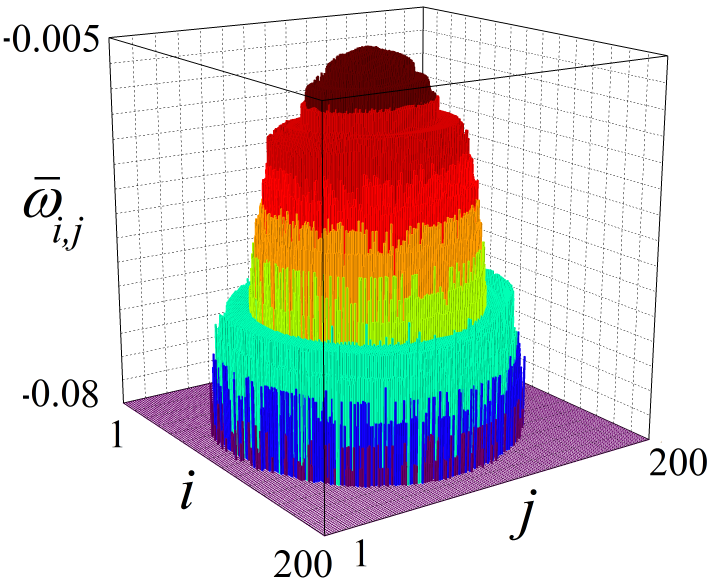} \includegraphics[width=0.23\linewidth]{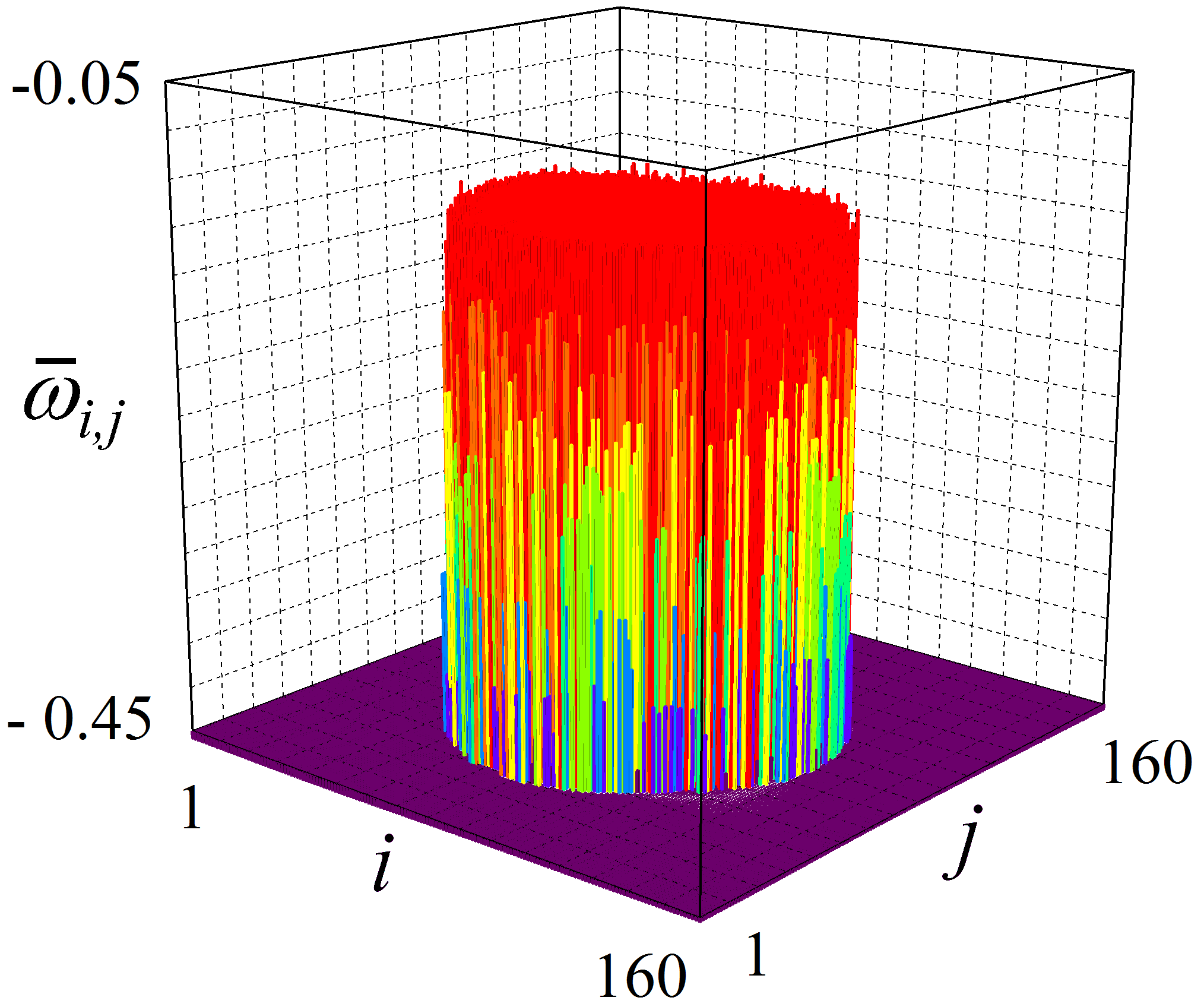} \includegraphics[width=0.23\linewidth]{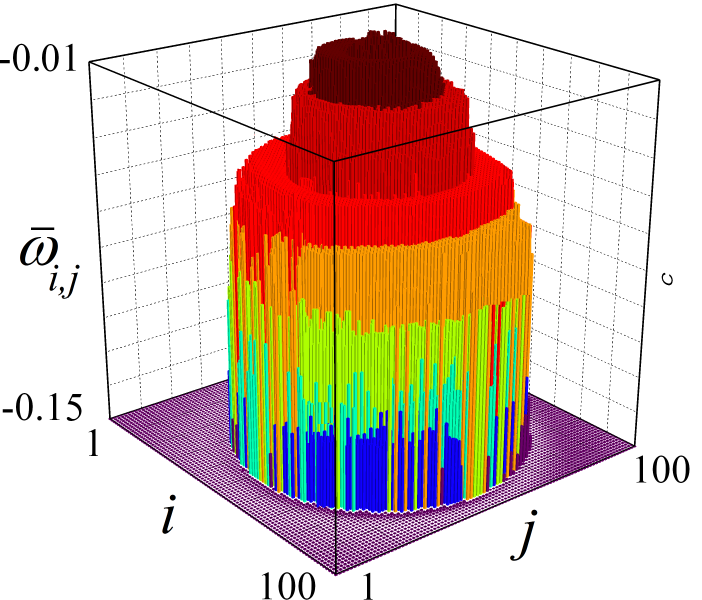} \includegraphics[width=0.035
\linewidth]{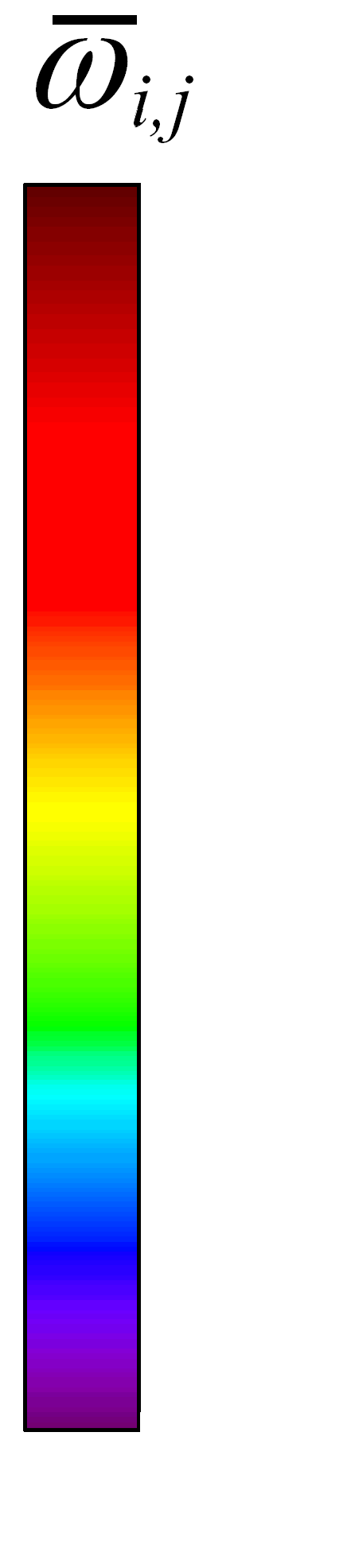} 
\end{center}
\vspace*{-0.1cm}
\includegraphics[width=0.23\linewidth]{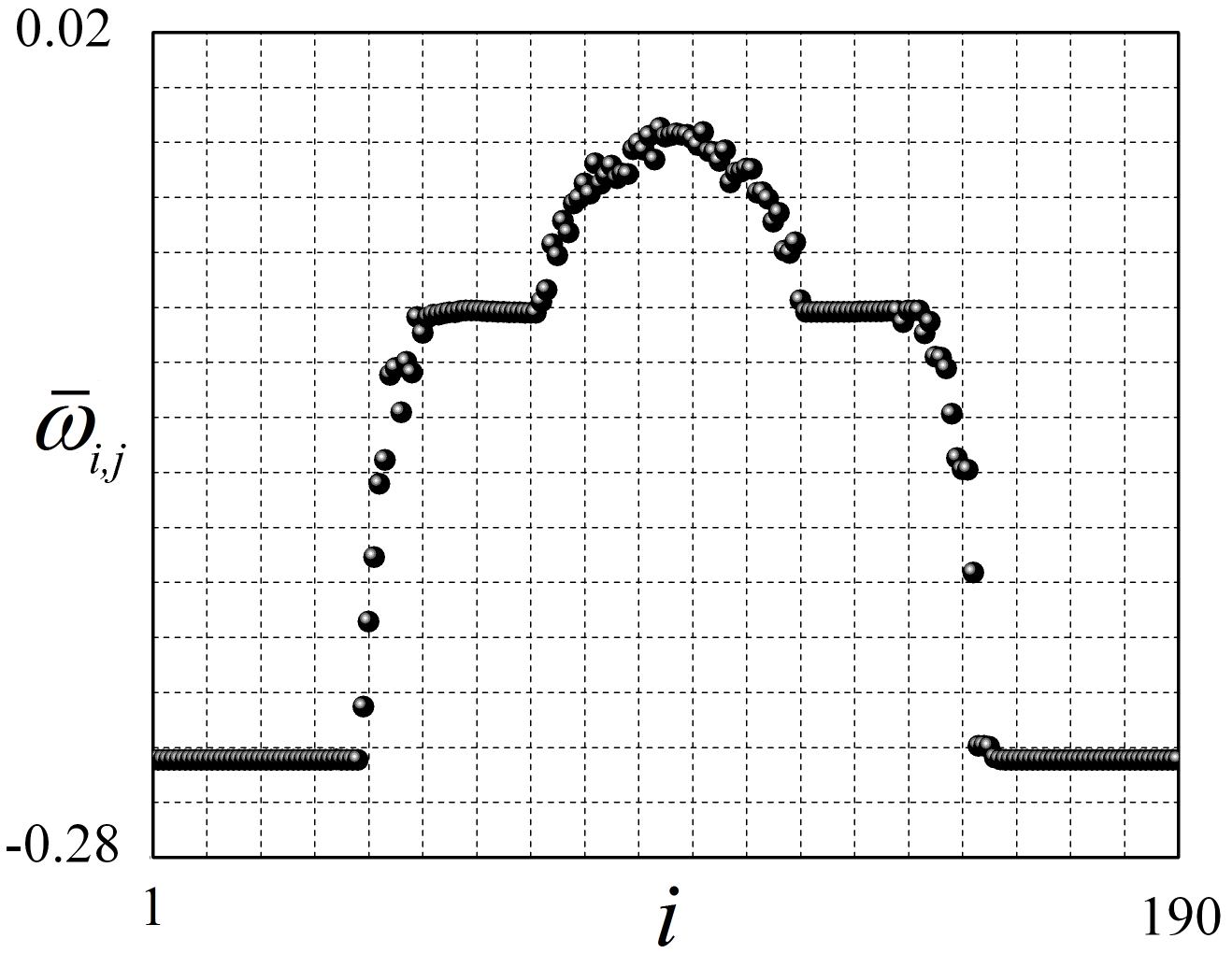}  \ \includegraphics[width=0.226\linewidth]{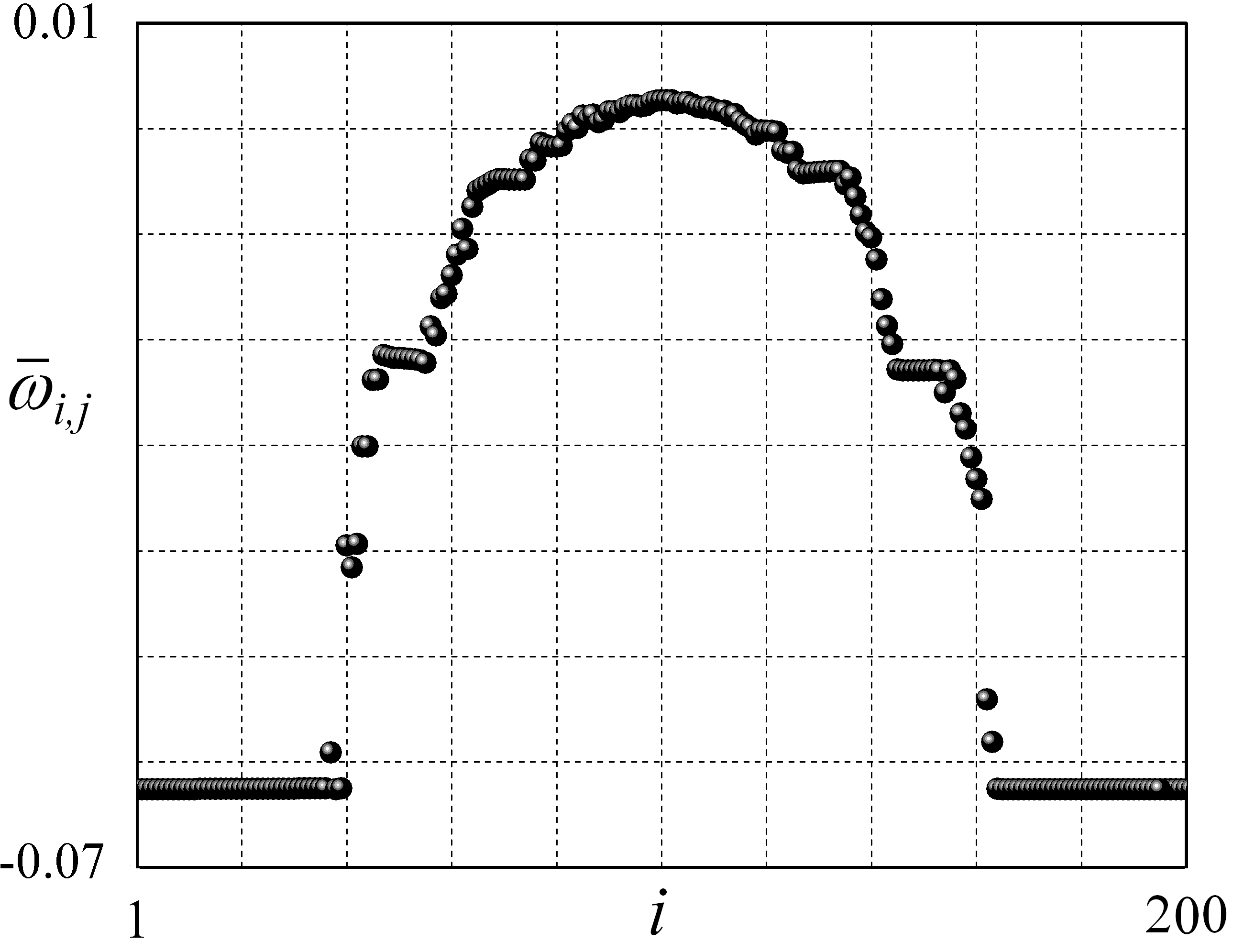}    \includegraphics[width=0.24\linewidth]{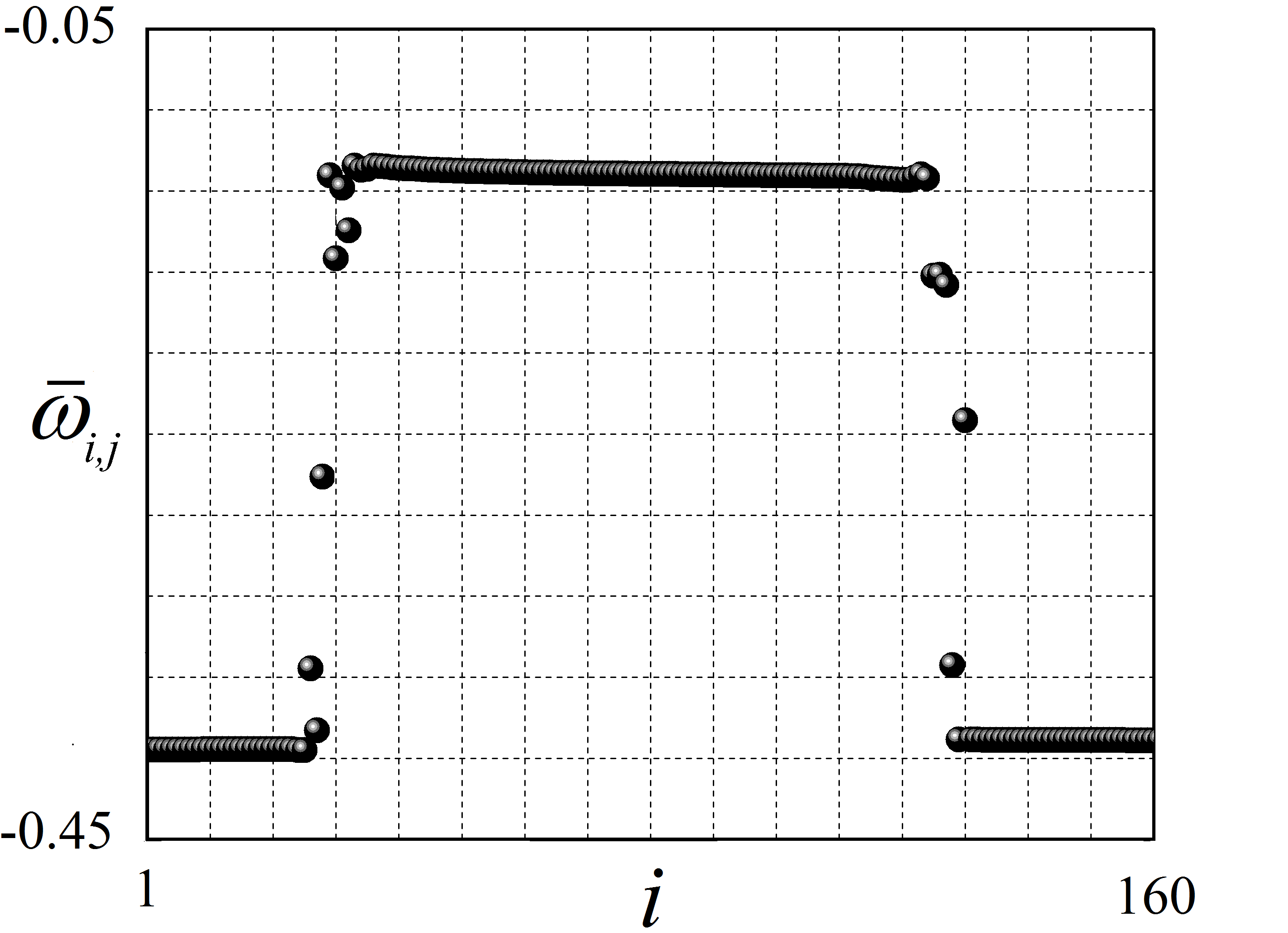}  \includegraphics[width=0.222\linewidth]{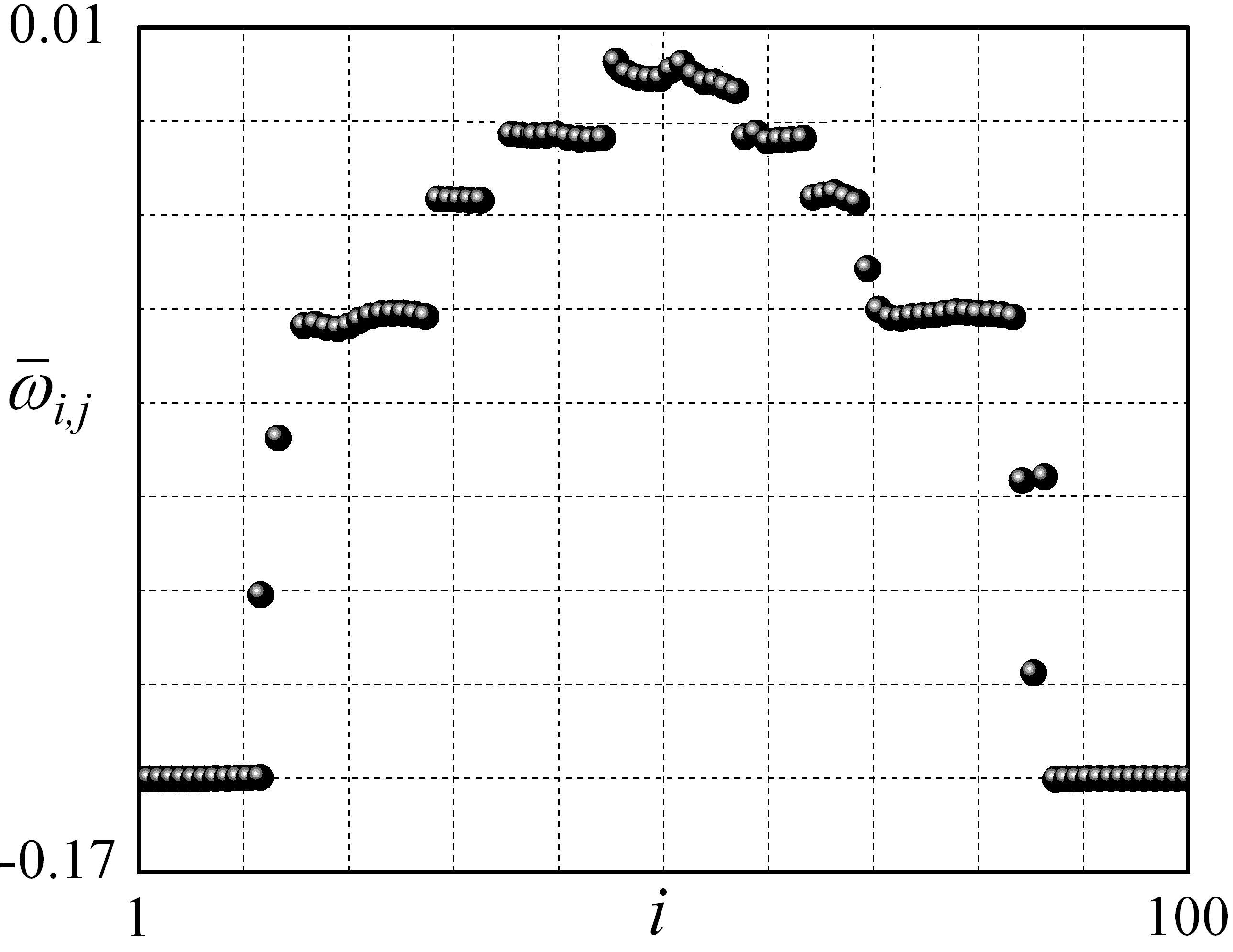}
  \caption{Examples of 4-core spiral wave chimeras with partially coherent cores (only one core is shown). 
In the rows from up to down: phase snapshots, time-averaged frequencies of the oscillators, 3D representation of the frequencies, and cross-sections of the frequencies. 
Parameters:
(a)  $\alpha=0.62, \mu=0.055$,   (b) $\alpha=0.72, \mu=0.027$,  (c) $\alpha=0.47, \mu=0.115$ , 
 (d) $\alpha=0.55, \mu=0.032$  . Other parameters:  (a,b,c) $N=600$, $P=96$, and the simulation time $t=10^4$ ; (d)   $N=800$, $P=56$, and the simulation time $t=2\times10^4$ .  Time-averaged frequency interval $\Delta T = 10^3$.}
 \label{f6} 
\end{figure*}

\vspace{-0.6cm}
\section{Conclusion. Diversity of the emerging network dynamics }
As it was demonstrated in Ch.2,  collective network dynamics becomes much richer, if inertia is added in the standard Kuramoto model. Solitary states arise making the system behavior highly multistable. They affect the spiral chimeras adding a solitary cloud to the core background. 
Chimera cores which are normally incoherent with a bell-shaped frequency distribution can contain then concentric rings of coherence with different averaged frequencies, i.e. become quasiperiodic \cite{OWK2018}. The most peculiar behavior arises when the cores become frequency-coherent, and we find that this kind patterns exists in a wide enough region of the system parameters.

\begin{figure}[ht! ]
\vspace{-0.6cm}
\begin{center}
 \includegraphics[width=0.35\linewidth]{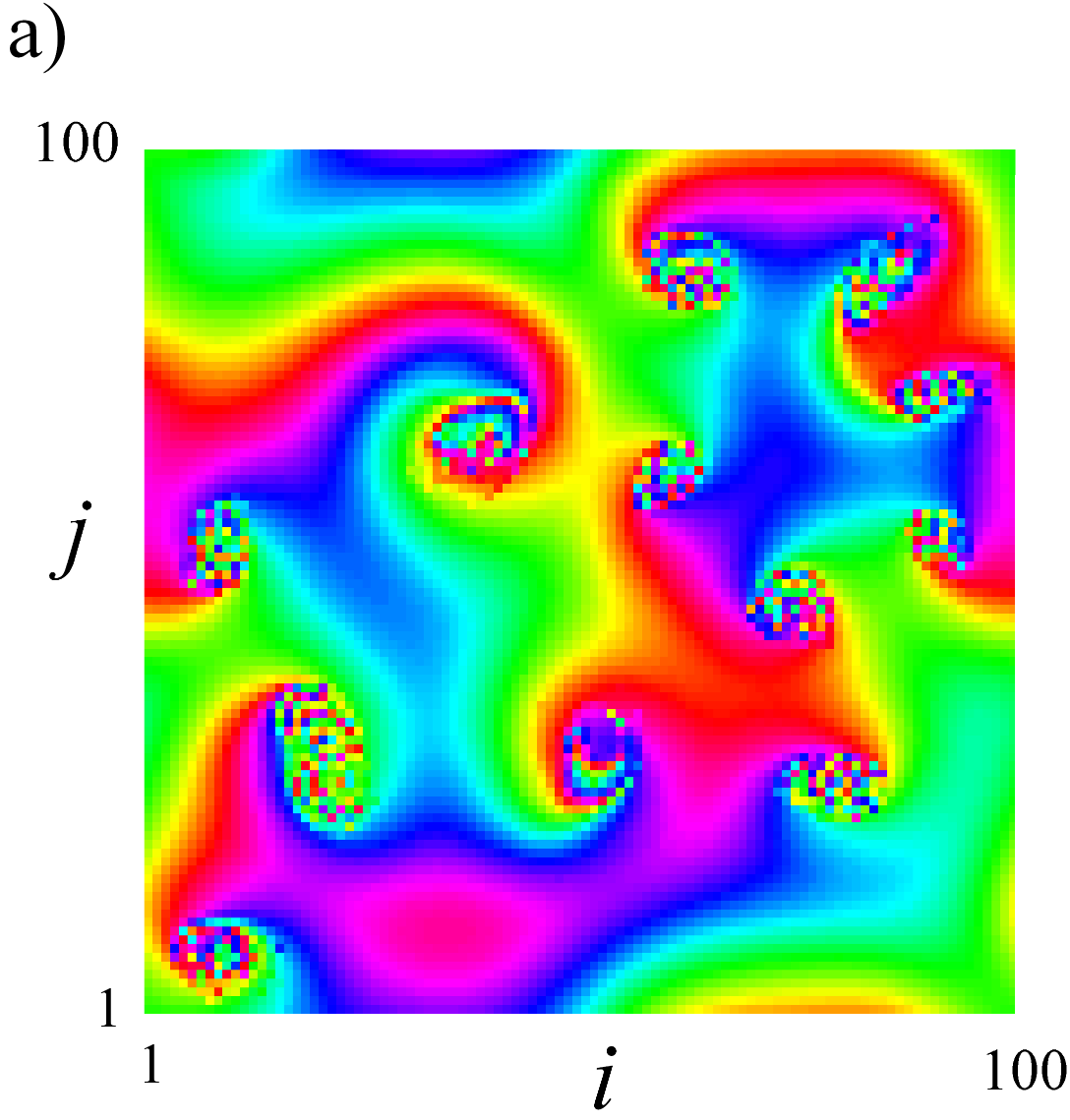}  \qquad  \includegraphics[width=0.35\linewidth]{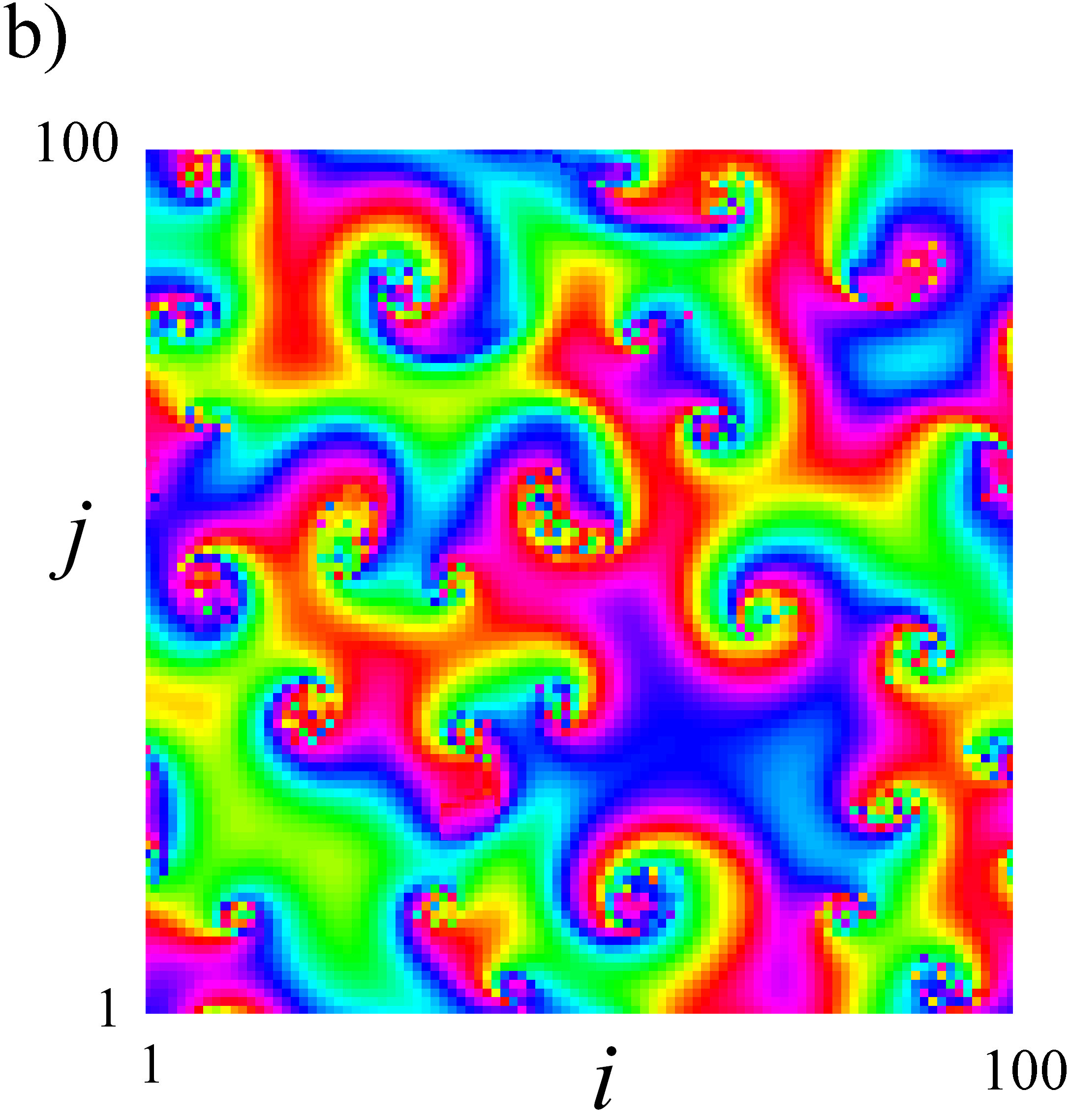} \ \ \includegraphics[width=0.045\linewidth]{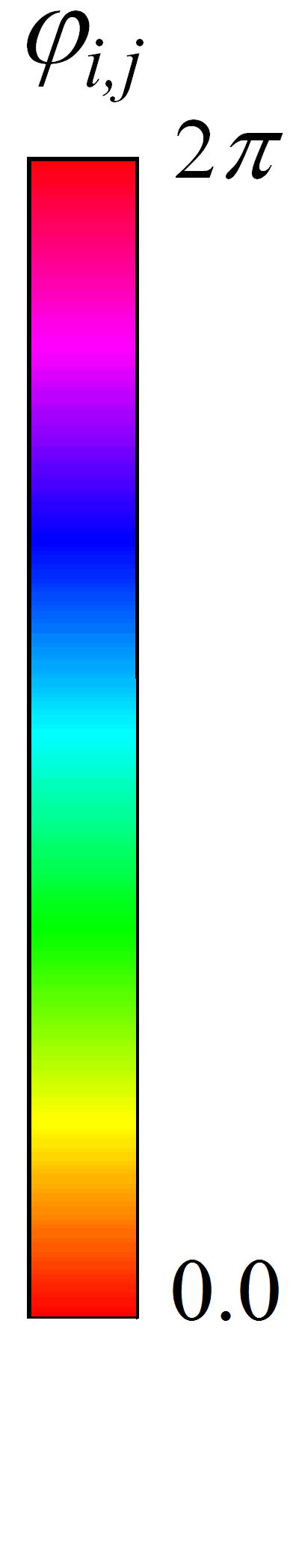} \\
 \includegraphics[width=0.35\linewidth]{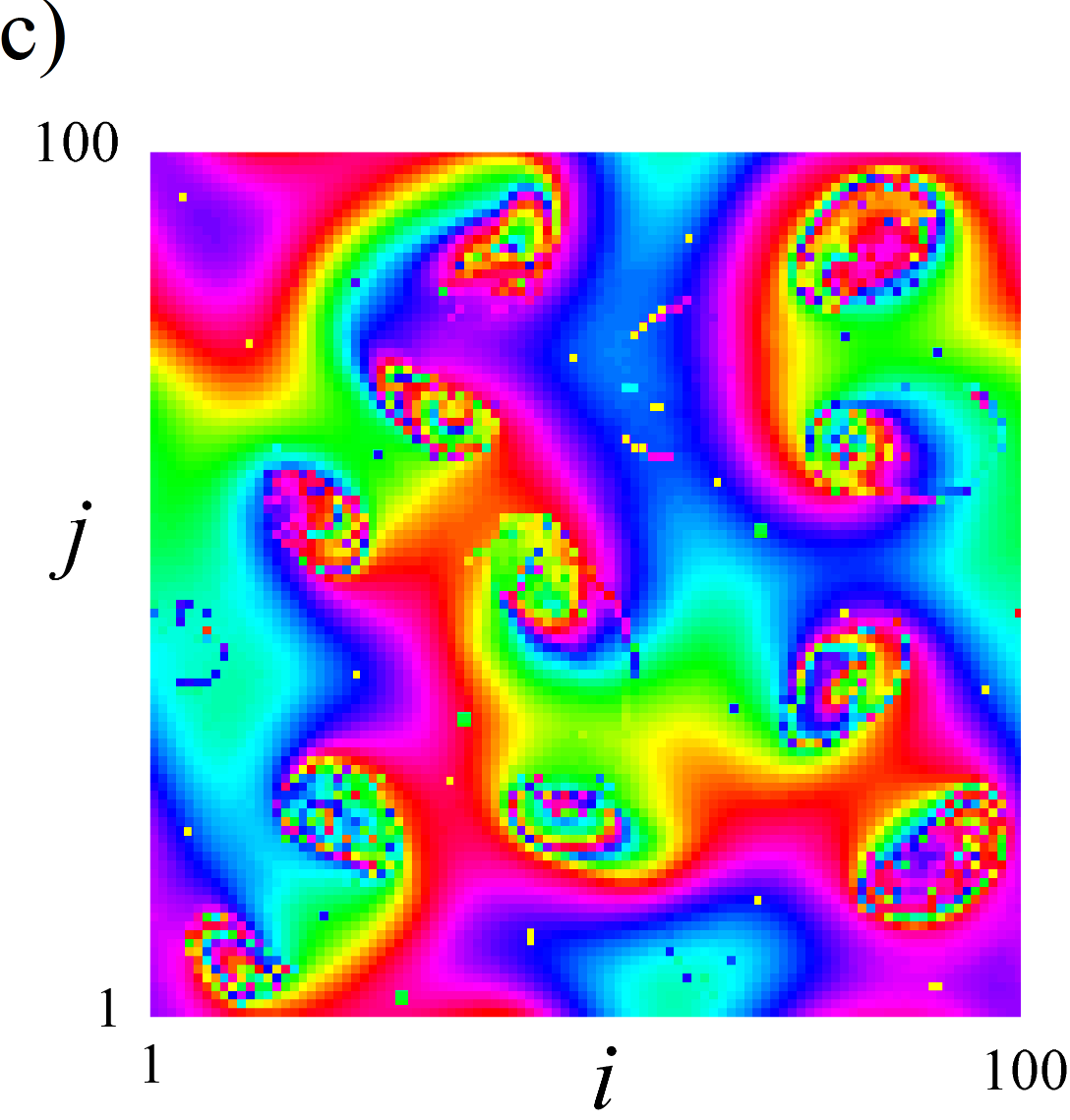}  \qquad \includegraphics[width=0.35\linewidth]{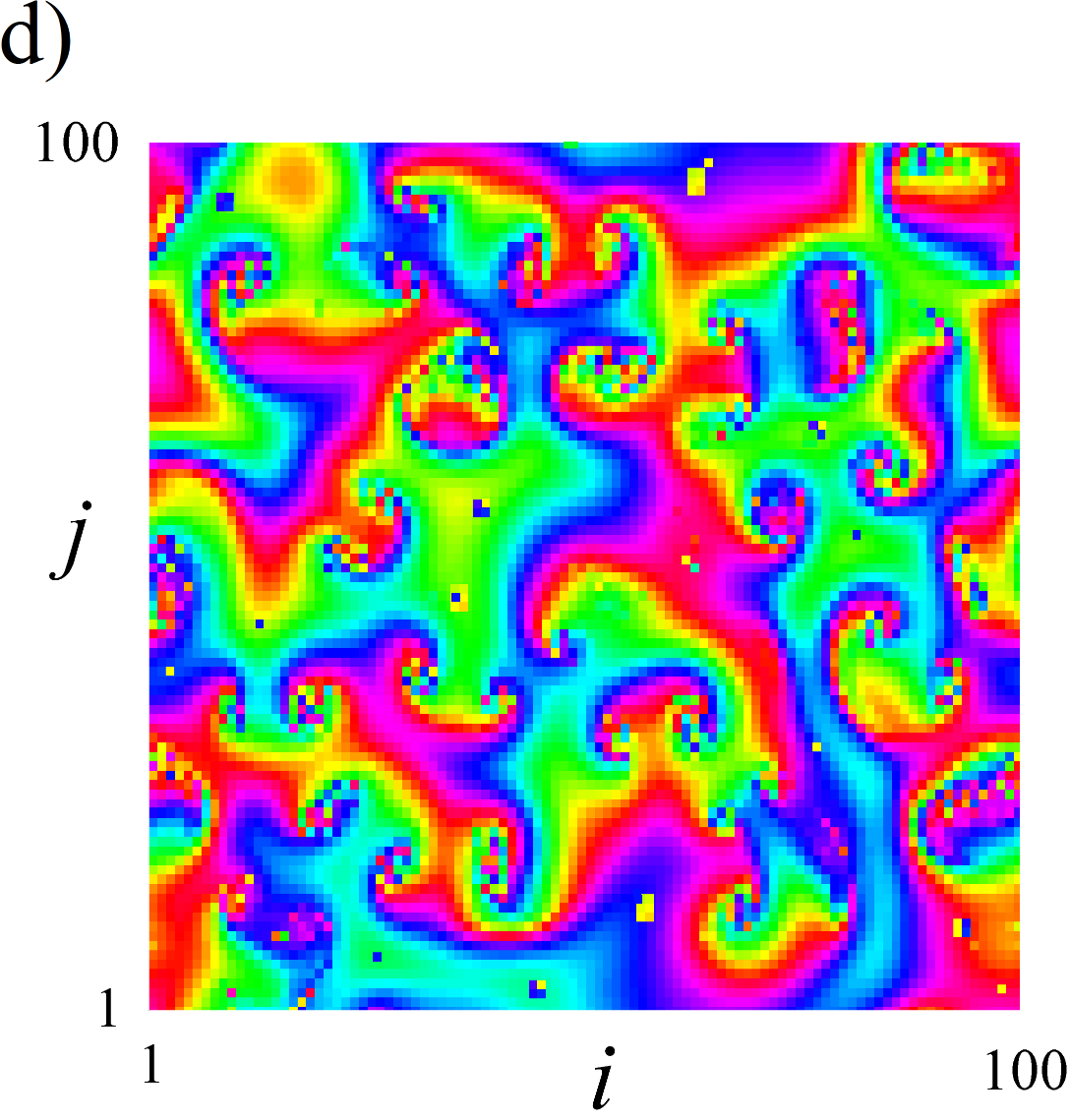} \ \ \includegraphics[width=0.045\linewidth]{Faza-insert-Fig7.png}
 \includegraphics[width=0.35\linewidth]{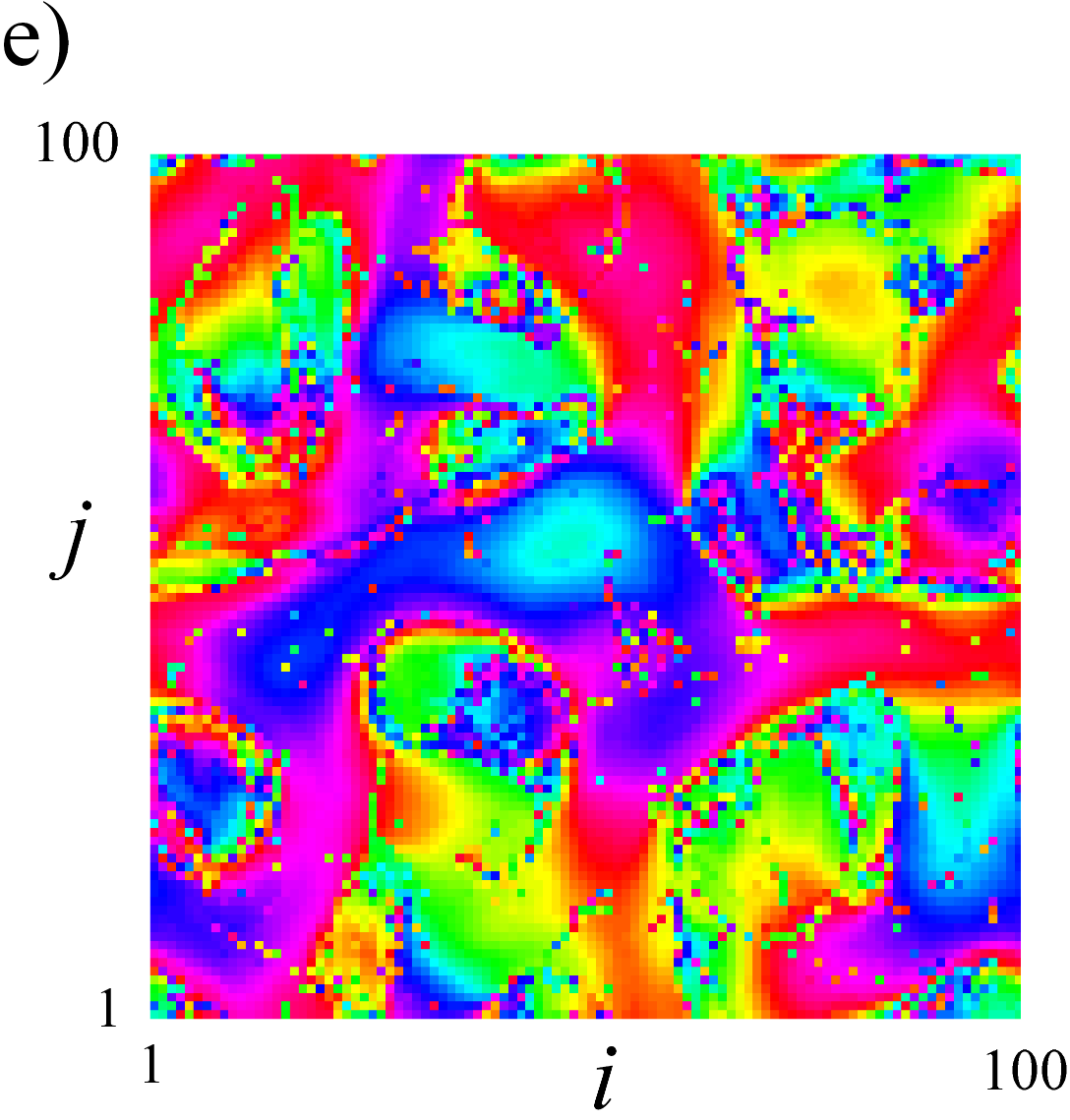}  \qquad \includegraphics[width=0.35\linewidth]{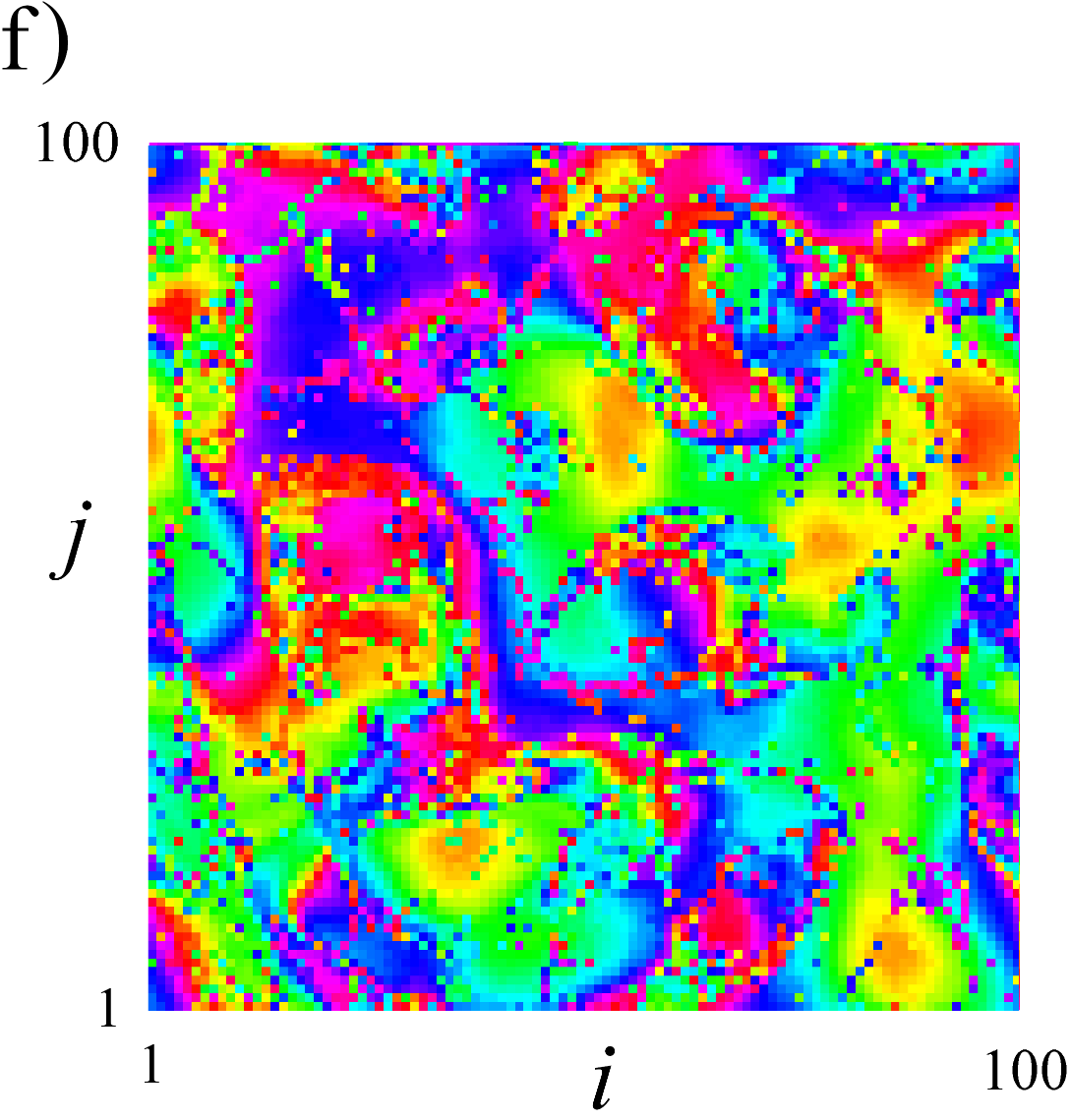} \ \ \includegraphics[width=0.045\linewidth]{Faza-insert-Fig7.png}
\vspace{-0.2cm}
\end{center}
 \caption
{Typical phase snapshots of multiple spiral wave chimeras outside (a, b) and inside  (c, d) of the solitary region,  and spatiotemporal patterns (e, f) at a further increase of the parameter $\mu$. Parameters in the left panel: $P=7$,  $\alpha=0.8$, and  $\mu=0.018,  0.032$, and $0.065$ in a), c), and e), respectively. The right panel: $P=5$, $\alpha=0.9$, and  $\mu=0.025, 0.030$, and $0.055$ in b), d), and f), respectively.  $N=100$.}
 \label{f7}
\end{figure}

In this chapter, we present examples of other peculiar behaviors in  model (1). First, we illustrate in Fig.7 (a) and (b) that the number of incoherent cores in spiral wave chimeras can vary in a wide range, from the minimal 2 and 4  (as in Ch.2) to much larger numbers. There are 12 chimera cores in Fig. 7(a) and 26 in 
(b), respectively.  In both cases, as our simulations confirm, the cores are not stationary standing, 
in contrary, they are fast moving throughout the phase space in a visually chaotic manner. Then the resulting collective behavior may be referred to as  a spiral chaos.
Note that  both examples, (a) and (b), are obtained for parameters outside the solitary region. There are no solitary oscillators on the spiraling background, and, hence, they resemble
 standard 2D chimeric patterns known for the Kuramoto model without inertia, see \cite{ks2003,sk2004,kkjm2004}. 

\begin{figure*}[ht! ]
\vspace*{-0.8cm} 
\begin{center}
\hspace*{-1.0cm}
{ \bf Phase snapshot} \\
 \includegraphics[width=0.28\linewidth]{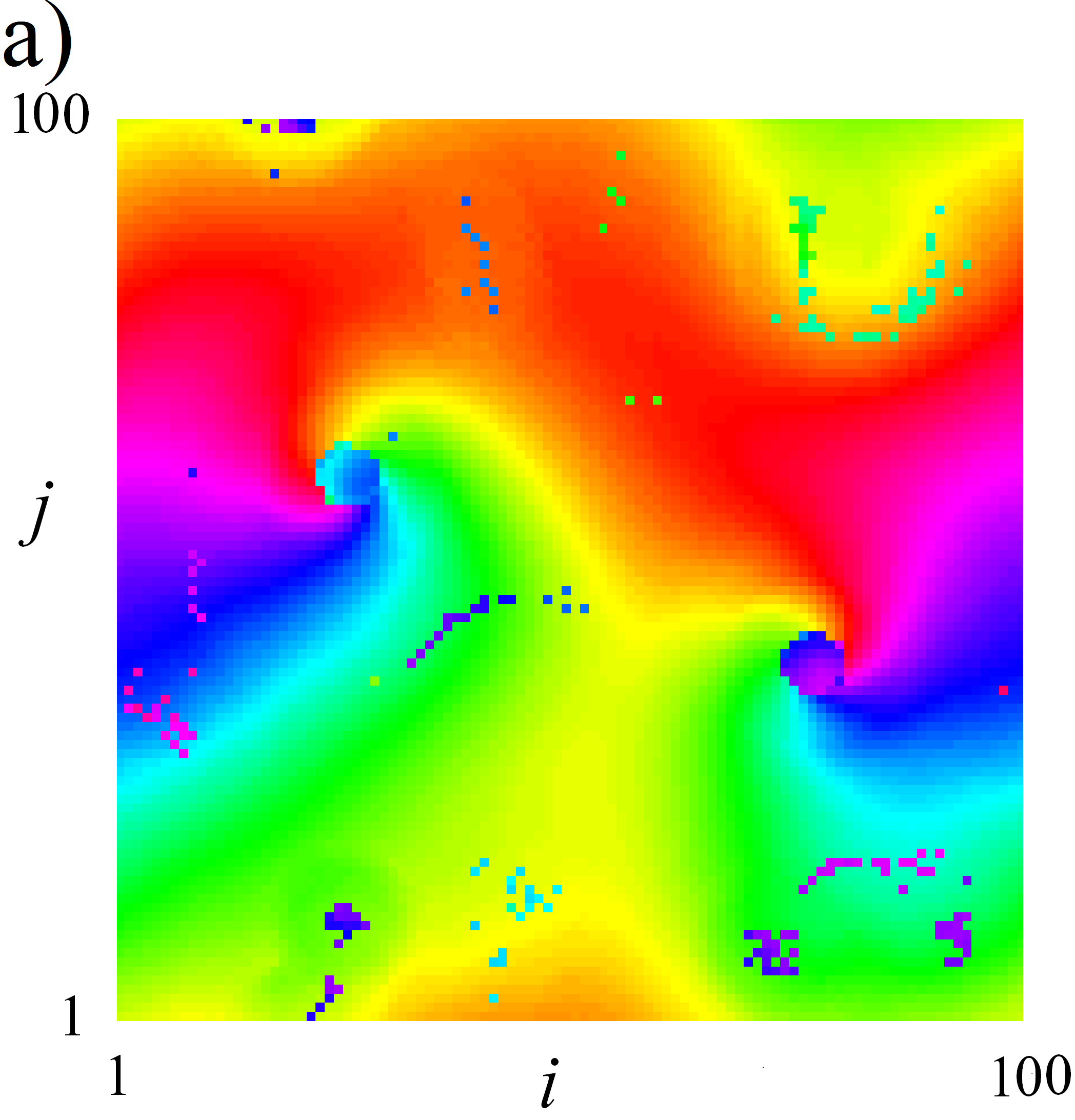}  \quad 
 \includegraphics[width=0.28\linewidth]{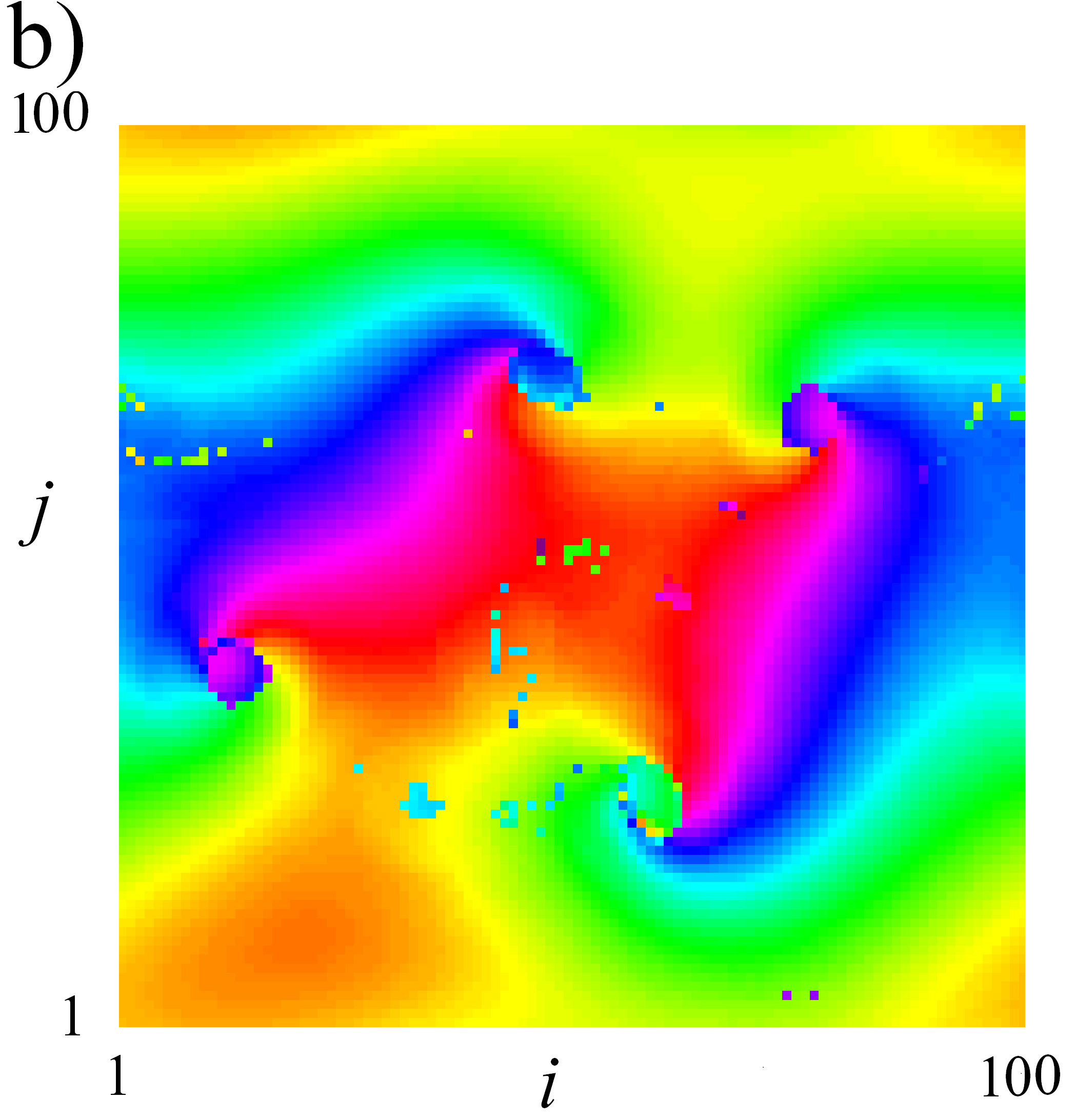}   \quad 
 \includegraphics[width=0.28\linewidth]{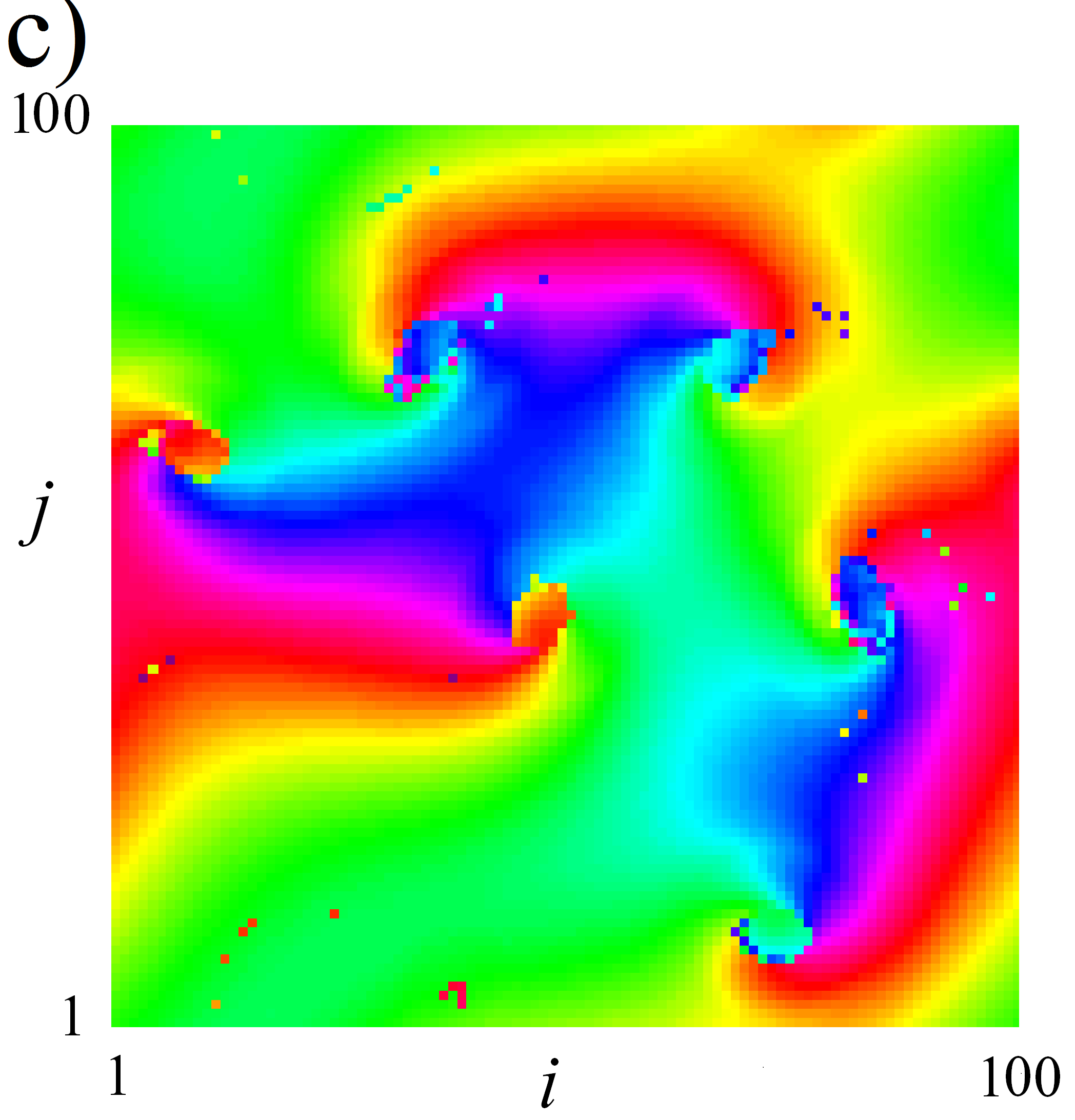}  
\includegraphics[width=0.06\linewidth]{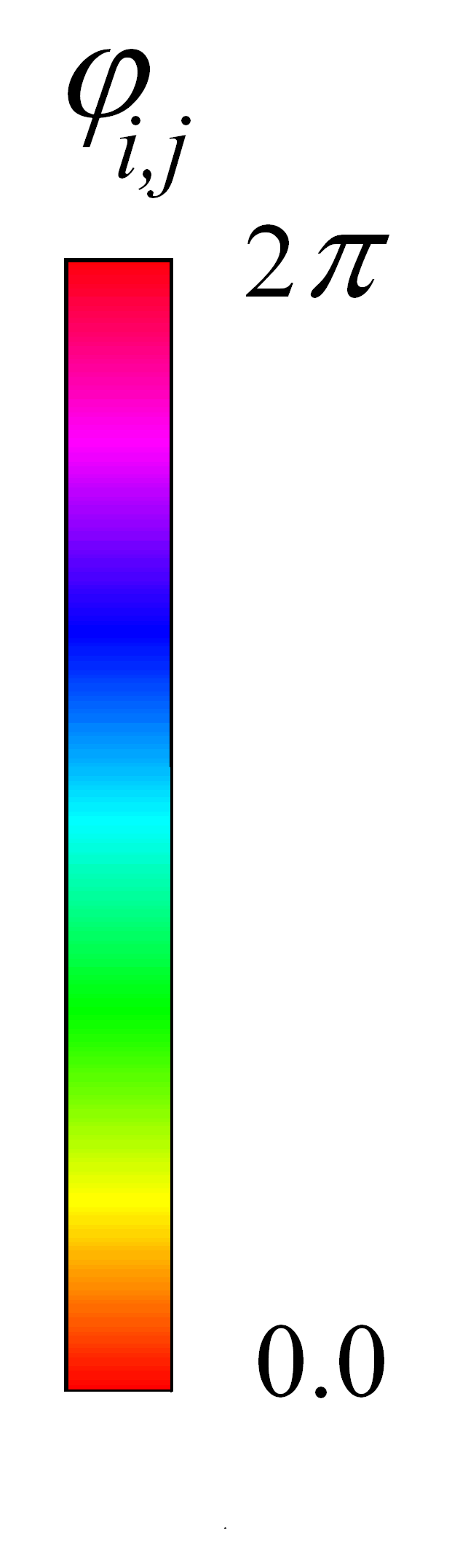} \\
\vspace*{0.1cm} 
\hspace*{-0.9cm}
{ \bf Time-averaged frequencies} \\
   \includegraphics[width=0.28\linewidth]{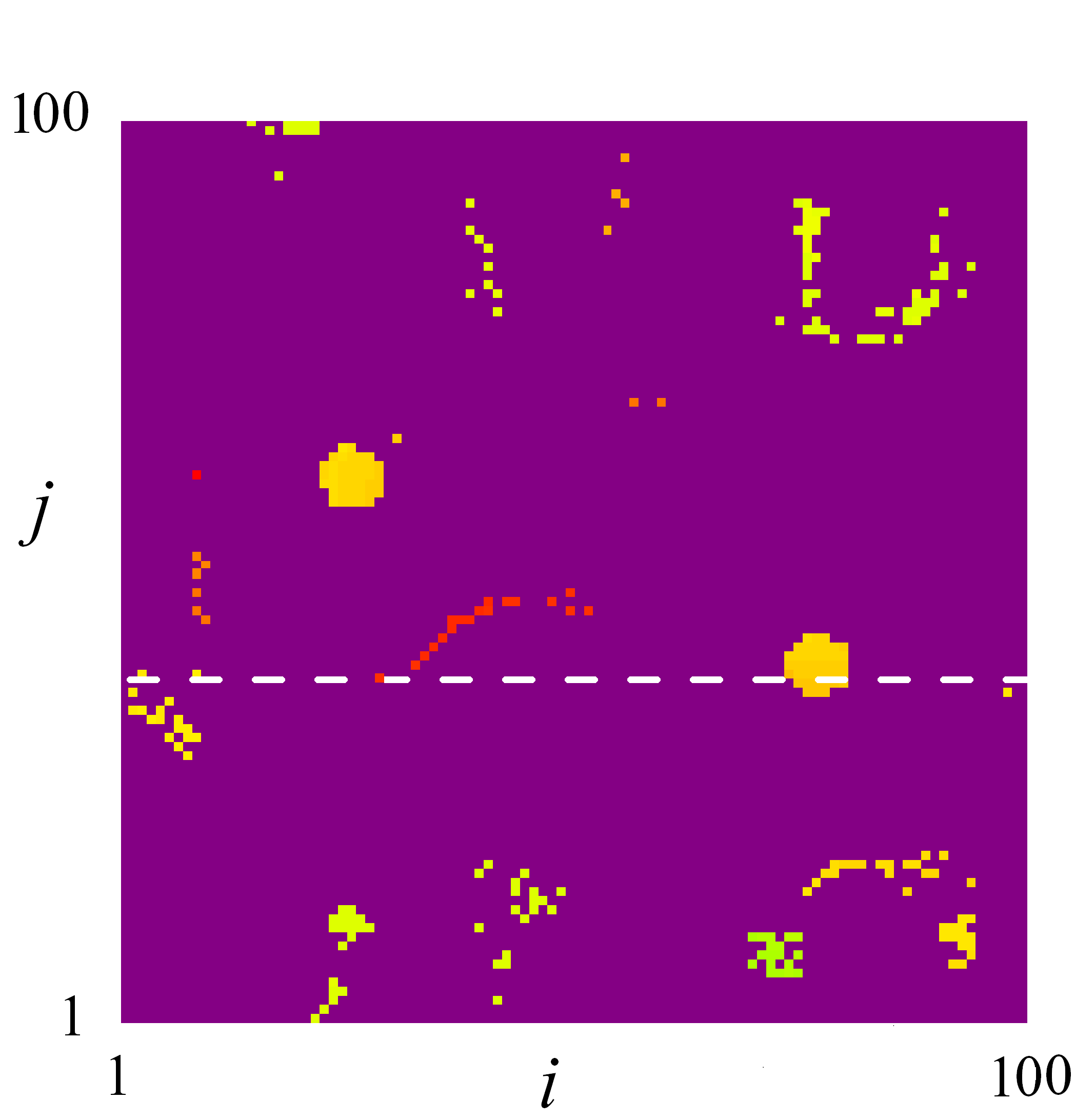}  \quad 
 \includegraphics[width=0.28\linewidth]{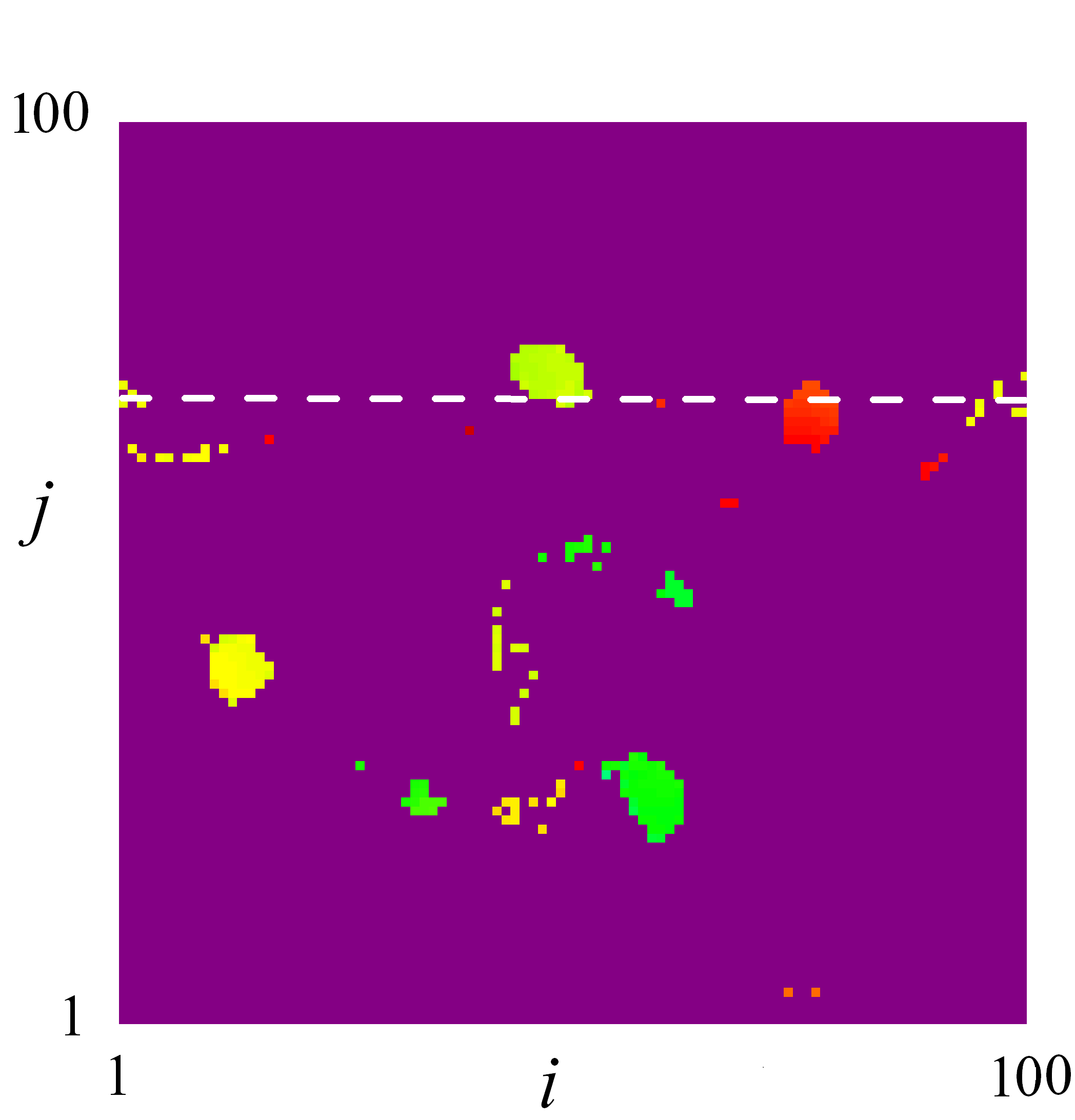}   \quad 
 \includegraphics[width=0.28\linewidth]{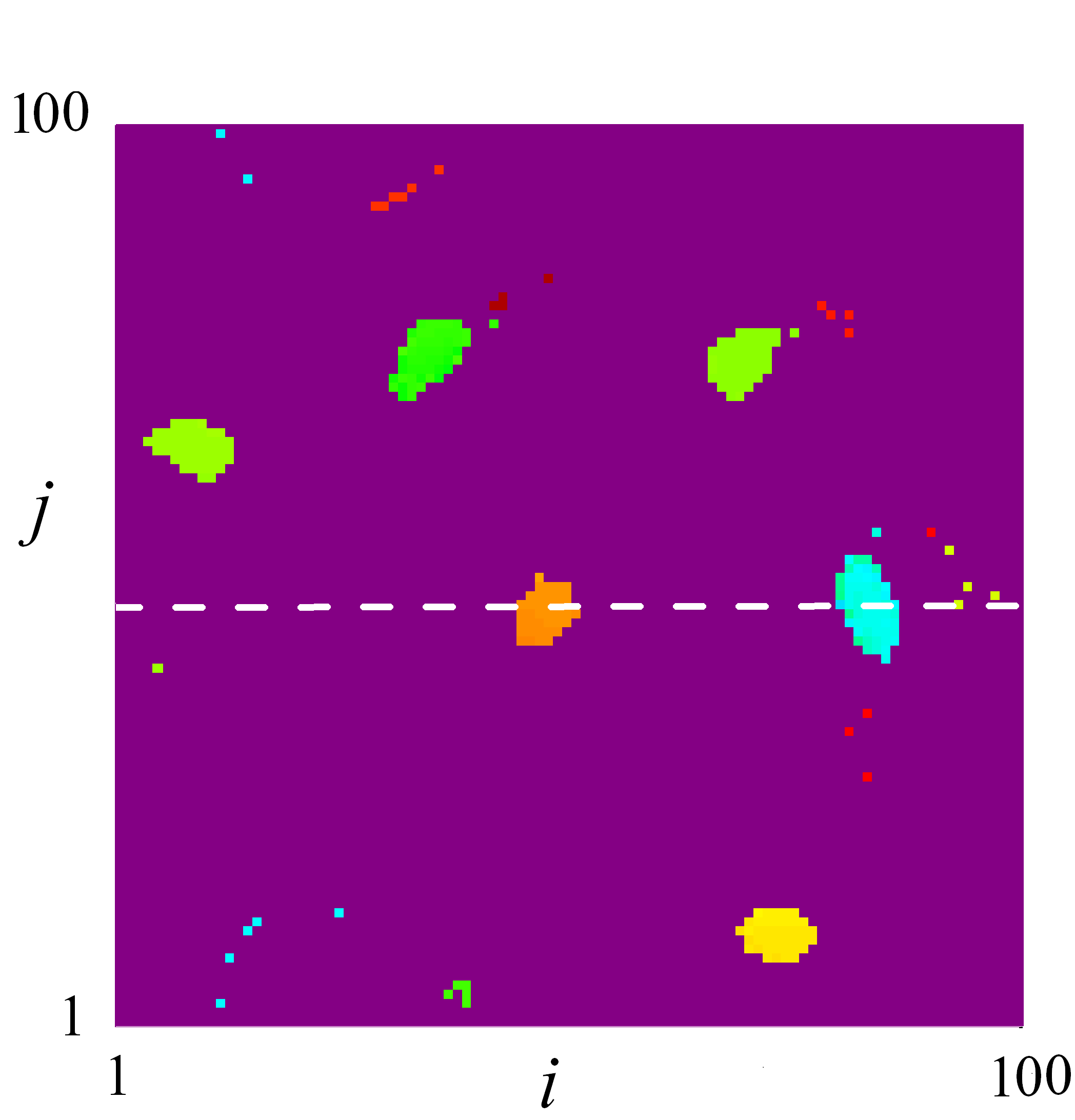}  
\includegraphics[width=0.06\linewidth]{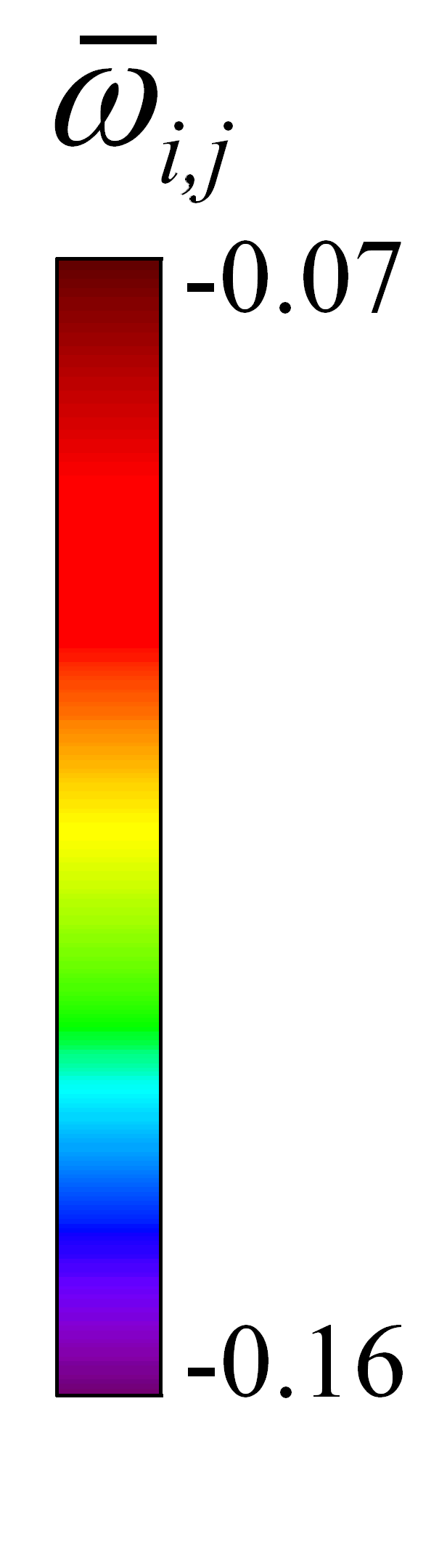}  \\
\vspace*{0.1cm} 
\hspace*{-0.9cm}
{ \bf Slice of the averaged frequencies} \\
 \end{center}
\vspace*{-0.3cm}
\hspace*{0.05cm}
 \includegraphics[width=0.28\linewidth]{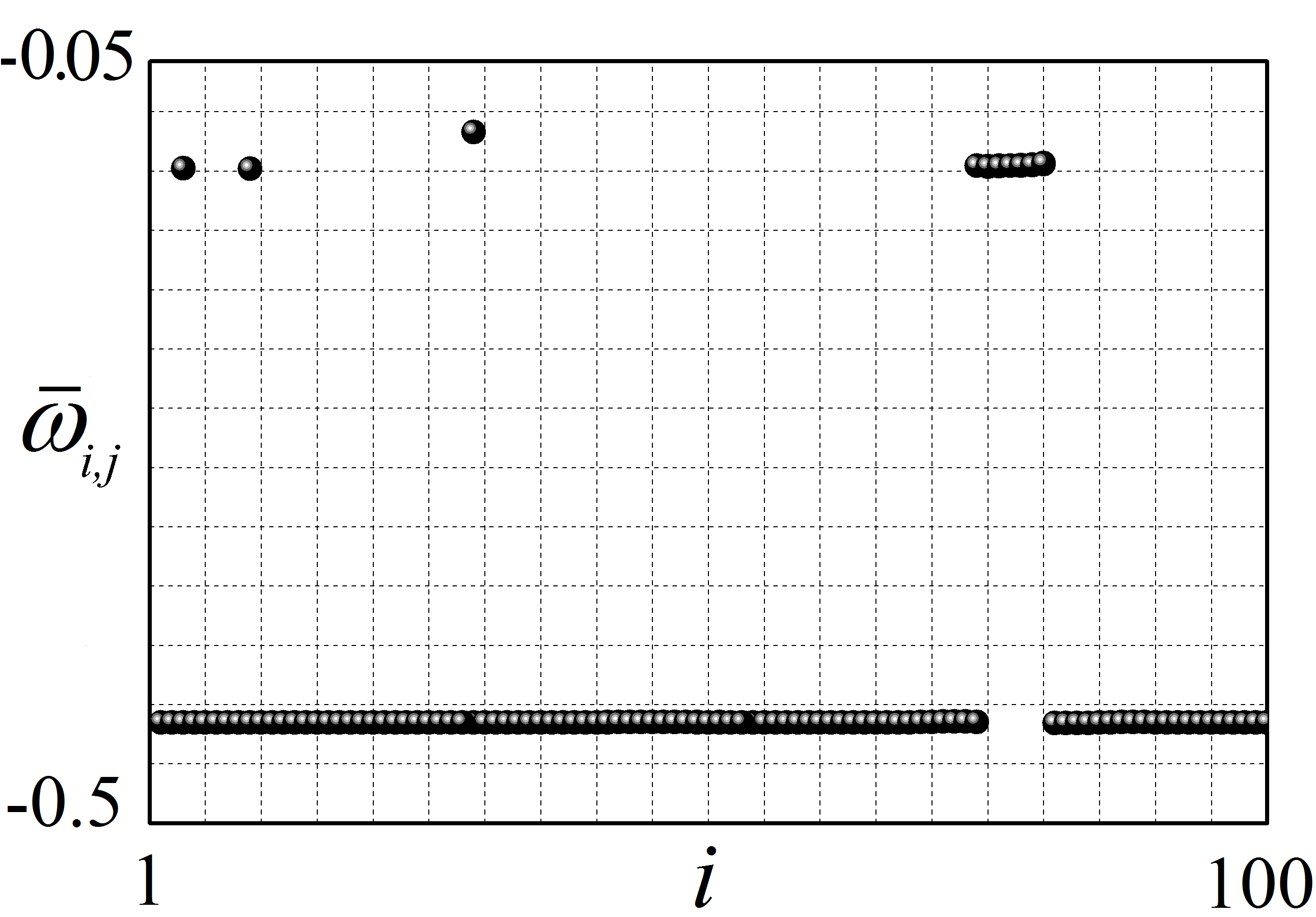}  \quad 
 \includegraphics[width=0.28\linewidth]{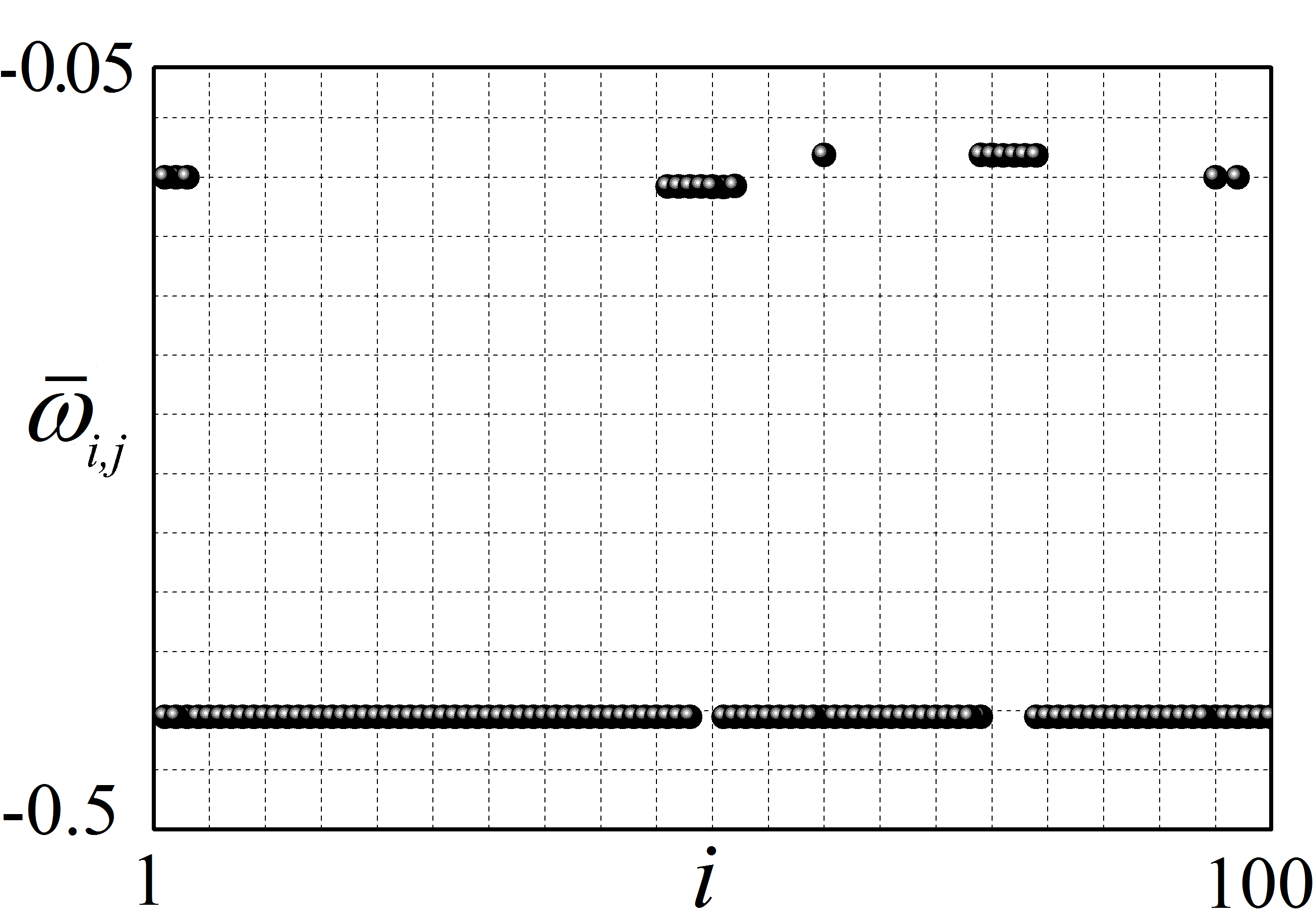}  \quad 
\includegraphics[width=0.28\linewidth]{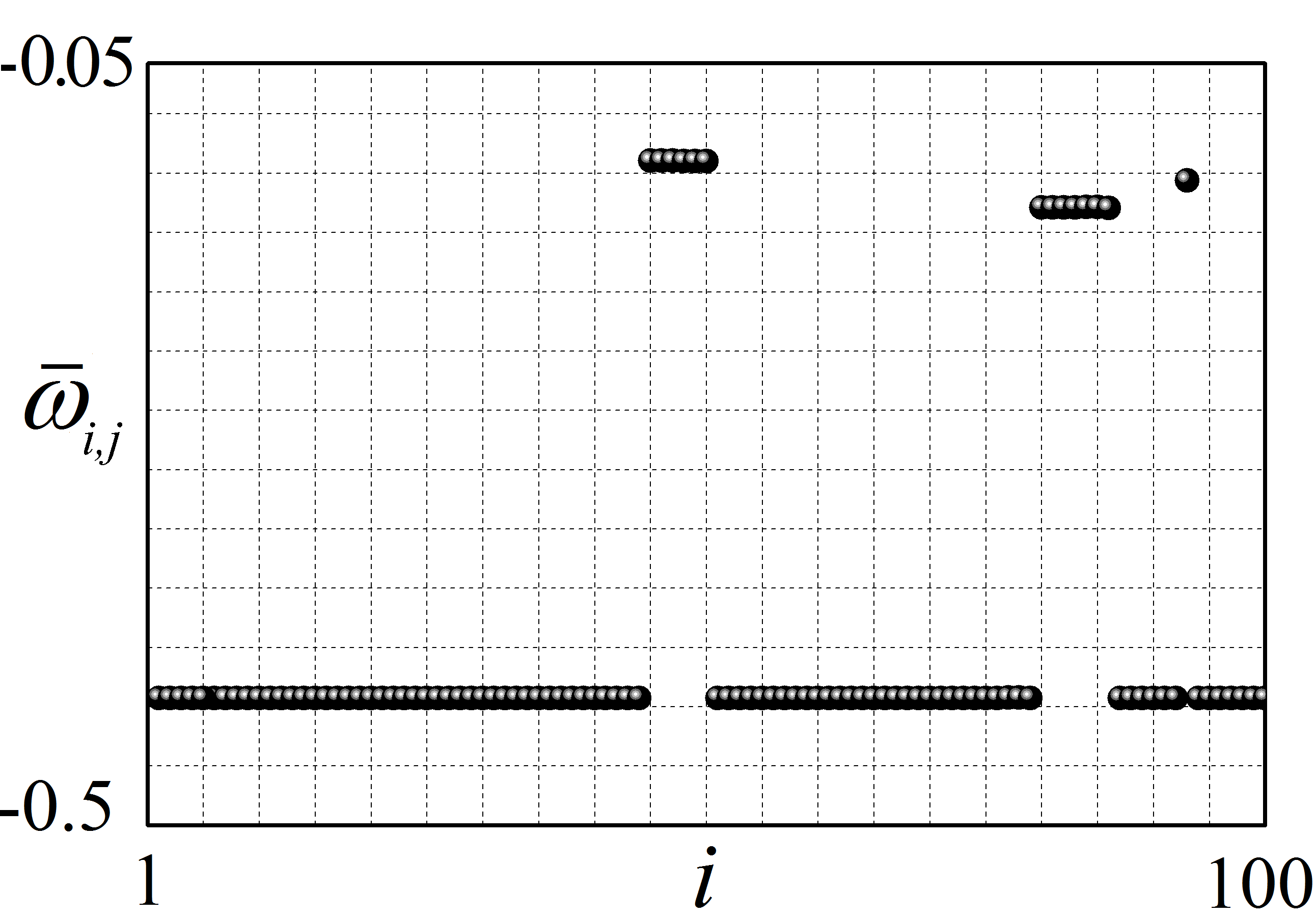}    \\
\vspace*{-0.4cm}
\begin{center}
\hspace*{-0.9cm}
{ \bf  Averaged frequencies in ordered oscillator index} \\
 \end{center}
\vspace*{-0.2cm}
\hspace*{0.002cm}
   \includegraphics[width=0.285\linewidth]{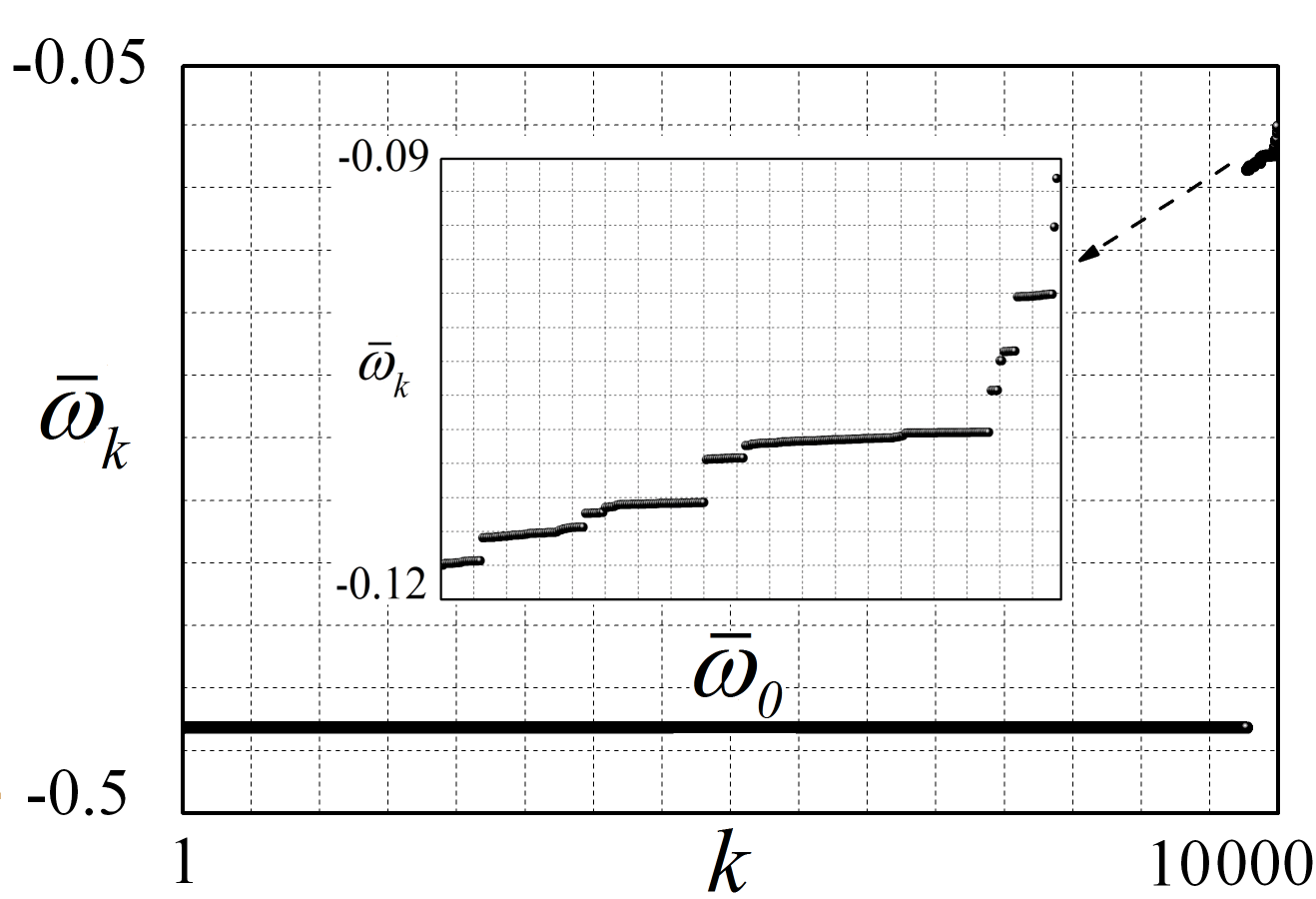} \ \ \ 
  \includegraphics[width=0.285\linewidth]{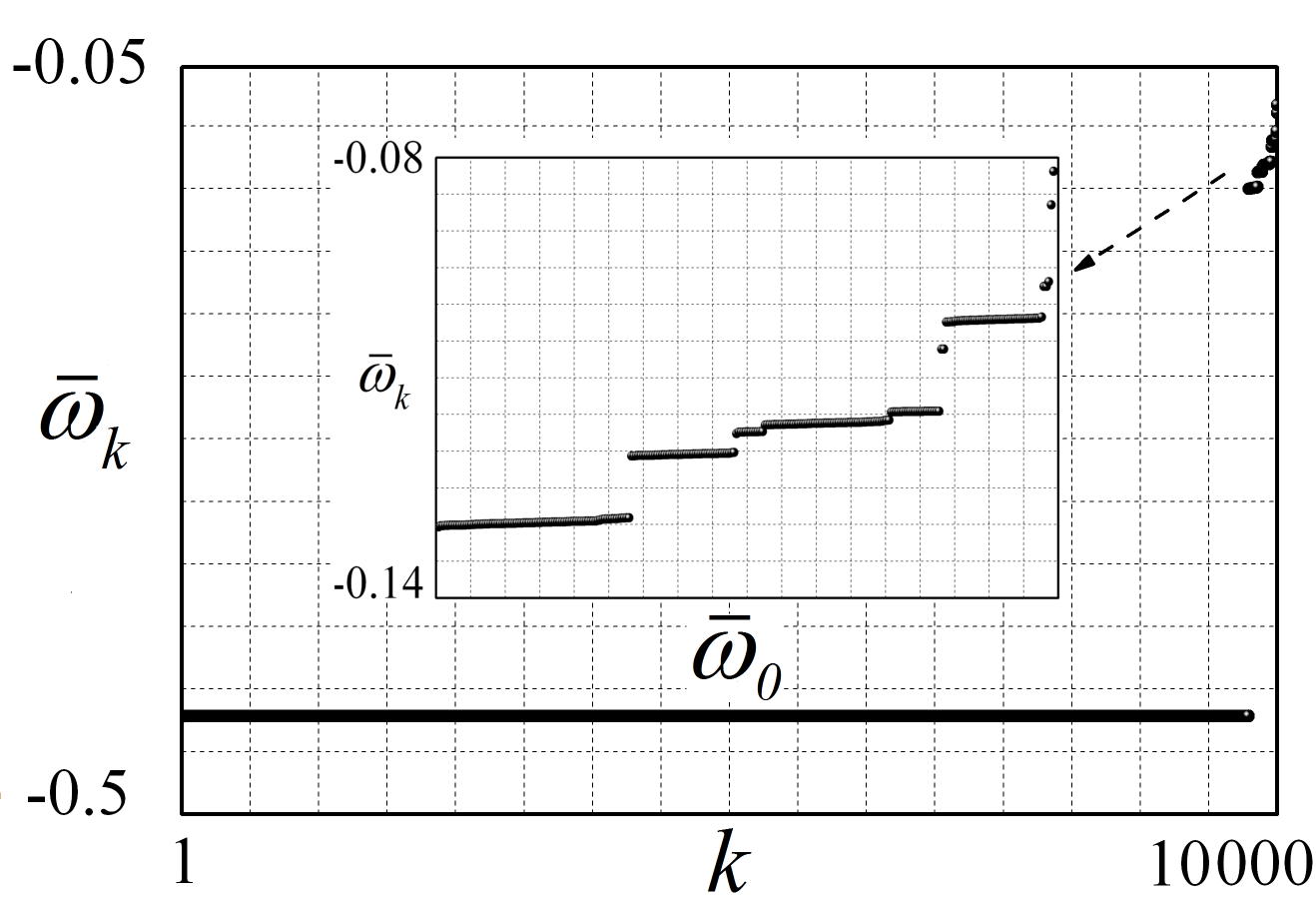}   \ \ 
  \includegraphics[width=0.285\linewidth]{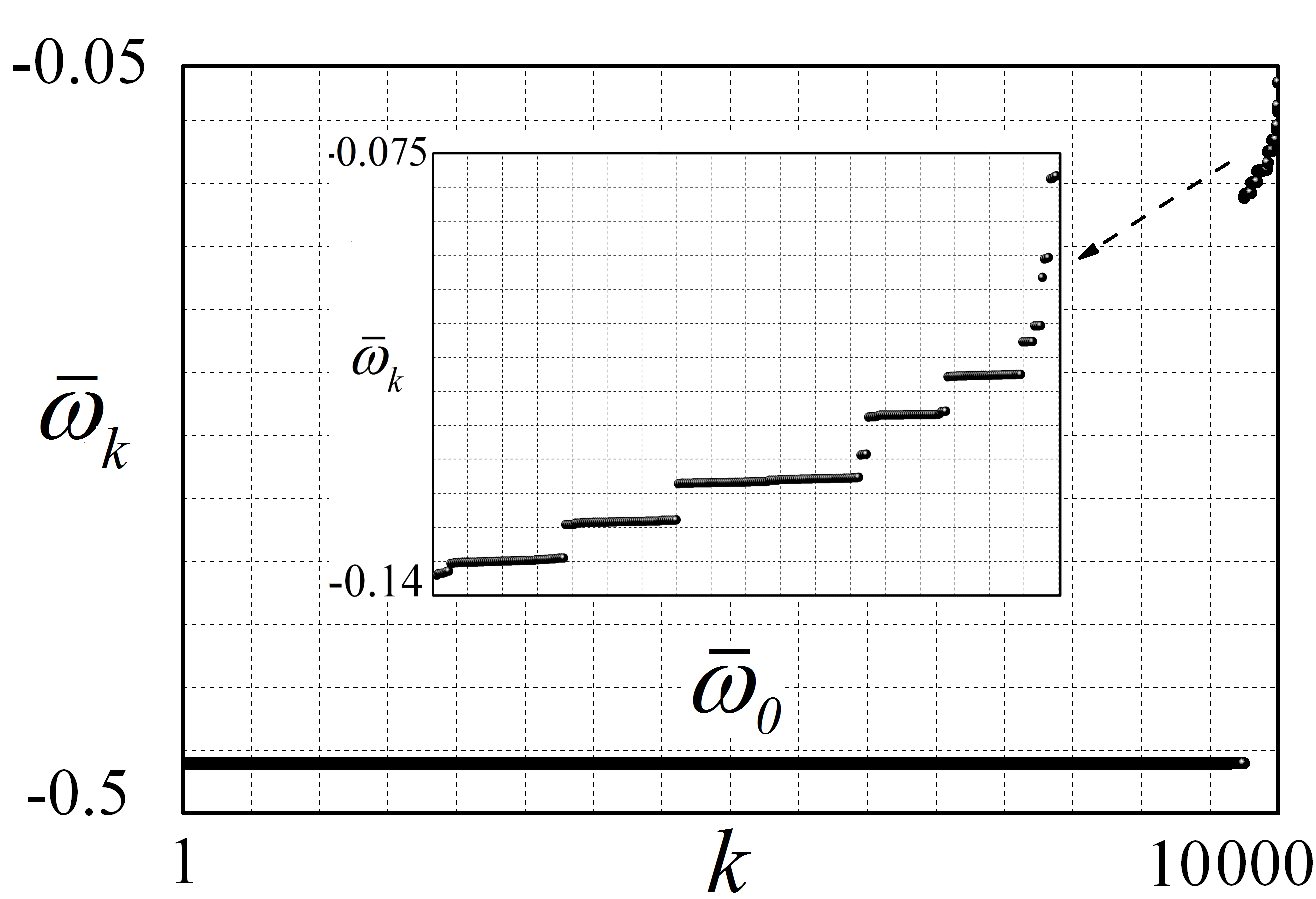}   
\vspace*{-0.1cm}
\caption
{Pattern variability in the solitary region of model (1).  Examples  of  2-, 4-, and 6-core spiral wave chimera states  obtained from random initial conditions at the same parameter values $\alpha=0.45$ and $\mu=0.11$ (left, middle, and right columns, respectively).  Other parameters $N=100$, $P=7$. Simulation time $t=2\times10^{5}$. Frequency averaging interval $\Delta T = 1000$.}
\end{figure*}

\begin{figure*}[ht! ]
\begin{center}
\vspace*{-0.4cm} 
 \hspace*{-4.2cm}
{ \bf  {\large a)} \qquad  \qquad  \qquad Phase snapshot } \\

   \includegraphics[width=0.54\linewidth]{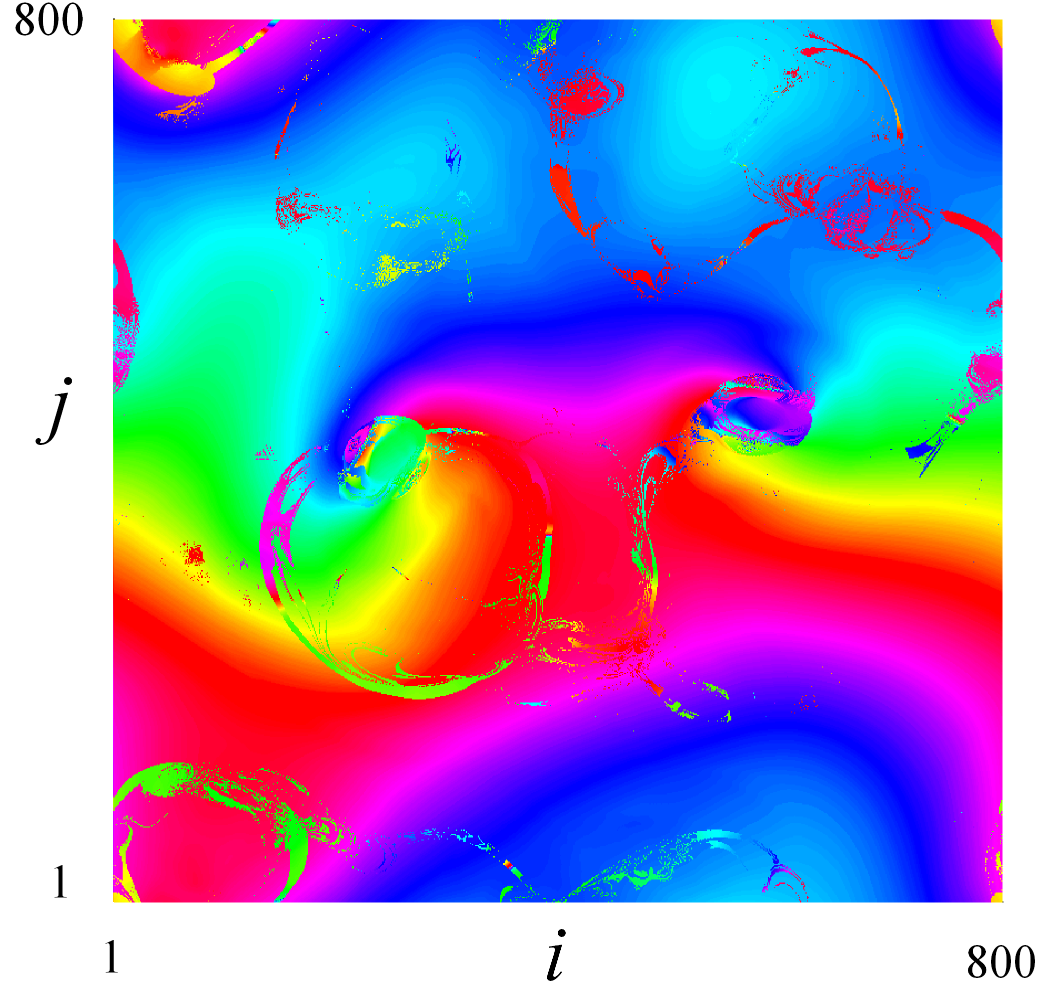}   
\includegraphics[width=0.08\linewidth]{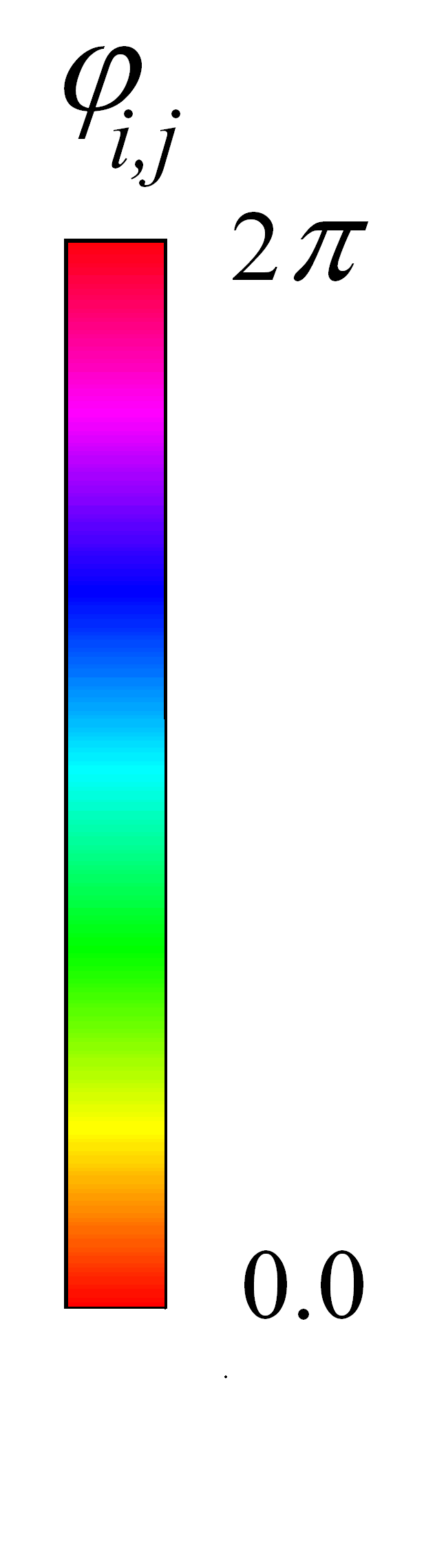}  \\
 \hspace*{-3.2cm}
{ \bf     {\large b)}   \qquad  \qquad Time-averaged frequencies} \\
\vspace*{-0.05cm} 
   \includegraphics[width=0.54\linewidth]{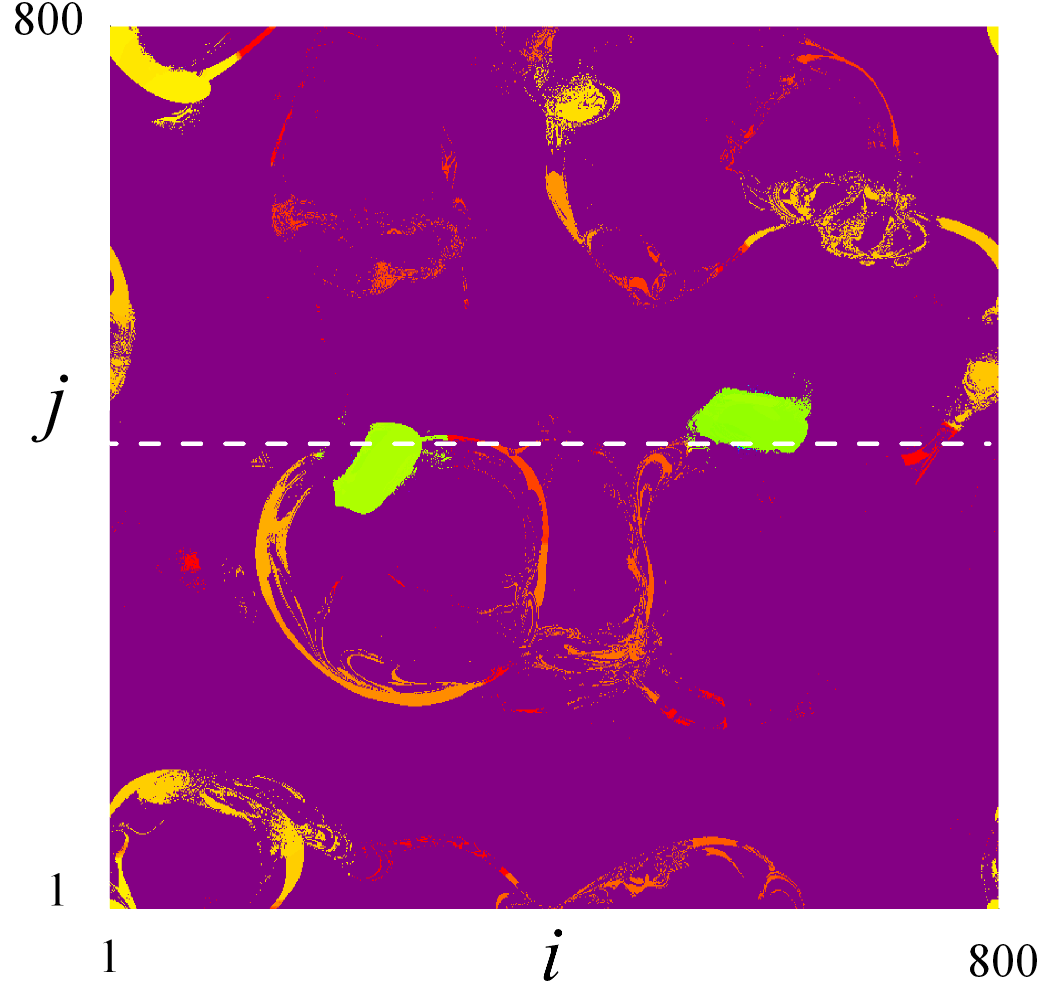}  
\includegraphics[width=0.08\linewidth]{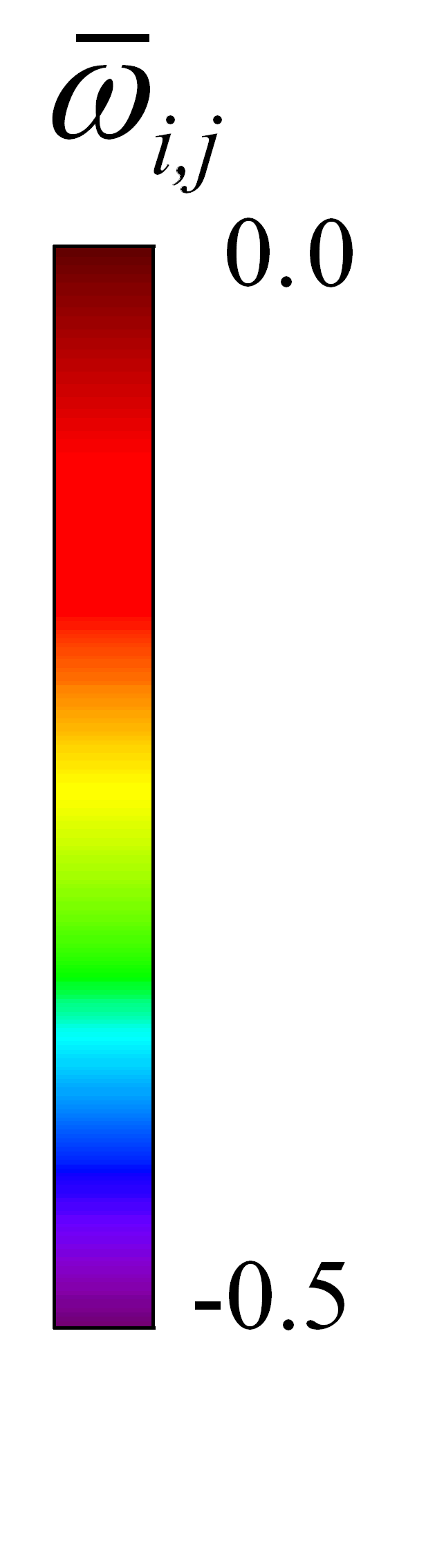}  \\
 \hspace*{-2.85cm}
{ \bf     {\large c)}     \qquad   \qquad   Frequency slice along  $j=419$} \\

\hspace*{-1.0cm} 
   \includegraphics[width=0.5\linewidth]{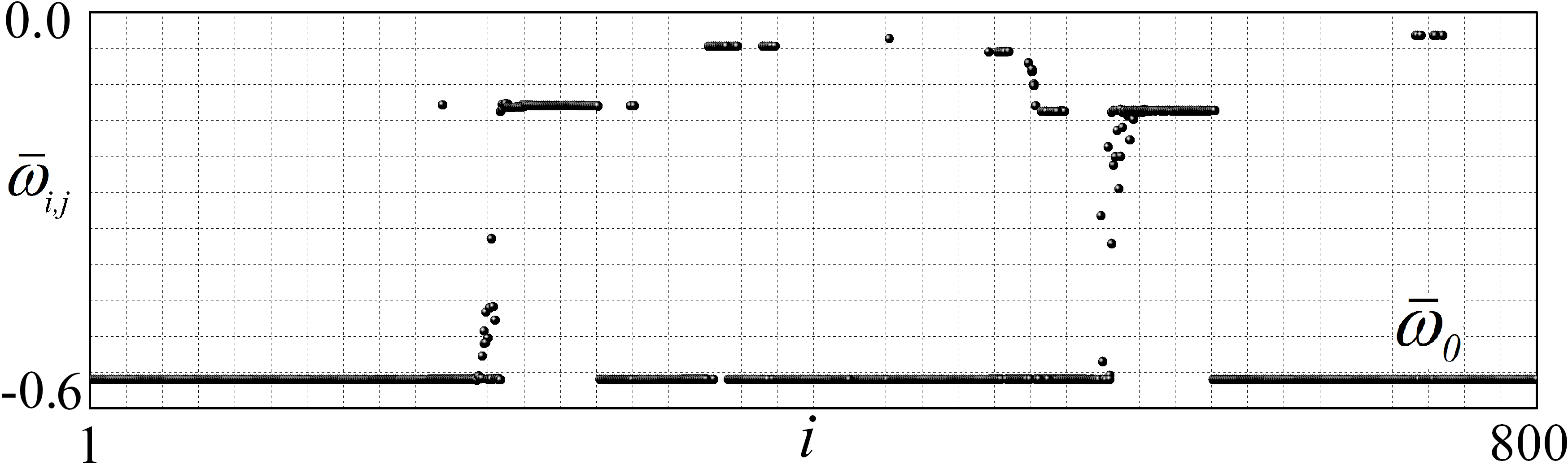} 

 \hspace*{-2.1cm}
{ \bf    {\large d)}    \qquad   Frequency in ordered oscillator index} \\
\hspace*{-1.0cm} 
\includegraphics[width=0.5\linewidth]{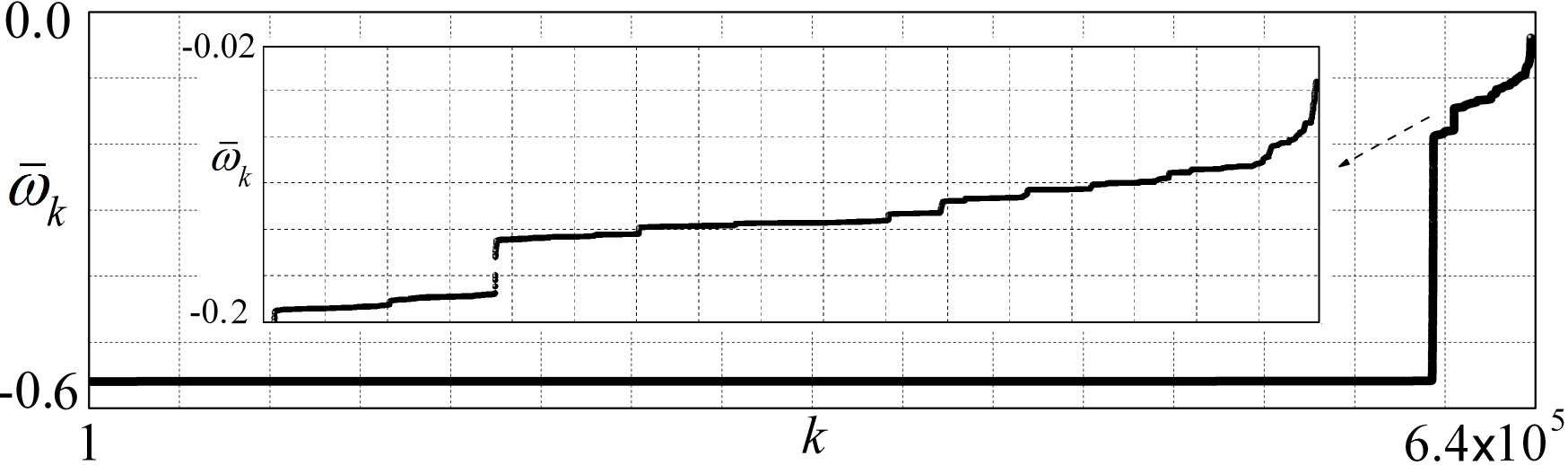}
\end{center}
 \vspace*{-0.3cm} 
\caption
 {Two-core spiral wave chimera with a solitary cloud: 
(a) snapshots of the phase distributions  $\varphi_{i,j}$, (b) time-averaged  frequencies $\bar{\omega}_{i,j}$, (c) cross-section of time-averaged  frequencies $\bar{\omega}_{i,j}$ along the dashed white line in (b) at $j=419$, (d)  ordered oscillator index   $\alpha=0.45, \mu=0.15, N=800$, $P=56$. $t=10^{4}$. Time-averaged frequency interval $\Delta T = 1000$.}
 \label{f9}
\end{figure*}

When entering the solitary region, as illustrated in Fig.7 (c) and (d), numerous solitary oscillators arise outside the cores forming a solitary cloud, and the cores themselves undergo transformations loosing the circular shape. With further increase of the coupling strength $\mu$, the solitary cloud becomes denser, and the spiraling is terminated as illustrated in Fig.7(e) and (f). The arising spatiotemporal patterns can be considered as solitary states with chaotic dynamics, similar to the state in Fig. 4(e). Chaotization of the states is confirmed by positive Lyapunov exponents. Our simulations confirm numerous positive Lyapunov exponents, which number correlates as we assume with the total number of solitary oscillators visually seen in the figures; with that, the maximal Lyapunov exponents there are about $0.05$ for (e) and $0.06$ for (f). 

To illustrate the spiral chimera variability in the solitary region, three more examples are shown in Fig.8  with spiral chimeras including 2, 4, and 6 cores (left, middle, and right columns, respectively). The states were obtained in simulations of model (1) at the fixed parameters $\alpha=0.45$ and $\mu=0.11$, but with different, randomly chosen  initial conditions.  The time-averaged frequencies of the oscillators are shown in the second panel, they testify that all spiral cores are coherent.  The core frequencies are equal in the first, 2-core example (left panel), but do not in the  4- and 6-core examples (middle and right panels).  

 Note also that the solitary cloud around the cores appears to be rather poor, as the parameter point in all examples lies quite close to the solitary boundary (see \cite{JMK2015,JBLDKM2018} for the discussion of this issue in the 1D case).  The cross-sections of the time-averaged frequencies along the white blank line are shown in the third panel. As it can be seen there, oscillators from the clouds try to follow the core-frequencies. A more detailed inspection  assures, however, that, in fact, their frequencies are slightly different,  forming a striking stepwise constant shape shown in the  low panel of the figure in the ordered oscillator index.

Our last example shown in Fig.~9 is a result of long simulations of the model (1) with the parameters $N=800$ and $P=56$ (which represent a high resolution analog of the $N=100, P=7$ case, mostly considered above), and starting from random initial conditions. The phase lag parameter $\alpha$ is chosen the same as in Fig. 8, $\alpha=0.45$, while the coupling strength is taken stronger, $\mu=0.15$. The stronger coupling causes, as one can observe, a more developed structure of the solitary cloud, as well as a richer frequency scattering of the solitary oscillators. More than 15 different frequencies are detected, as it can be concluded from (d) where the frequencies are arranged by the ordered oscillator index.

In summary, we have identified a novel scenario for the spiral chimera transition in networks of coupled oscillators with inertia. The transition starts from the 'classical' Kuramoto-Shima's spiral chimera state with bell-shaped frequency characteristics of the incoherent cores, develops through the states with in-core frequency plateaus (quasiperiodic chimeras), and goes eventually into a spiral chimera with the completely coherent cores. 

This peculiar state with coherent cores exists in a wide enough parameter region. At the boundary of the region, it drops (in a crisis bifurcation) to a sort of the spatiotemporal chaos in the form of a chaotic mixture of the numerous solitary oscillators with laminar lakes. During the whole transition, after entering the solitary region, the coherent core chimera is normally surrounded by a solitary cloud formed by isolated oscillators randomly distributed in the spiraling domain of the pattern, exhibiting different average frequencies.  The solitary cloud, first sparse, becomes denser with a deeper penetration inside the solitary region. We expect that this transition indicates a common, probably universal phenomenon in networks of very different nature, due to Newtonian dynamics.

\section*{Acknowledgements}
We thank E. Knobloch and  A. Pikovsky for illuminating  discussions, and the unknown Referees for the constructive criticism which helped us to improve the clarity of the presentation.  We also thank the Ukrainian Grid Infrastructure for providing the computing cluster resources and the parallel and distributed software used this work.

\section*{Authors' contribution}
V.M. and Y.M.  supervised the study and guided the investigation.
V.M. and O.S. performed the numerical simulations. V.M. and Y.M. interpreted the results and wrote the manuscript.
 All authors contributed to editing and revising the manuscript.

\end{document}